\documentclass[12pt,preprint]{aastex}

\shorttitle{Multiwavelength Observations of Mrk\,421}
\shortauthors{Whipple}

\begin{document}
\bibliographystyle{plainnat}
\title{Multiwavelength Observations of Markarian 421 in 2005\,-\,2006}

\author{D. Horan,$^{1}$
V. A. Acciari,$^{2,3}$
S. M. Bradbury,$^{4}$
J. H. Buckley,$^{5}$
V. Bugaev,$^{5}$
K. L. Byrum,$^{1}$
A. Cannon,$^{6}$
O. Celik,$^{7}$
A. Cesarini,$^{8}$
Y. C. K. Chow,$^{7}$
L. Ciupik,$^{9}$
P. Cogan,$^{10}$
A. D. Falcone,$^{11}$
S. J. Fegan,$^{7}$
J. P. Finley,$^{12}$
P. Fortin,$^{13}$
L. F. Fortson,$^{9}$
D. Gall,$^{12}$
G. H. Gillanders,$^{8}$
J. Grube,$^{6}$
G. Gyuk,$^{9}$
D. Hanna,$^{10}$
E. Hays,$^{1,14,15}$
M. Kertzman,$^{16}$
J. Kildea,$^{3}$
A. Konopelko,$^{17}$
H. Krawczynski,$^{5}$
F., Krennrich,$^{18}$
M. J. Lang,$^{8}$
K. Lee,$^{5}$
P. Moriarty,$^{2}$
T. Nagai,$^{18}$
J. Niemiec,$^{18,19}$
R. A. Ong,$^{7}$
J. S. Perkins,$^{3}$
M. Pohl,$^{18}$
J. Quinn,$^{6}$
P. T. Reynolds,$^{20}$
H. J. Rose,$^{4}$
G. H. Sembroski,$^{12}$
A. W. Smith,$^{1}$
D. Steele,$^{9}$
S. P. Swordy,$^{14}$
J. A. Toner,$^{8}$
V. V. Vassiliev,$^{7}$
S. P. Wakely,$^{14}$
T. C. Weekes,$^{3}$
R. J. White,$^{4}$
D. A. Williams,$^{21}$
M. D. Wood,$^{7}$
B. Zitzer,$^{12}$
\\
and
\\
H. D. Aller,$^{22}$ 
M. F. Aller,$^{22}$ 
M. Baker,$^{23}$ 
D. Barnaby,$^{24}$
M. T. Carini,$^{24}$ 
P. Charlot,$^{25,26}$ 
J. P. Dumm,$^{23}$ 
N. E. Fields,$^{23}$ 
T. Hovatta,$^{27}$
B. Jordan,$^{28}$
Y. A. Kovalev,$^{29}$ 
Y. Y. Kovalev,$^{29,30}$ 
H. A. Krimm,$^{31,32}$
O. M. Kurtanidze,$^{33}$
A. L\"ahteenm\"aki,$^{27}$ 
J. F. Le Campion,$^{25,26}$
J. Maune,$^{24}$
T. Montaruli,$^{23}$
A. C. Sadun,$^{34}$ 
S. Smith,$^{24}$
M. Tornikoski,$^{27}$
M. Turunen,$^{27}$
R. Walters,$^{24}$
}

\email{deirdre.horan@gmail.com}

\altaffiltext{1}{High Energy Physics Division, Argonne National Laboratory, 9700 S. Cass Avenue, Argonne, IL 60439, USA}
\altaffiltext{2}{Department of Physical and Life Sciences, Galway-Mayo Institute of Technology, Dublin Road, Galway, Ireland}
\altaffiltext{3}{Fred Lawrence Whipple Observatory, Harvard-Smithsonian Center for Astrophysics, P.O. Box 97, Amado, AZ 85645-0097, USA}
\altaffiltext{4}{School of Physics and Astronomy, University of Leeds, Leeds, LS2 9JT, UK}
\altaffiltext{5}{Department of Physics, Washington University, St. Louis, MO 63130, USA}
\altaffiltext{6}{School of Physics, University College Dublin, Belfield, Dublin 4, Ireland}
\altaffiltext{7}{Department of Physics and Astronomy, University of California, Los Angeles, CA 90095, USA}
\altaffiltext{8}{School of Physics, National University of Ireland, Galway, Ireland}
\altaffiltext{9}{Astronomy Department, Adler Planetarium and Astronomy Museum, Chicago, IL 60605, USA}
\altaffiltext{10}{Physics Department, McGill University, Montr\'{e}al, QC H3A 2T8, Canada}
\altaffiltext{11}{Department of Astronomy and Astrophysics, 525 Davey Lab., Penn. State University, University Park, PA 16802, USA}
\altaffiltext{12}{Department of Physics, Purdue University, West Lafayette, IN 47907, USA}
\altaffiltext{13}{Department of Physics and Astronomy, Barnard College, Columbia University, NY 10027, USA}
\altaffiltext{14}{Enrico Fermi Institute, University of Chicago, Chicago, IL 60637, USA}
\altaffiltext{15}{N.A.S.A./Goddard Space-Flight Center, Code 661, Greenbelt, MD 20771, USA}
\altaffiltext{16}{Department of Physics and Astronomy, DePauw University, Greencastle, IN 46135-0037, USA}
\altaffiltext{17}{Physics Department, Pittsburg State University, 1701 South Broadway, Pittsburg, KS 66762, USA}
\altaffiltext{18}{Department of Physics and Astronomy, Iowa State University, Ames, IA 50011, USA}
\altaffiltext{19}{Instytut Fizyki J\c{a}drowej PAN, ul. Radzikowskiego 152, 31-342 Krak\'ow, Poland}
\altaffiltext{20}{Department of Applied Physics and Instrumentation, Cork Institute of Technology, Bishopstown, Cork, Ireland}
\altaffiltext{21}{Santa Cruz Institute for Particle Physics and Department of Physics, University of California, Santa Cruz, CA 95064, USA}
\altaffiltext{22}{Department of Astronomy, University of Michigan, Ann Arbor, MI 48109-1042, USA}
\altaffiltext{23}{University of Wisconsin, Madison, WI 53706, USA}
\altaffiltext{24}{Western Kentucky University, 1 Big Red Way, Bowling Green, KY 42104, USA}
\altaffiltext{25}{Universit\'e de Bordeaux, Observatoire Aquitain des Sciences de l'Univers, 2 rue de l'Observatoire, BP 89, 33271 Floirac Cedex, France}
\altaffiltext{26}{CNRS, Laboratoire d'Astrophysique de Bordeaux -- UMR 5804, 2 rue de l'Observatoire, BP 89, 33271 Floirac Cedex, France}
\altaffiltext{27}{Mets\"ahovi Radio Observatory, TKK, Helsinki University of Technology, Mets\"ahovintie 114, 02540 Kylm\"al\"a, Finland}
\altaffiltext{28}{School of Cosmic Physics, Dublin Institute For Advanced Studies, Dublin, Ireland}
\altaffiltext{29}{Astro Space Center of Lebedev Physical Institute, Profsoyuznaya 84/32, 117997 Moscow, Russia}
\altaffiltext{30}{Max-Planck-Institut f\"ur Radioastronomie, Auf dem H\"ugel 69, 53121 Bonn, Germany}
\altaffiltext{31}{CRESST and NASA Goddard Space Flight Center, Greenbelt, MD 20771, USA}
\altaffiltext{32}{Universities Space Research Association, 10211 Wincopin Circle, Suite 500, Columbia, MD 21044-34342, USA}
\altaffiltext{33}{Abastumani Observatory, 383762 Abastumani, Georgia}
\altaffiltext{34}{Department of Physics, University of Colorado Denver, Denver, CO  80208, USA}

\begin{abstract} 

Since September 2005, the Whipple 10m Gamma-ray Telescope has been
operated primarily as a blazar monitor. The five Northern Hemisphere
blazars that have already been detected at the Whipple Observatory,
Markarian 421, H1426+428, Markarian 501, 1ES\,1959+650 and
1ES\,2344+514, are monitored routinely each night that they are
visible. We report on the Markarian 421 observations taken from
November 2005 to June 2006 in the gamma-ray, X-ray, optical and radio
bands. During this time, Markarian 421 was found to be variable at all
wavelengths probed. Both the variability and the correlations among
different energy regimes are studied in detail here. A tentative
correlation, with large spread, was measured between the X-ray and
gamma-ray bands, while no clear correlation was evident among the
other energy bands. In addition to this, the well-sampled spectral
energy distribution of Markarian 421 (1101+384) is presented for three
different activity levels. The observations of the other blazar
targets will be reported separately.

\end{abstract}

\keywords{gamma rays: blazars --- BL Lacertae Objects: individual
(Markarian 421) --- gamma rays: observations}

\section{Introduction}
\label{intro}

Among active galactic nuclei (AGN), blazars are the most extreme and
powerful sources known and are believed to have their jets more
aligned with the line of sight than any other class of radio loud AGN.
They are high luminosity objects, characterized by large, rapid,
irregular amplitude variability in all accessible spectral bands. They
have a core-dominated radio morphology with flat ($\alpha_r$,
$\propto \nu^{-\alpha_r}$ $\lesssim$ 0.5) radio spectra, which join
smoothly to the infra-red (IR), optical and ultra-violet spectra. In
all of these bands the flux exhibits high and variable
polarization. Blazars are visible across the entire electromagnetic
spectrum having a broad continuum extending from the radio through the
gamma-ray regime. Although much is known about the characteristics of
their broadband emission, there are still many uncertainties about the
underlying blazar emission mechanisms and many different models can
explain their observed properties (\citealp{Boettcher:07};
\citealp{Sambruna:07}). Their high variability and their broadband
emission make long term, well-sampled, multiwavelength observations of
blazars very important for constraining and understanding their
emission mechanisms and characteristic timescales.

When plotted as $\nu$F$_{\nu}$ versus frequency, the blazar spectral
energy distribution (SED) has a double-peaked structure. Both peaks
are found to vary, often both in strength and in peak frequency, as
the activity level of the blazar changes. The first peak is usually
referred to as the synchrotron peak because in both leptonic and
hadronic models for blazar emission, it is believed generally to be
the result of incoherent synchrotron emission from relativistic
electrons and positrons, presumed to be present in the magnetic fields
of the jet. The origin of the second peak, usually referred to as the
inverse-Compton peak, is less well determined. In synchrotron
self-Compton (SSC) models it is assumed that the synchrotron photons
are up-scattered to higher energies by the electrons while in external
Compton (EC) models, these seed photons can come from the accretion
disk, the broad-line region, the torus, the local infrared background,
the cosmic microwave background, the ambient photons from the central
accretion flow or some combination of these sources. Hadronic models
have also been invoked to explain the broadband spectra of blazars
(e.g. \citealp{Mannheim:93}; \citealp{Mucke:03}). \citet{Aharonian:00}
proposed that the X-ray to gamma-ray emission is synchrotron radiation
from protons accelerated in highly magnetized compact regions of the
jet. Other authors propose that proton-proton collisions, either
within the jet itself or between the jet and ambient clouds, give rise
to neutral pions which then decay to gamma rays (\citealp{Dar:97};
\citealp{Beall:99}; \citealp{Pohl:00}).

As of this writing, 22 of the high-confidence sources in the rapidly
growing catalog of very-high-energy (VHE; E $>$ 100 GeV) sources are
AGN\footnote{http://tevcat.uchicago.edu}. All but one of these, the
Fanaroff-Riley Type I galaxy M87, are blazars.  Of the blazars, one is
a flat spectrum radio quasar, three are low-frequency-peaked BL Lacs
and the remainder are high-frequency-peaked BL Lacs (HBLs). Five of
the TeV blazars have been detected with the Whipple 10\,m Gamma-ray
Telescope. The VHE emission from four of these blazars was first
discovered at the Whipple Observatory, which has been in operation
with an imaging camera from 1982 until the present day (Kildea et
al. 2007).


Imaging atmospheric Cherenkov telescopes (IACTs), such as the Whipple
Telescope, have sufficient sensitivity to sample the short-term
variability of blazars. However, IACT duty cycles are low ($\sim$
10\%) because they can only operate on clear, moonless
nights. Typically, less than 1000 hours per year are available for
observing and therefore, the observing programs are
over-subscribed. In September 2005, the first telescopes of the
VERITAS array \citep{Weekes:01:VERITAS} came on line and the observing
program at the Whipple Telescope was redefined to dedicate it almost
exclusively to nightly blazar monitoring. Since that time, Markarian
421, H1426+428, Markarian 501, 1ES\,1959+650 and 1ES\,2344+514 have
been monitored routinely each night that they are visible (dark clear
skies, elevation $>$ 55$^\circ$, when possible). For the first time,
this has provided the opportunity to obtain long-term and well-sampled
VHE light-curves of these highly variable objects. Part of the
motivation for these observations was to provide a trigger for more
sensitive VHE observations of these AGN by the new generation of IACT
telescopes (CANGAROO-III, HESS, MAGIC and VERITAS) and to provide
baseline observations for similar observations with GLAST.

The results of the monitoring campaign on Markarian\,421 (Mrk\,421;
1101+384) are described here. At a redshift of $z$=0.031, Mrk\,421 was
the first source of VHE gamma rays to be discovered
\citep{Punch:92:Mrk421}. Since that pivotal discovery, Mrk\,421 has
been studied extensively in the VHE band and many periods of intense
variability have been observed. Mrk\,421 is the archetypal
HBL. Historically, it has exhibited shifts in both the peak frequency
and in the power of the first and second components of the SED
\citep{Fossati:08}. In April 2001, Mrk\,421 entered an extremely
active phase. It was observed intensively with the Whipple 10m
Telescope during this time and, for the first time in a HBL, the
spectrum in the VHE regime was found to harden as the flux of VHE
gamma rays increased (\citealp{Krennrich:02}; \citealp{Fossati:08}).

Often, for multiwavelength (MWL) campaigns in which VHE telescopes
participate, the call for full multiwavelength coverage is invoked
when large flares are observed in the TeV band
(\citealp{Krawczynski:00}; \citealp{Krawczynski:04};
\citealp{Rebillot:06}; \citealp{Fossati:08}). Like the observations
described here, the campaign carried out by \citet{Blazejowski:05} in
2003\,-\,2004 was not instigated due to flaring activity from
Mrk\,421. That campaign, which concentrated mainly on the X-ray and
gamma-ray data, revealed the correlation between the emission in the
two wavebands to be fairly loose. Rapid variability was detected in
both wavebands, and the X-ray emission was found to peak days after
the gamma-ray emission for one giant flare. Evidence for TeV orphan
flaring, similar to that found in 1ES\,1959+650 by
\citet{Krawczynski:04}, was also seen. In addition to significant
X-ray and gamma-ray coverage, the dataset described here includes
excellent coverage of Mrk\,421 in the optical and radio bands.

To encourage and coordinate observations of the target AGN at other
wavelengths, a web page containing the observing timetable and the
preliminary light curves at TeV wavelengths was made publicly
accessible and continuously maintained via the Whipple link on the
VERITAS webpages: http://veritas.sao.arizona.edu. Emails were
distributed throughout 2005\,-\,2006 to coordinate the Mrk\,421
observations. Data from twelve different wavebands, spanning more than
18 orders of magnitude in energy, were obtained over a period of
approximately 230 days. Because this campaign was not triggered by
activity at any wavelength, there were no {\it{a priori}} assumptions
about flux levels and we can look for correlations on longer
timescales and under different source conditions. Since these Mrk\,421
data were gathered over a long period of time and, in the case of some
wavebands, from a number of different instruments for a particular
energy range, it was important to ensure that the data were
well-calibrated so that the variability of Mrk\,421 over the course of
the campaign was not contaminated by inaccurate intra-night and/or
intra-telescope normalizations.

In Section~\ref{observation} we describe the observations presented in
this paper. This section is subdivided by wavelength band. In
Section~\ref{results} the results of the observations are presented
while in Section~\ref{discussionAndConclusion} we discuss the
implications of these results in the context of emission models.

\section{Observations and Data Reduction}
\label{observation}

\begin{figure}
\resizebox*{0.853\textwidth}{!}{\includegraphics[draft=false]{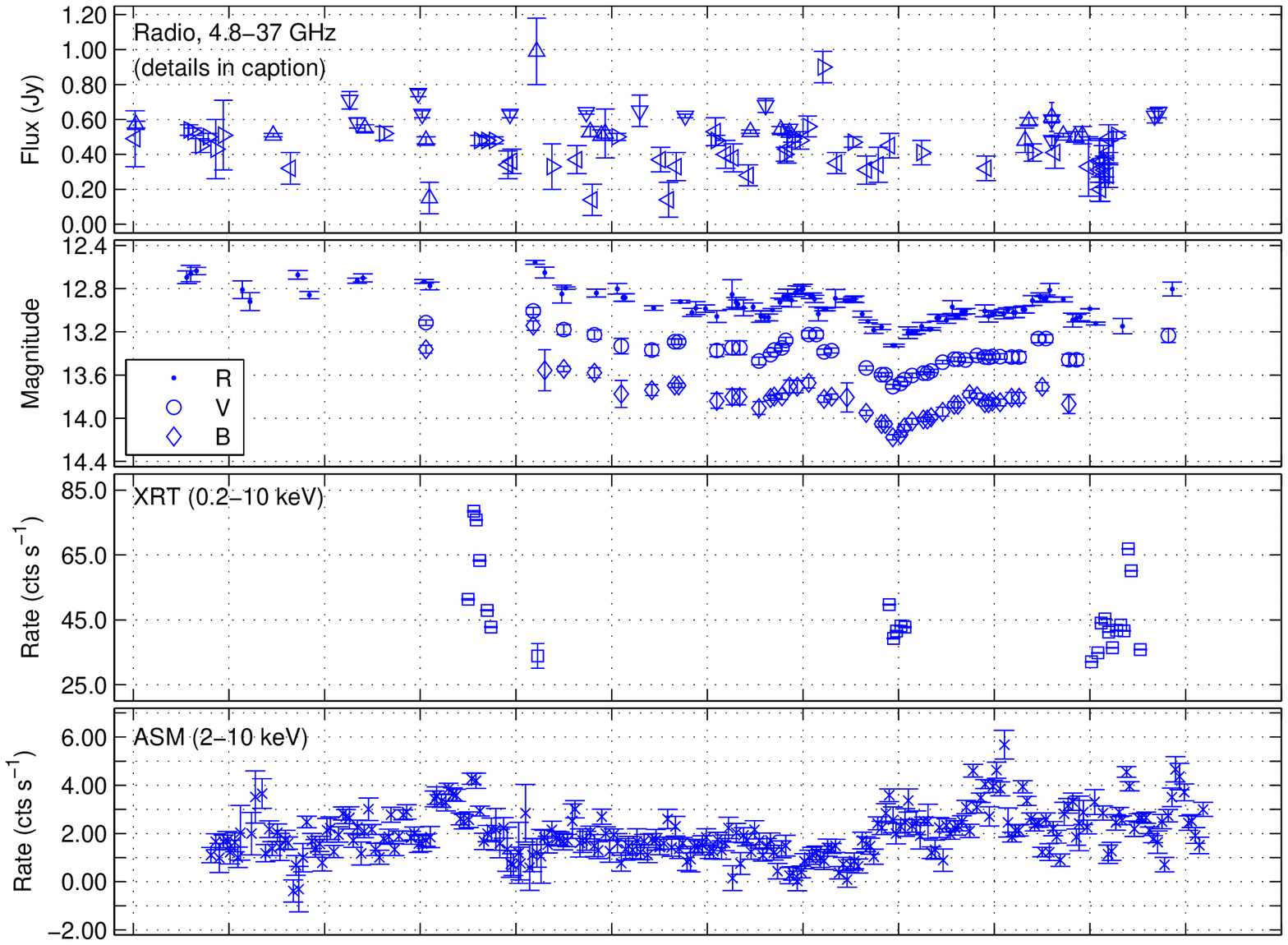}}\\
\resizebox*{0.85\textwidth}{!}{\includegraphics[draft=false]{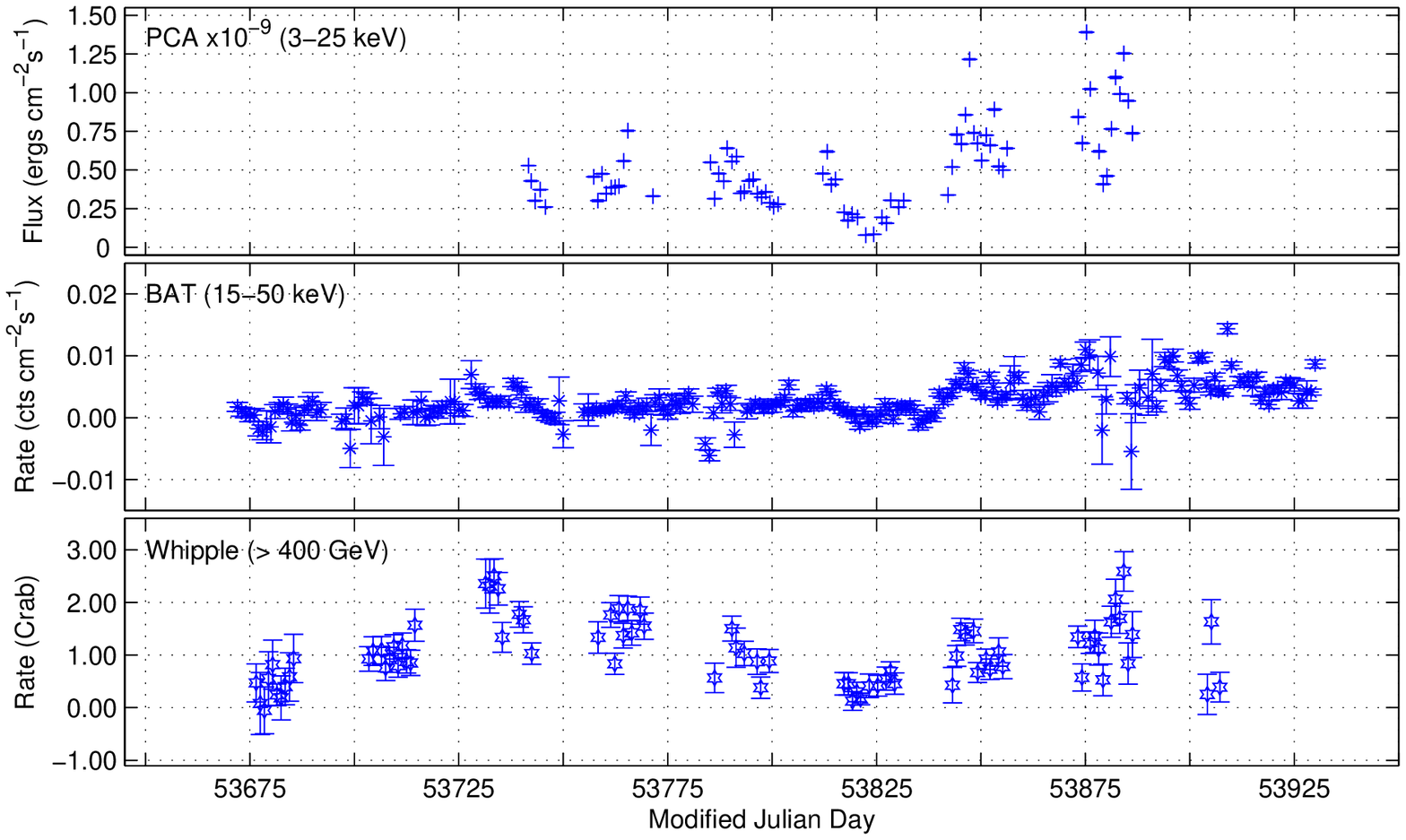}}\\
\caption{\label{FIG:ENTIRE-DATASET}The entire database, as summarized
in Table~\ref{TAB:ALL}. Each data point represents all data obtained
on that night as part of this campaign. The radio data at four
different frequencies are plotted in the top panel: 4.8 GHz
($\bigtriangledown$), 8 GHz ($\bigtriangleup$), 14.5 GHz
($\triangleright$) and 37 GHz ($\triangleleft$). The optical data are
combined from many different observatories (Table~\ref{TAB:OPTICAL}).}
\end{figure}

The complete dataset presented in this paper is summarized in
Table~\ref{TAB:ALL} and the lightcurves are shown together in
Figure~\ref{FIG:ENTIRE-DATASET}. In the following section, the data
collection procedures and the Mrk\,421 data gathered at each of the
instruments are described.

\subsection{VHE Gamma-Ray Data}
\label{gamma-ray}

All the TeV gamma-ray observations presented here were made with the
10\,m Gamma-ray Telescope at the Fred Lawrence Whipple Observatory
\citep{Kildea:07}. Although sensitive in the energy range from
200\,GeV to 20\,TeV, the peak response energy of the telescope to a
Crab-like spectrum during the observations reported upon here was
approximately 400\,GeV. This is the energy at which the telescope is
most efficient at detecting gamma rays and, as is typical for IACTs
such as Whipple, is subject to a 20\% uncertainty.

The Mrk\,421 data were analyzed using the imaging technique and
analysis procedures developed by the Whipple Collaboration
(\citealp{Hillas:85}; \citealp{Reynolds:93}). A Quicklook analysis was
performed at the conclusion of each observation and the results were
posted daily on a public web
page\footnote{http://veritas.sao.arizona.edu/content/blogsection/6/40}.
Three offline analyses (independent but using the standard Supercuts
methodology) were used to derive the TeV light-curves presented here.

Two different modes of observation are employed at the Whipple
Telescope, ``{\it{On\,-\,Off}}'' and ``{\it{Tracking}}''
\citep{Catanese:98}. In both modes, the data are usually taken in
28-minute scans. Unlike data taken in the {\it{On\,-\,Off}} mode, the
scans taken in the {\it{Tracking}} mode do not have independent
control data which can be used to establish the background level of
gamma-ray like events during the scan. These control data are
essential in order to estimate the number of events passing all cuts
that would have been detected during the scan in the absence of the
candidate gamma-ray source. In order to perform this estimate, a
tracking ratio is calculated by analyzing ``darkfield data''
\citep{Horan:02}. These consist of {\it{Off-Source}} data taken in the
{\it{On\,-\,Off}} mode and of observations of objects found not to be
sources of gamma rays. A large database of these scans is analyzed
and, in this way, the background level of events passing all gamma-ray
selection criteria can be characterized as a function of zenith angle.
The data presented here were mostly taken in the {\it{Tracking}} mode.

During the period MJD 53676 to MJD 53908, a total of 144.1 hours of
data were taken on Mrk\,421. Since IACTs can only operate during
moonless conditions, there is a period centered on full moon each
month when gamma-ray observations are not possible. From the total of
328 runs on the source ({\it{On}} and {\it{Tracking}} modes), 275 runs
(84\%), comprising 122.57 hours, survived the quality selection
process. Apart from any problems reported by observers in the nightly
logs, the main parameter used to determine whether the data run should
be included in the final analysis was the stability of the raw trigger
rate. Any data run whose raw trigger rate deviated significantly from
being steady for its duration was discarded.

\begin{figure}
\resizebox*{0.95\textwidth}{!}{\includegraphics[draft=false]{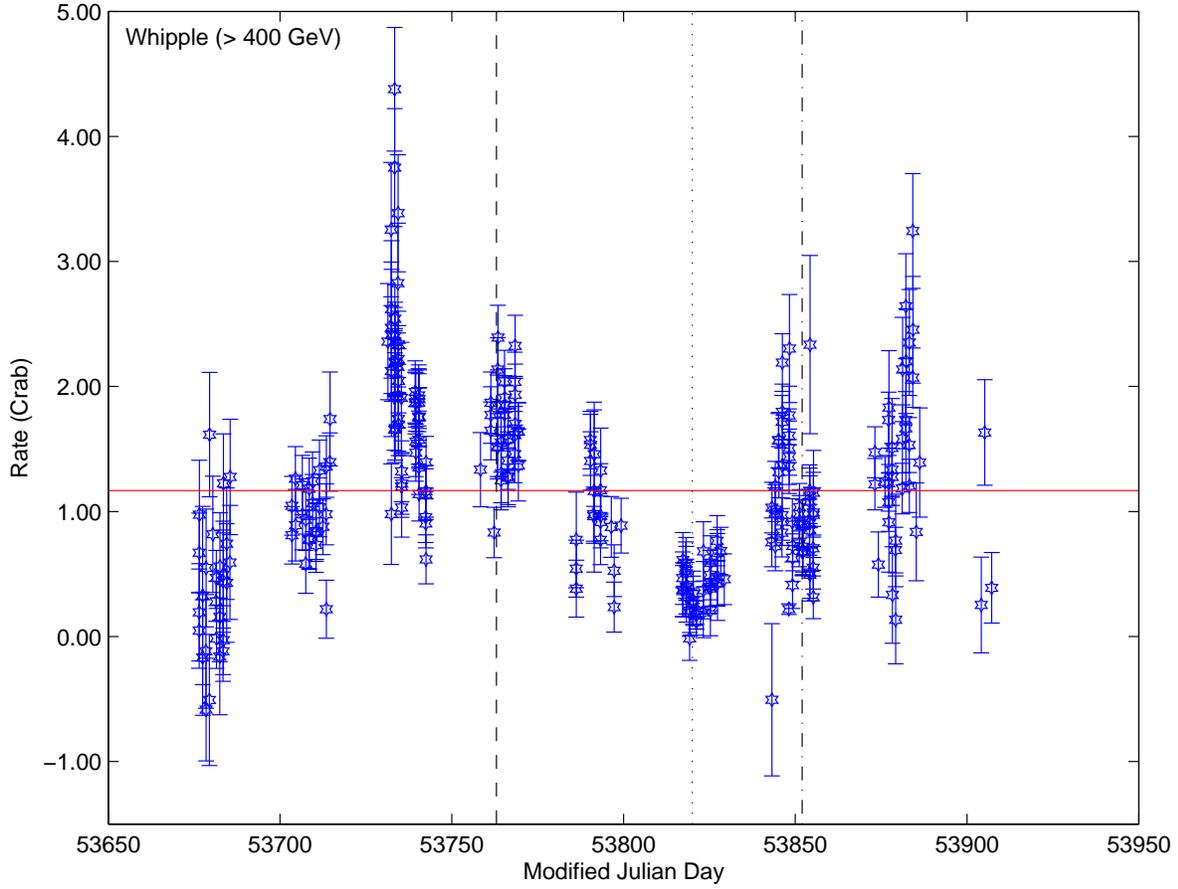}}\\%
\caption{\label{FIG:GAMMA}The gamma-ray lightcurve, with each exposure
(typically 28-minute) binned separately, for the duration of the
observing campaign. The nights for which the SED was calculated are
marked: low state by the dotted line; medium state by the dash-dotted
line; high state by the dashed line. The horizontal red line shows the
mean rate detected during a single exposure on Mrk\,421.}
\end{figure}

When possible, the AGN data were interspersed with observations of the
Crab Nebula, the standard candle in this energy regime. A total of
44.41 hours of these were taken and subjected to the same analysis
procedure. In addition, nightly runs were taken with the telescope
pointing to the zenith; these data were used to calculate the
telescope throughput \citep{LeBohec:03} so that data could be
corrected for inter-nightly changes in atmospheric transparency. Both
the AGN and the Crab data were corrected for variations in the
throughput. In addition to this, a correction was applied to
compensate for elevation effects in the data due to the increasing
volume of atmosphere through which the Cherenkov light propagates as
the zenith angle of the observations increases. These corrections
ensured that data taken on different nights, under different
atmospheric conditions and at different elevations could be
compared. The Mrk\,421 rates were then converted into equivalent Crab
rates. It should be noted that this simplistic scaling is strictly
only valid for a TeV spectrum near that of the Crab Nebula (spectral
index of -2.6). Although Mrk\,421 has been known to display spectral
variability in the past, the measurement uncertainties on the flux
points are such that the effect of spectral variability should not be
significant. The gamma-ray lightcurve containing the rate from each
approximately 28-minute scan on Mrk\,421 is shown in
Figure~\ref{FIG:GAMMA}.

\subsection{X-ray Data}
\label{X-ray}

\begin{figure}
\resizebox*{0.95\textwidth}{!}{\includegraphics[draft=false]{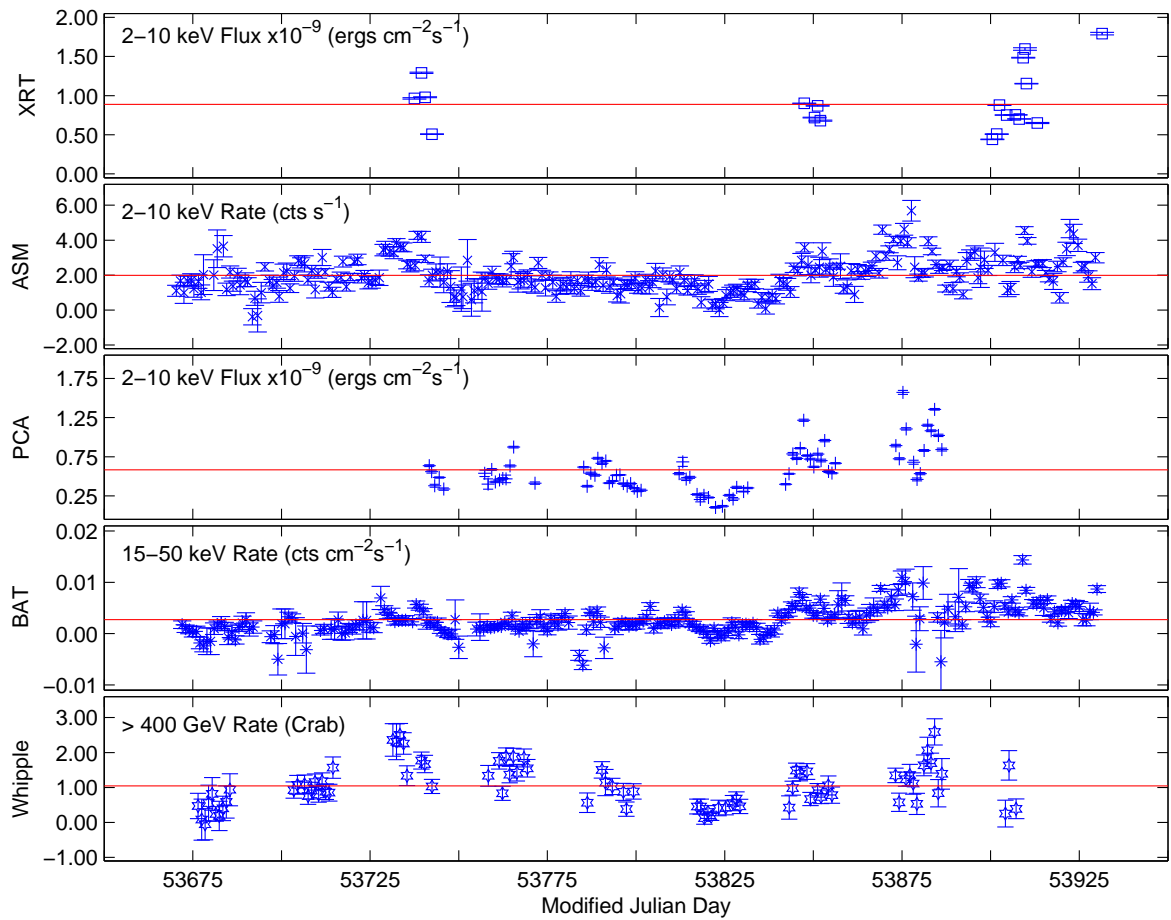}}

\caption{\label{FIG:X-GAMMA}The X-ray and gamma-ray lightcurves.  The
mean flux in each waveband is shown by a horizontal red
line. {\it{Top:}} XRT; {\it{2nd From Top:}} ASM; {\it{Middle:}} PCA;
{\it{2nd from Bottom:}} BAT; {\it{Bottom:}} Gamma-ray. Each data point
represents all data obtained on that night as part of this campaign.}

\end{figure}

The X-ray data for this multiwavelength campaign span the energy range
from 0.2 to 50 keV. These data come from four different instruments,
two on the Rossi X-ray Timing Explorer and two on the Swift
Satellite. The X-ray lightcurves are shown together with the gamma-ray
data in Figure~\ref{FIG:X-GAMMA}. The details of the analysis of each
X-ray dataset are given in the following subsections.

\subsubsection{The All Sky Monitor (ASM)}
\label{asm}

The All-Sky Monitor (ASM), on board the Rossi X-ray Timing Explorer
(RXTE) \citep{Swank:94} operates in the 1.5 to 12 keV energy band and
scans most of the sky every 1.5 hours. It consists of three coded
aperture cameras, scanning shadow cameras, which can be rotated to
view different regions of the sky \citep{Levine:96}. Each scanning
shadow camera is a sealed proportional counter filled with a
xenon-CO$_2$ mixture. The data presented here are the one-day averages
from the ASM quicklook pages
\footnote{http://http://xte.mit.edu/XTE}. Each one-day average data
point represents the one-day average of the fitted source fluxes from
a number (typically 5 to 10) of individual ASM dwells. A total of 256
nights of ASM data on Mrk\,421 are presented here from MJD 53670 - MJD
53930. During this time, the mean nightly flux from Mrk\,421 in the
2-10 keV band was 1.99 counts s$^{-1}$.

\subsubsection{The Proportional Counter Array (PCA)}
\label{pca}

The primary instrument on board NASA's X-ray satellite RXTE is the
Proportional Counter Array (PCA). It is an array of five proportional
counters with a total net area of 6250 cm$^2$ and is effective over
the energy range 2 to 60 keV. Mrk\,421 was observed with the RXTE PCA
from MJD 53741 to MJD 53887 in short daily exposures ranging from 2.1
to 48.0 minutes, with a mean exposure of 19 minutes. The PCA exposures
were scheduled to coincide with the Whipple 10 m telescope
observations, so that the data gaps of $\approx$ 12 days in the X-ray
database coincide with the bright moon periods when no gamma-ray
observations were possible. Due to various observing constraints, the
X-ray and gamma-ray observations were not always simultaneous, with
the PCA observations starting on average 99 minutes behind the Whipple
10 m observations. Data analysis was carried out with the HEAsoft 6.1
tools and current calibration files, following the standard
procedure. The version number is consistent with the Swift analysis.
The PCA data presented here comprise 75 exposures from MJD 53741 to
MJD 53887 with a mean flux of 5.18 x 10$^{-10}$ erg cm$^{-2}$
s$^{-1}$.

\subsubsection{The X-Ray Telescope (XRT)}
\label{xrt}

The X-Ray Telescope (XRT), one of the instruments on the Swift
Satellite \citep{Gehrels:04}, is a focusing X-ray telescope with a 110
cm$^2$ effective area and a 23 arcmin field of view
\citep{Burrows:05}. It is sensitive to X-rays in the 0.2 to 10 keV
band. The XRT data presented here were reduced using the latest
HEAsoft tools (version 6.1.0), including Swift software version 2.0,
and the latest response (version 8) and ancillary response files
(created using xrtmkarf) available in CALDB at the time of analysis.
Data were screened with the standard parameters, including the
elimination of time periods when the CCD temperature was warmer than
-48$^\circ$ C.  Due to the high rates of Mrk\,421 during the XRT
observations, only Windowed Timing (WT) mode data were used in this
analysis, and only grades 0 - 2 were included.  Since the count rate
stayed below $\approx$100 c/s, the WT mode data is free of significant
pile-up effects. The data were corrected for effects due to bad
columns and bad pixels. Source and background regions were both chosen
in a way that avoids overlap with serendipitous sources in the image.

For the analysis presented here, data from the XRT were summed in
one-day bins resulting in a total of 24 measurements from MJD 53737 to
MJD 53913. As can be seen in Figure~\ref{FIG:X-GAMMA}, the XRT
lightcurve is the most sparsely sampled of the X-ray lightcurves.
During all XRT observations reported on here however, Mrk\,421 was
detected above 5$\sigma$. The mean flux from Mrk\,421 during these XRT
observations in the energy range from 0.2 to 10 keV was 47.23 counts
sec$^{-1}$.

\subsubsection{The Burst Alert Telescope (BAT)}
\label{bat}

The Burst Alert Telescope (BAT; \citet{Barthelmy:00}), also on Swift,
is a large field of view (1.4 steradian) X-ray telescope with imaging
capabilities in the energy range from 15 to 150 keV. It has a coded
aperture mask with 0.52\,m$^2$ CdZnTe detectors. The BAT typically
observes 50\% to 80\% of the sky each day and has accumulated light
curves for 530 non-GRB sources \citep{Krimm:06}. The BAT data shown
here are the Swift/BAT transient monitor results provided by the
Swift/BAT team \citep{BAT:lightcurves} and span the energy range from
15 to 50 keV. As part of its on-board GRB monitor, the BAT flight
software produces ``scaled maps'' in the energy band 15 to 50 keV and
on timescales $\geq$ 64 seconds. The data were first verified to
ensure that any corrupted or incomplete data were rejected. Then, they
were analyzed using the standard BAT software tools. Full details of
this analysis are given at the BAT transients webpage
\citep{BAT:analysis}. BAT data are available for 240 nights during
this multiwavelength campaign. When these were filtered to remove data
that were flagged as bad, 234 nights remained. These data were taken
between MJD 53670 and MJD 53930 and, during this time, the mean flux
from Mrk\,421 recorded by the BAT in the energy range from 15 to 50
keV was 2.72 x 10$^{-3}$ count sec$^{-1}$.

\subsection{Optical Data}
\label{optical}

Eight optical observatories contributed data sets to this
campaign. They are the Fred Lawrence Whipple Observatory (FLWO) 1.2 m
telescope (located adjacent to the Whipple 10 m gamma-ray telescope on
Mt. Hopkins), the Tenagra 0.8 m telescope in Tenagra, Arizona, USA,
the Bradford Robotic Telescope in Tenerife, Canary Islands, Spain, the
WIYN 0.9 m telescope on Kitt Peak, Arizona, USA, the 0.7 m telescope
at Abastumani Observatory in Abastumani, Georgia, the 0.6 m Bell
Observatory at Western Kentucky University, USA, Bordeaux Observatory
in Floriac, France, and the 1.05 m REOSC telescope at Osservatorio
Astronomico di Torino, Italy.  Table \ref{TAB:OPTICAL} gives a brief
description of each telescope, its filter system and bandpass, and its
contribution to the optical dataset.

The data from the various observatories were reduced and the
photometry performed independently by different analysts using
different strategies. The FLWO, Bradford, and Tenagra data, for
example, usually consisted of a single or a few images per filter per
night.  Relative aperture photometry was performed using standard
routines in IRAF\footnote{IRAF is written and supported by the IRAF
programming group at the National Optical Astronomy Observatories
(NOAO) in Tucson, Arizona. http://iraf.noao.edu/}. Magnitudes were
calculated with respect to star 1 of Villata \citep{Villata:98} with a
photometry aperture of 10 arcsec.  The host galaxy light was not
subtracted at this stage of the analysis.  Other optical analysts used
slightly different strategies: the photometry for the Bell, WIYN and
Abastumani data were quoted with respect to stars 1-3 of Villata, and
the Torino photometry was performed using Gaussian fitting rather than
by counting within a fixed aperture.

Combining the various optical data to produce a composite lightcurve
for each spectral band is complicated by the fact that different
observatories use different photometric systems. Furthermore,
photometric apertures and the definition of the reported measurement
error for each nightly-averaged flux is inconsistent across
datasets. Therefore we have adopted a simple approach for the
construction of the composite lightcurves whereby a unique flux offset
is found for each spectral band of every instrument based on
overlapping observations \citep{Steele:07}.

\begin{figure}
\resizebox*{0.95\textwidth}{!}{\includegraphics[draft=false]{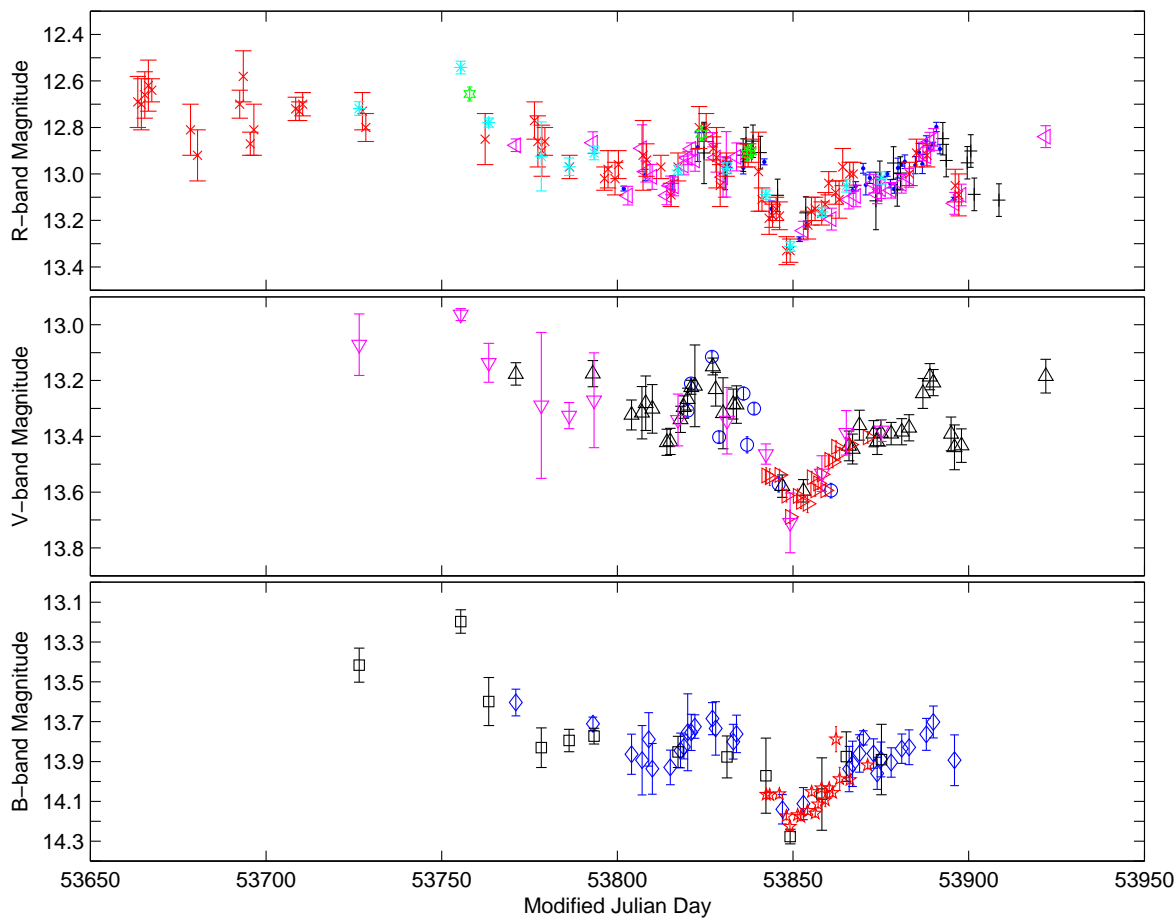}}\\%
\caption{\label{FIG:OPTICAL}{\it{Top:}} The composite R-band
lightcurve with the following legend: Abastamani - blue dots, Bell -
black plus symbols, Bradford - magenta $\triangleleft$ symbols, FLWO -
red x symbols, Tenagra - cyan asterisks, Torini - green hexagrams;
{\it{Middle:}} The composite V-Band lightcurve with the following
legend: Bordeaux - blue circle, Bradford - black $\bigtriangleup$
symbols, WIYN - red $\triangleright$ symbols, Tenagra - magenta
$\bigtriangledown$ symbols; {\it{Bottom:}} The composite B-Band
lightcurve with the following legend: Bradford - blue diamonds,
Tenagra - black squares, WIYN - red pentagrams.}
\end{figure}

\begin{figure}
\resizebox*{0.95\textwidth}{!}{\includegraphics[draft=false]{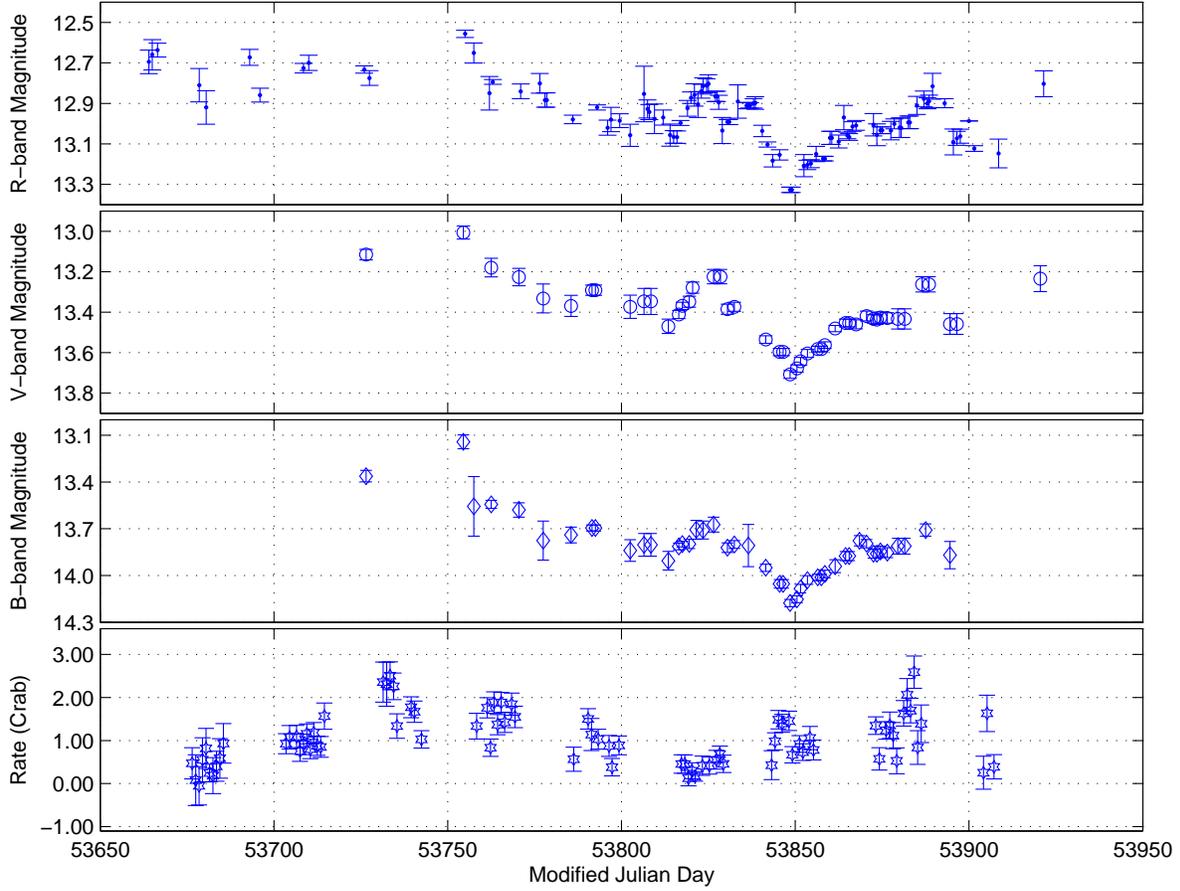}}
\caption{\label{FIG:OPT-GAMMA-LIGHTCURVES}The normalized optical
lightcurves and the gamma-ray lightcurve. The data were binned by MJD:
{\it{Top:}} R-band, {\it{Second from Top:}} V-band, {\it{Second from
Bottom:}} B-band {\it{Bottom:}} gamma-ray.}
\end{figure}

In the R-band, for example, we first find the initial average flux
offset for each R-band lightcurve with respect to that of the FLWO
(since it spans the largest range of dates) based on observations
overlapping by less than 0.8 days (except for the Bell and Torino
lightcurves, whose coincidence windows were widened to 1.8 days to
ensure a sufficient number of overlaps).  Then, for each lightcurve, a
comparison lightcurve is constructed using the lightcurves from the
other observatories with the initial flux offsets applied. A new
offset is then found for each lightcurve based on overlapping
observations in the respective comparison lightcurve. The process is
repeated iteratively until the variance of the offset residuals for
each lightcurve is minimized. The final composite lightcurve for each
spectral band is then made by simply combining the nightly-averaged
flux from each observatory with the final offsets applied and these
are shown in Figure~\ref{FIG:OPTICAL}. Of the three wavelengths, the
R-band is the best-sampled, comprising 166 data points on 97 nights
spanning a period of 260 days. Figure~\ref{FIG:OPT-GAMMA-LIGHTCURVES}
shows the very well-sampled optical lightcurves, after all the data
from different instruments have been combined and normalized, along
with the gamma-ray data during this campaign. The mean magnitude of
Mrk\,421 during this campaign was 12.95 in the R-band, 13.40 in the
V-band and 13.82 in the B-band.

\subsection{Radio}
\label{radio}

\begin{figure}
\resizebox*{0.95\textwidth}{!}{\includegraphics[draft=false]{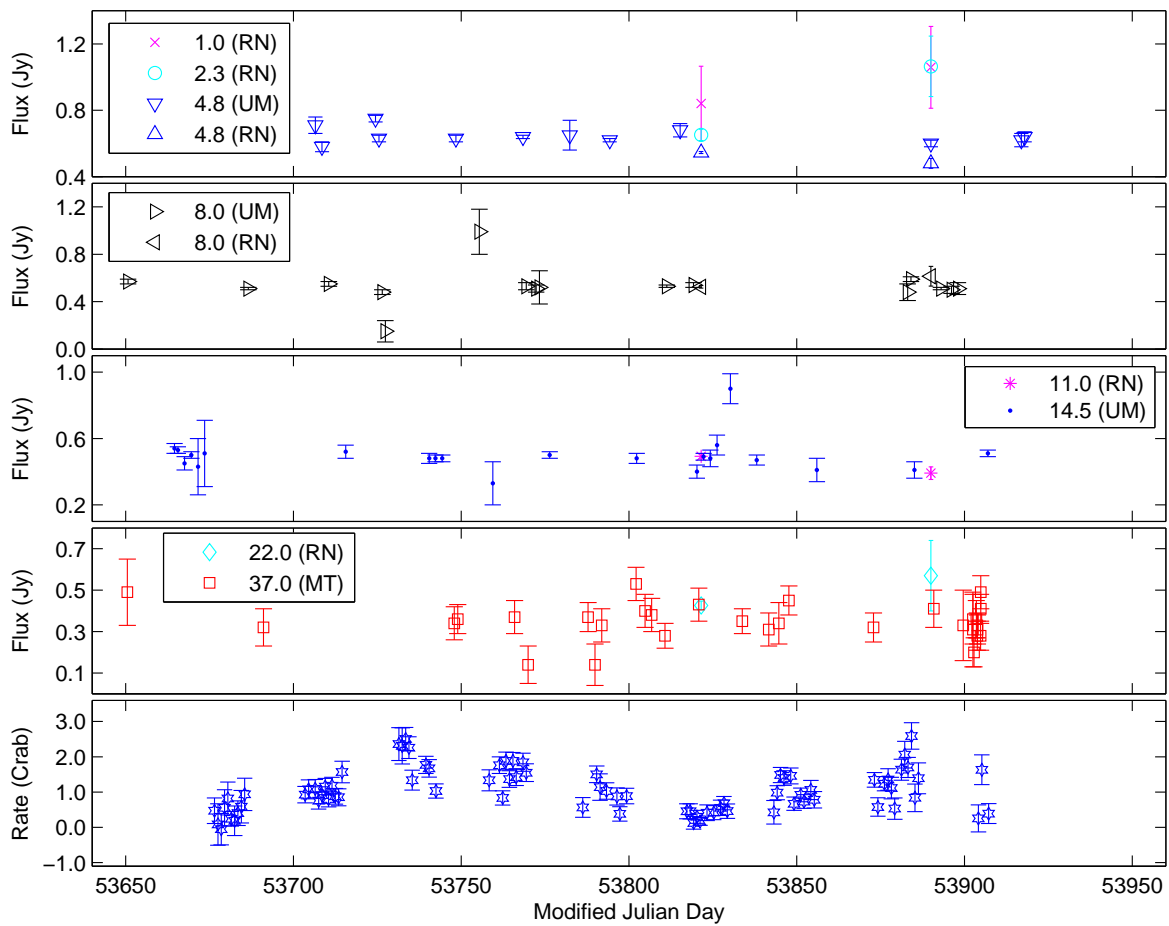}}\\%
\caption{\label{FIG:RADIO}All of the radio data taken during this
observing campaign. {\it{Top:}} The data from 1.0 to 4.8 GHz from
RATAN (RN) and the University of Michigan Radio Astronomy Observatory
(UM); {\it{Second from Top:}} The data at 8.0 GHz from RATAN (RN) and
the University of Michigan Radio Astronomy Observatory (UM);
{\it{Middle:}} The data from 11.0 to 14.5 GHz from RATAN (RN) and the
University of Michigan Radio Astronomy Observatory (UM); {\it{Second
from Bottom:}} The data from 22.0 to 37.0 GHz from RATAN (RN) and
Metsahovi (MT). {\it{Bottom:}} The gamma-ray data from the Whipple 10m
Telescope. Each data point represents all data obtained on that night
as part of this campaign.}
\end{figure}

Typically, the variations in AGN are slower at radio frequencies than
at the higher frequency bands with the most variability being observed
at the higher radio frequencies (see, e.g., \citet{Kovalev:02}). In
general, because of this slower long-term variability, AGN are not
observed with as high temporal coverage in this regime. The radio data
presented here were taken at eight frequencies at three different
radio observatories but, for four of these frequency bands the radio
coverage was very sparse. The fluxes and their associated standard
errors are given in janskys, so they have already been normalized for
the bandwidth of their receivers. Generally, Mrk\,421 is found to vary
on monthly timescales. The radio lightcurves are shown along with the
gamma-ray lightcurve in Figure~\ref{FIG:RADIO}.  In the subsections
that follow, the radio data and analysis from the participating
observatories are described.

\subsubsection{Mets\"ahovi Radio Observatory}

The 37 GHz observations were made with the 13.7 m diameter Mets\"ahovi
radio telescope, which is a radome enclosed paraboloid antenna
situated in Finland (24 23' 38''E, +60 13' 05''). The measurements
were made with a 1 GHz-band dual beam receiver centered at 36.8
GHz. The HEMPT (high electron mobility pseudomorphic transistor) front
end operates at room temperature.  The observations are ON--ON
observations, alternating the source and the sky in each feed horn. A
typical integration time to obtain one flux density data point is
1200--1400 s. The detection limit of the telescope at 37 GHz is on the
order of 0.2 Jy under optimal conditions. Data points with a
signal-to-noise ratio $<$ 4 are handled as non-detections.

The flux density scale is set by observations of DR 21. Sources 3C\,84
and 3C\,274 are used as secondary calibrators. A detailed description
on the data reduction and analysis is given in \citet{Terasranta:98}.
The error estimate in the flux density includes the contribution from
the measurement rms and the uncertainty of the absolute calibration.

\subsubsection{University of Michigan Radio Astronomy Observatory}
Most of the data at 4.8, 8.0 and 14.5 GHz came from the University of
Michigan Radio Astronomy Observatory (UMRAO) using the 26\,m diameter
parabolic reflector. Both on-source and background flux measurements
were performed. To provide improved sensitivity, the receivers have
relatively broad bandwidth of 500, 780 and 2000 MHz at the three
frequencies respectively. The adopted flux density scale is based upon
the flux of Cassiopeia A, the primary flux calibrator, using its
measured decay rate at centimeter wavelengths. A grid of secondary
flux calibrators distributed in RA are also observed. In order to
correct for environmental variations, a calibration source was
observed approximately every 1-2 hours. The integration times at UMRAO
are typically 30 minutes and measurements are usually taken within
$\pm$2.5 hours of the meridian.

\subsubsection{Radio Astronomical Telescope of the Academy of Sciences}

The 1--22~GHz instantaneous radio spectrum of Mrk\,421 was observed
with the 600-meter ring radio telescope RATAN-600 \citep{Korolkov:79}
of the Special Astrophysical Observatory, Russian Academy of Sciences,
located in Zelenchukskaya, Russia, on March 26, 27, \& 28, and May 3
\& 4, 2006.  The continuum spectrum was measured every time
quasi-simultaneously (within several minutes) in a transit mode at six
different bands with the following central frequencies (and
bandwidths): 0.95~GHz (0.03~GHz), 2.3~GHz (0.25~GHz), 4.8~GHz
(0.6~GHz), 7.7~GHz (1.0~GHz), 11.2~GHz (1.4~GHz), 21.7~GHz (2.5~GHz).
Details on the method of observation, data processing, and amplitude
calibration are described in \citep{Kovalev:99}. The March data were
collected using the Northern sector of the telescope while the June
spectrum was observed at the Southern sector with the Flat
reflector. Since no significant time variations were found during
three days in March and two days in May, two averaged RATAN total flux
density spectra for March and May 2006 are provided and used in this
paper.

\subsubsection{VLBA: MOJAVE program}

\begin{figure}
\resizebox*{0.95\textwidth}{!}{\includegraphics[draft=false]{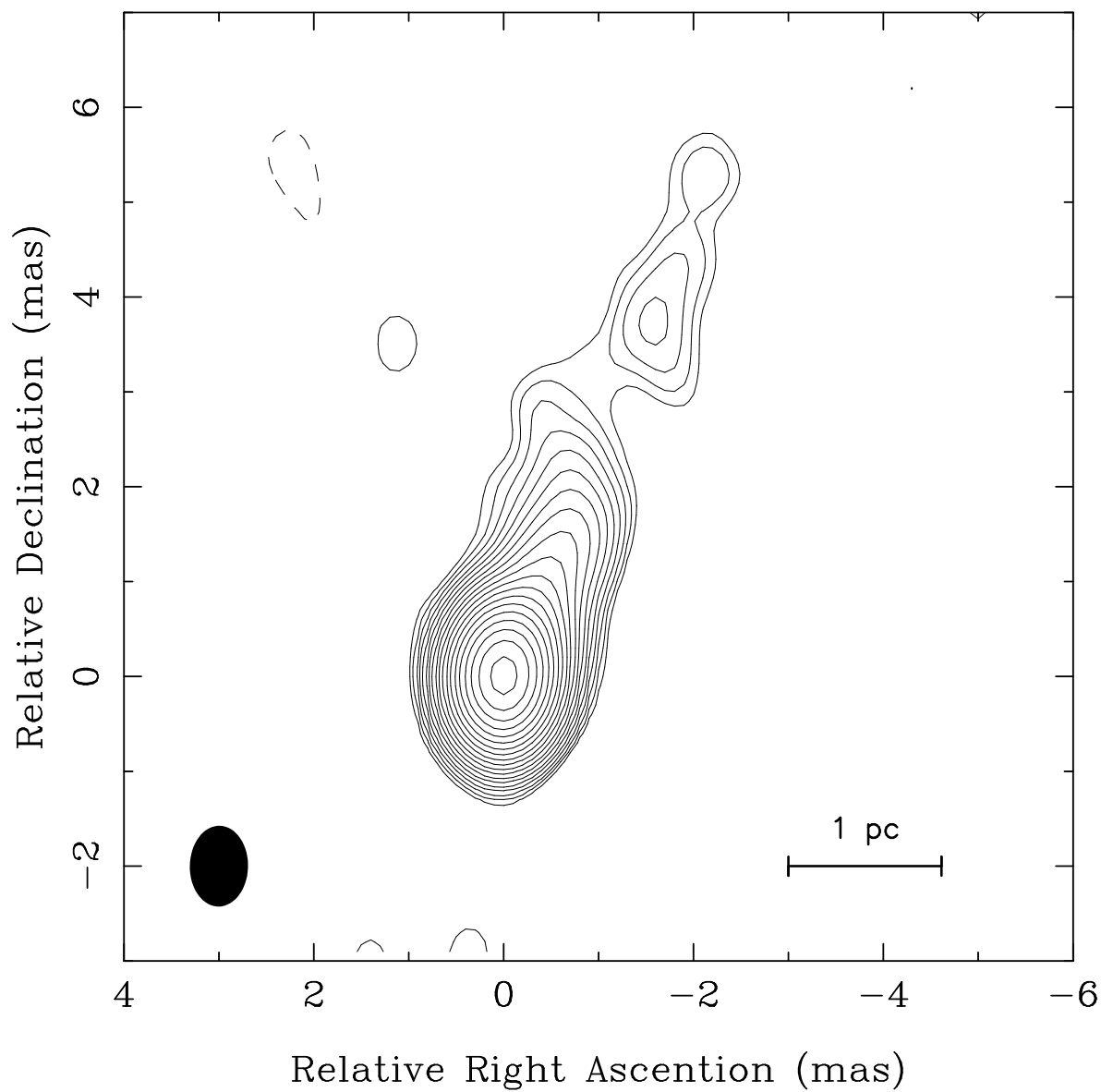}}\\%

\caption{\label{FIG:VLBI}Naturally weighted VLBA Stokes I 15.3 GHz
image of Mrk\,421. The epoch of observation is 5\,April\,2006
(MJD\,53830). The contours are plotted in successive powers of
$\sqrt{2}$ times the lowest contour of 0.5\,mJy. The peak intensity is
296~mJy\,beam$^{-1}$, The synthesized beam is shown in the left
corner. This plot is made from an image FITS file provided by the
2\,cm survey / MOJAVE programs database (\citealp{Kellermann:04};
\citealp{Lister:05}).}
\end{figure}

In addition to these radio lightcurves, the data from the 2\,cm
VLBA/MOJAVE monitoring program (\citealp{Kellermann:04};
\citealp{Lister:05}) for the epoch 05\,April\,2006 (MJD\,53830) were
used (see the 15\,GHz VLBA image in Figure~\ref{FIG:VLBI}). The total
flux density (Stokes I) at 15 GHz integrated over this image is 336
mJy. The flux density from the core region, when it is modeled using a
circular Gaussian, is 288 mJy so Mrk\,421 was very core dominated at
this time implying that the bulk of the radio-band variability is
associated with the VLBI-imaged core. The angular size of the modeled
core on the half power level is 0.046 milliarcseconds. The core size
resolution limit was estimated for this dataset following
\citep{Kovalev:05} and appeared to be less than the measured core size
value. The linearly polarized flux density is 6 mJy. This polarized
flux is detected from the core region only. VLBI core brightness
temperature in the source frame is estimated to be 8$\times$10$^{11}$~K.

\section{Results}
\label{results}

The large, well-sampled multiwavelength lightcurves collected on
Mrk\,421 during 2005\,-\,2006 enabled us to carry out a number of
different analyses, which are described in the following
subsections. The variability across the entire spectrum was
investigated by computing both the fractional rms variability
amplitude and the point-to-point fractional variability amplitude for
each of the twelve wavebands. For the better-sampled lightcurves,
namely those in the optical, X-ray and gamma-ray wavebands, a more
detailed variability analysis was undertaken. Correlations were sought
between the emission at different energies and the discrete
correlation function was computed for the optical-gamma and the
X-ray-gamma datasets. Three different nights were chosen, when
Mrk\,421 was deemed to be in a high, medium and low emission state,
and broadband spectral energy distributions were constructed for each
of these nights. The high-flux spectral energy distribution was
compared with those from archival Mrk\,421 observations.

\subsection{Variability Amplitude}

The fractional rms variability amplitude, F$_{var}$ and the
point-to-point fractional rms variability amplitude, F$_{pp}$,
\citep{Zhang:05} were calculated for each of the energy bands where
a significant amount of data was gathered. They are defined as:

\begin{center}
\begin{equation}\label{EQ:FVAR}
F_{var} = \sqrt{ \frac{S^2 - {\overline{\sigma^2}}}{{\bar{F}^2} }}
\end{equation}
\end{center}

\noindent and

\begin{center}
\begin{equation}\label{EQ:FPP}
F_{pp} = \frac{1}{\bar{F}} \sqrt{\frac{1}{2(N-1)} \sum_{i=1}^{N-1} (F_{i+1} - F_i)^2 - \overline{\sigma^2}}
\end{equation}
\end{center}

\noindent where each flux measurement F$_i$ has a measurement error
$\sigma_i$, $\bar{F}$ is the arithmetic mean of the counts and:

\begin{center}
\begin{displaymath}
S^2 = \frac{1}{N-1} \sum_{i=1}^{N} (F_i - \bar{F})^2
\end{displaymath}
\end{center}

\noindent and $\overline{\sigma^2}$ is the mean error squared:

\begin{center}
\begin{displaymath}
\overline{\sigma^2} = \frac{1}{N} \sum_{i=1}^{N} \sigma_i^2
\end{displaymath}
\end{center}

\begin{figure}
\resizebox*{0.45\textwidth}{!}{\includegraphics[draft=false]{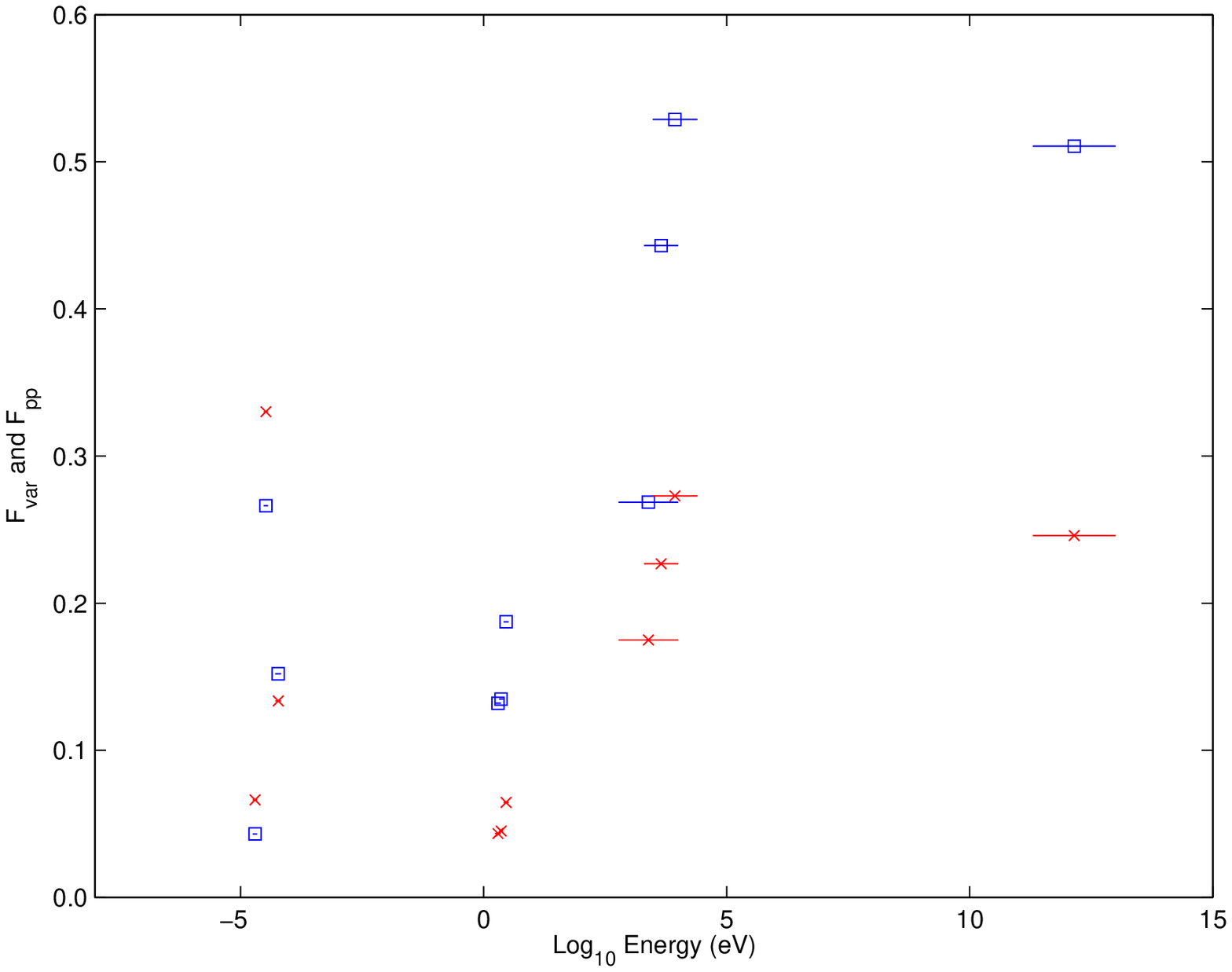}}%
\resizebox*{0.45\textwidth}{!}{\includegraphics[draft=false]{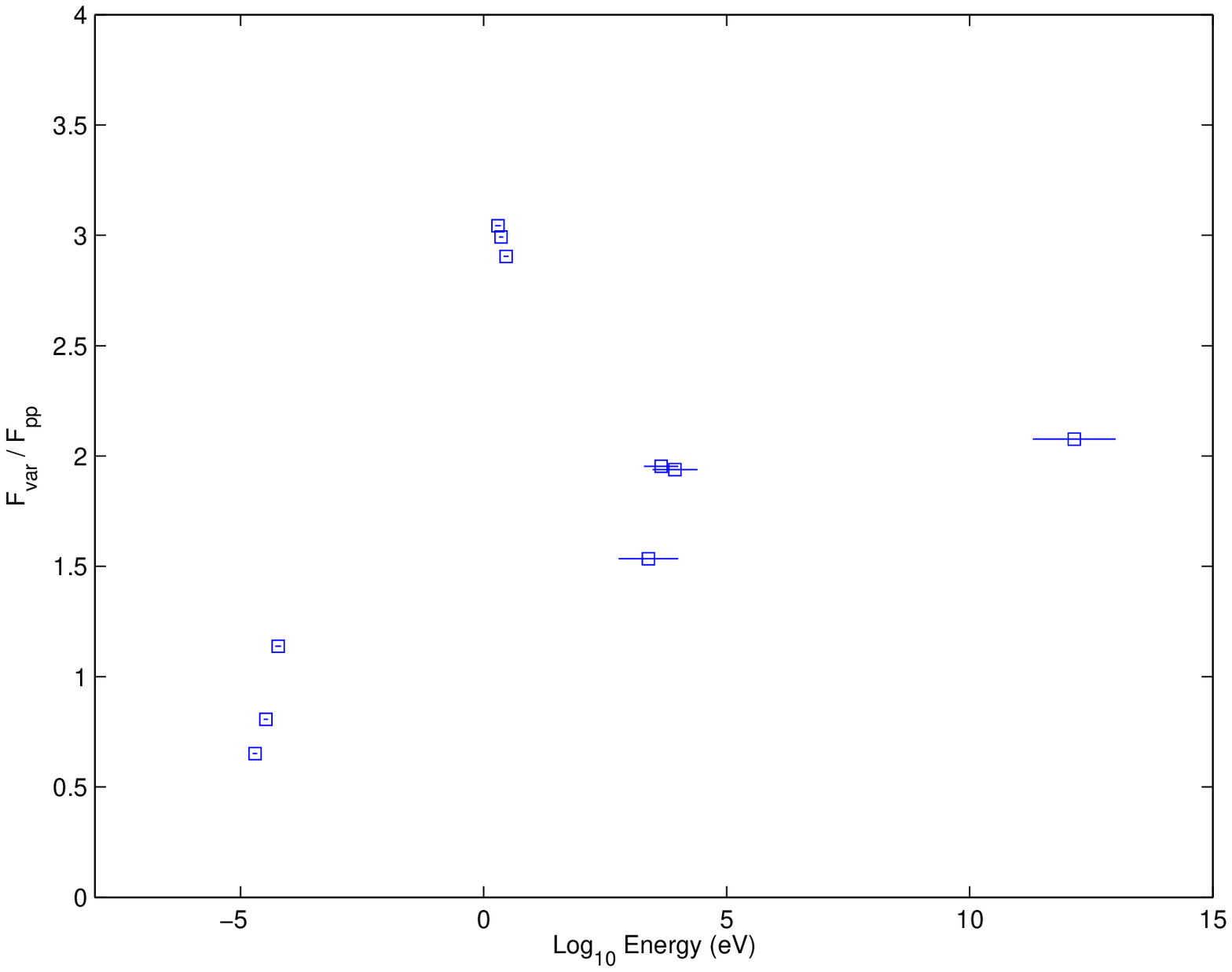}}\\%

\caption{\label{FIG:FVAR}{\it{Left:}} The fractional rms variability
amplitude (blue squares) and the point-to-point fractional rms
variability amplitude (red x's) for 10 of the 12 energy
bands. {\it{Right:}} The ratio of the fractional rms variability
amplitude to the point-to-point variability amplitude as a function of
energy. See text for more details.}
\end{figure}

These quantities were calculated using one-day binning for each of the
data sets. These results can be seen for all wavebands apart from the
37\,GHz radio dataset in Table~\ref{TAB:FVAR-FPP}. For the 37\,GHz
data, the difference between the terms to be subtracted in both
Equations~\ref{EQ:FVAR} and ~\ref{EQ:FPP} was negative, indicating
that the measurement errors were larger than the variability observed,
and therefore the calculation could not be completed. The fractional
rms variability amplitude, F$_{var}$ quantifies the integrated level
of variability present in a particular waveband while the
point-to-point fractional rms variability amplitude, F$_{pp}$, probes
the short-timescale variability by measuring the variations between
adjacent points in the lightcurve. Their ratio, taken as a function of
energy, provides information on the dependence of the power spectral
density slope on energy \citep{Zhang:05}. Figure~\ref{FIG:FVAR} shows
F$_{var}$ and the ratio of F$_{var}$ to F$_{pp}$ for ten of the twelve
wavebands. The values for the BAT data are not plotted because, as
will be discussed in Section~\ref{discussionAndConclusion}, the
significance level of the BAT detection was less than 3 sigma on most
nights ($\sim$ 58\%) during this campaign.

\subsection{Gamma-ray Flux Variability}

\begin{figure}
\resizebox*{0.45\textwidth}{!}{\includegraphics[draft=false]{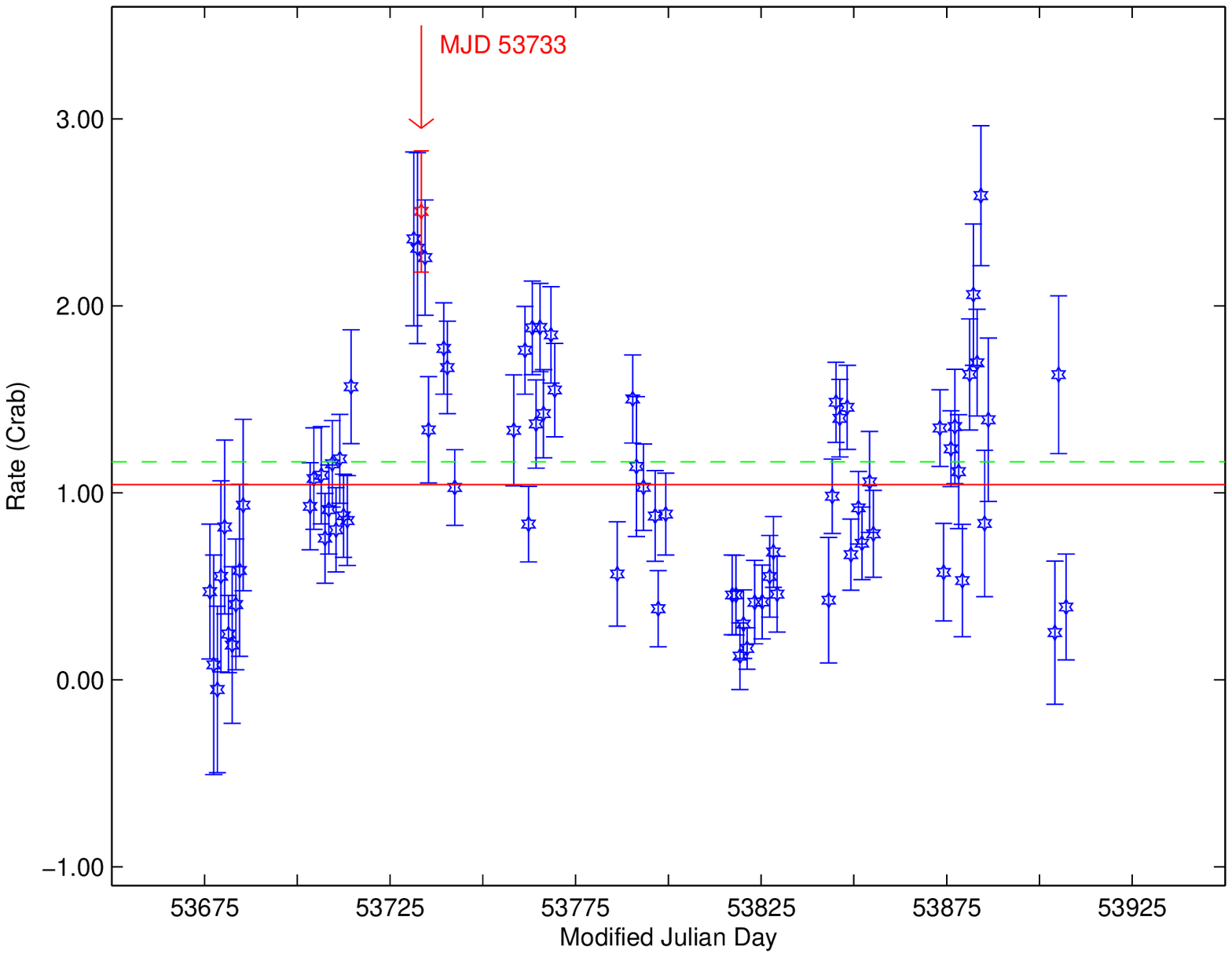}}%
\resizebox*{0.45\textwidth}{!}{\includegraphics[draft=false]{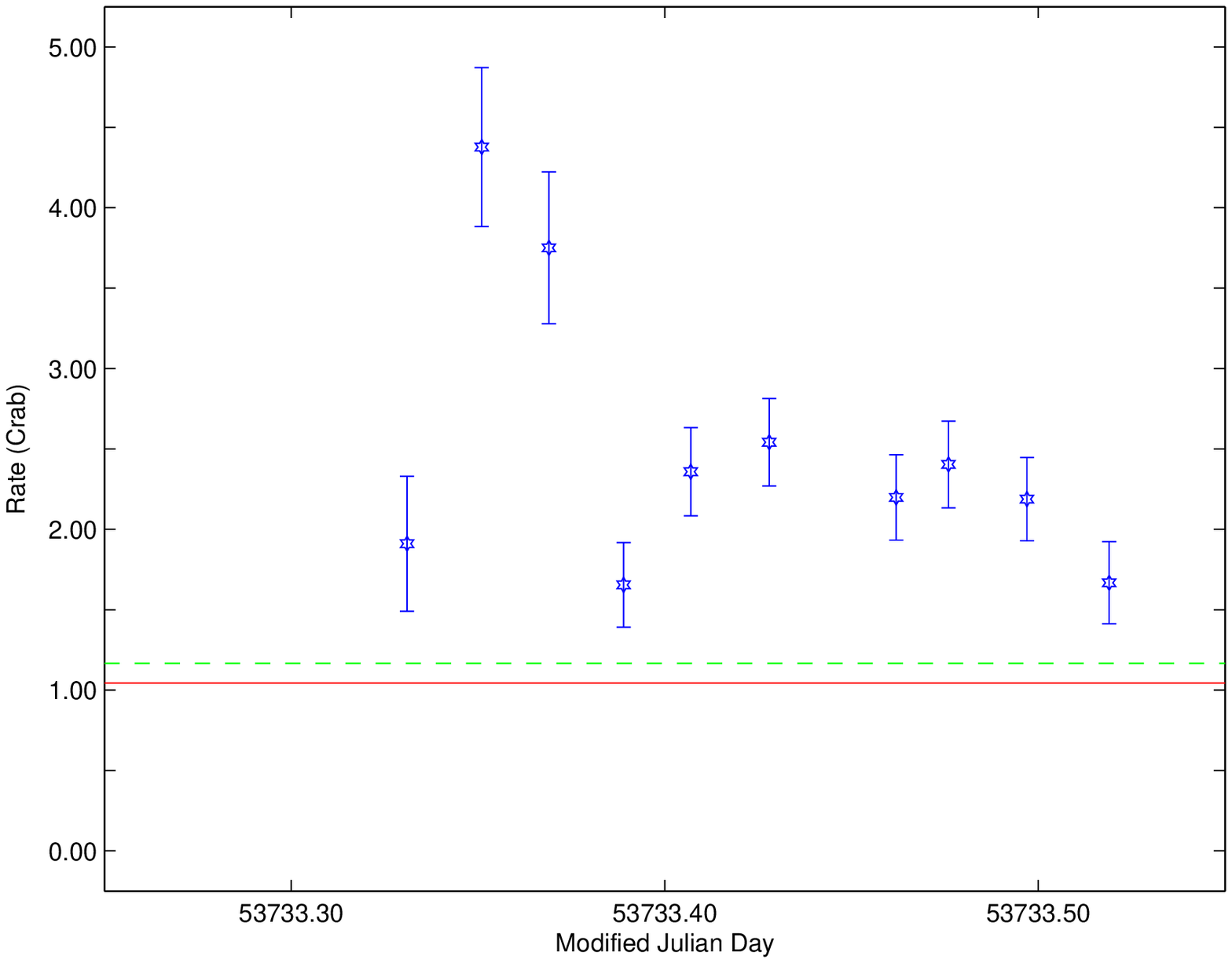}}\\%

\caption{\label{FIG:GAMMA-ONE-NIGHT}{\it{Left:}} The gamma-ray
lightcurve with the data binned in nightly bins. The mean nightly
gamma-ray rate is indicated by the horizontal, solid, red line. The
mean gamma-ray rate for an individual exposure (typically 28-minute)
is indicated by the horizontal, dashed, green line. The night (MJD
53733) during which the maximum rate for an individual exposure was
recorded is plotted in red and is also indicated by the
arrow. {\it{Right:}} The gamma-ray lightcurve for MJD 53733, the night
during which the maximum gamma-ray rate for an individual exposure was
recorded. The mean gamma-ray rate recorded during the entire campaign
for an individual exposure is indicated by the horizontal, dashed,
green line. Note that a different scale is used on the Y-axis in each
figure.}
\end{figure}

Throughout the observations presented here, the gamma-ray rate from
Mrk\,421 was found to be variable. This is evident from the
lightcurves shown in Figure~\ref{FIG:GAMMA-ONE-NIGHT}, in which the
average nightly gamma-ray rate is shown, and in
Figure~\ref{FIG:GAMMA}, where the gamma-ray rate is plotted for each
observing scan (typically of 28-minute duration). Mrk\,421 was
observed by the TACTIC gamma-ray telescope \citep{Yadov:07} during a
time period overlapping with these Whipple observations and, although
the energy ranges and sensitivities of the two instruments are
slightly different, similar overall trends are present in the two
datasets. The maximum gamma-ray rate for one of the Whipple 28-minute
exposures occurred on MJD 53733 when a rate of 4.38 $\pm$ 0.49 Crab
was recorded. This is a factor of 3.74 increase over the mean rate for
a 28-minute exposure during this campaign. The lightcurve for
MJD\,53733 is shown in Figure~\ref{FIG:GAMMA-ONE-NIGHT}. The mean
gamma-ray rate for this night was the second highest recorded during
this campaign at 2.50 $\pm$ 0.32 Crab. The flux from Mrk\,421 can be
seen to vary throughout this night, starting off at 1.91 $\pm$ 0.42
Crab before reaching its peak value and then decreasing to 1.67 $\pm$
0.26 Crab. The maximum average nightly gamma-ray rate of 2.59 $\pm$
0.37 Crab, a factor of 2.49 higher than the mean nightly rate, was
recorded on MJD\,53884.  A summary of the mean, minimum and maximum
gamma-ray rate from Mrk\,421 during this campaign is given in
Table~\ref{TAB:FVAR-FPP}. The fractional rms variability amplitude in
the gamma-ray band was found to be 0.511 with a point-to-point
fractional rms variability amplitude of 0.246. The ratio of these two
quantities, at $\sim$2, (see Figure~\ref{FIG:FVAR}) indicates that
red-noise variability is present \citep{Zhang:05}.

\subsection{X-ray Flux Variability}\label{SECTION:SF}

The X-ray flux from Mrk\,421, was monitored from 0.2 to 50 keV during
this observing campaign. Many episodic outbursts are seen in all X-ray
bands reported on here with the maximum emission reaching 1.7 times
the mean emission level in that band for the XRT data, 2.9 times the
mean emission level for that band in the ASM data, 2.7 times the mean
emission level for that band in the PCA data and, 5.3 times the mean
emission level for that band in the BAT data.  These data are
summarized in Table~\ref{TAB:FVAR-FPP}.  The data are plotted in
one-night bins in Figure~\ref{FIG:X-GAMMA} where inter-night
variability in addition to overall trends are evident in the X-ray
lightcurve. The X-ray emission in the ASM, PCA and BAT regimes is
well-correlated. The night of maximum emission was MJD 53877 for the
ASM data. No PCA data were taken this night and the maximum in its
emission occurred on MJD 53875, which was the second brightest night
for the BAT data. There is a trend towards increasing emission around
this time with the previous $\sim$ 30 days showing elevated emission
levels in the ASM, PCA and BAT data. For the XRT lightcurve, which is
not as well sampled as those in the other wavebands, evidence for
heightened emission over a period of two nights was seen from MJD
53739 to 53740. The rms variability amplitude is found to increase
with energy in the X-ray regime, from 0.269 for the XRT data to 0.529
for the PCA data. As can be seen in Figure~\ref{FIG:FVAR}, the ratio
of F$_{var}$ to F$_{pp}$ is found to be ~$\sim$ 2 for all X-ray data,
indicative of red-noise variability.

To further study the variability properties of the X-ray lightcurves,
the normalized, first-order structure function, SF$^{(1)}$($\tau$),
was computed for each lightcurve. SF$^{(1)}$($\tau$) is defined as:

\begin{center}
\begin{equation}\label{EQ:SF}
SF^{(1)}(\tau) = \frac{1}{N} \sum\left( \frac{F(t) - F(t+\tau)}{\overline{F(t)}}\right)^2
\end{equation}
\end{center}

\begin{figure}
\resizebox*{0.45\textwidth}{!}{\includegraphics[draft=false]{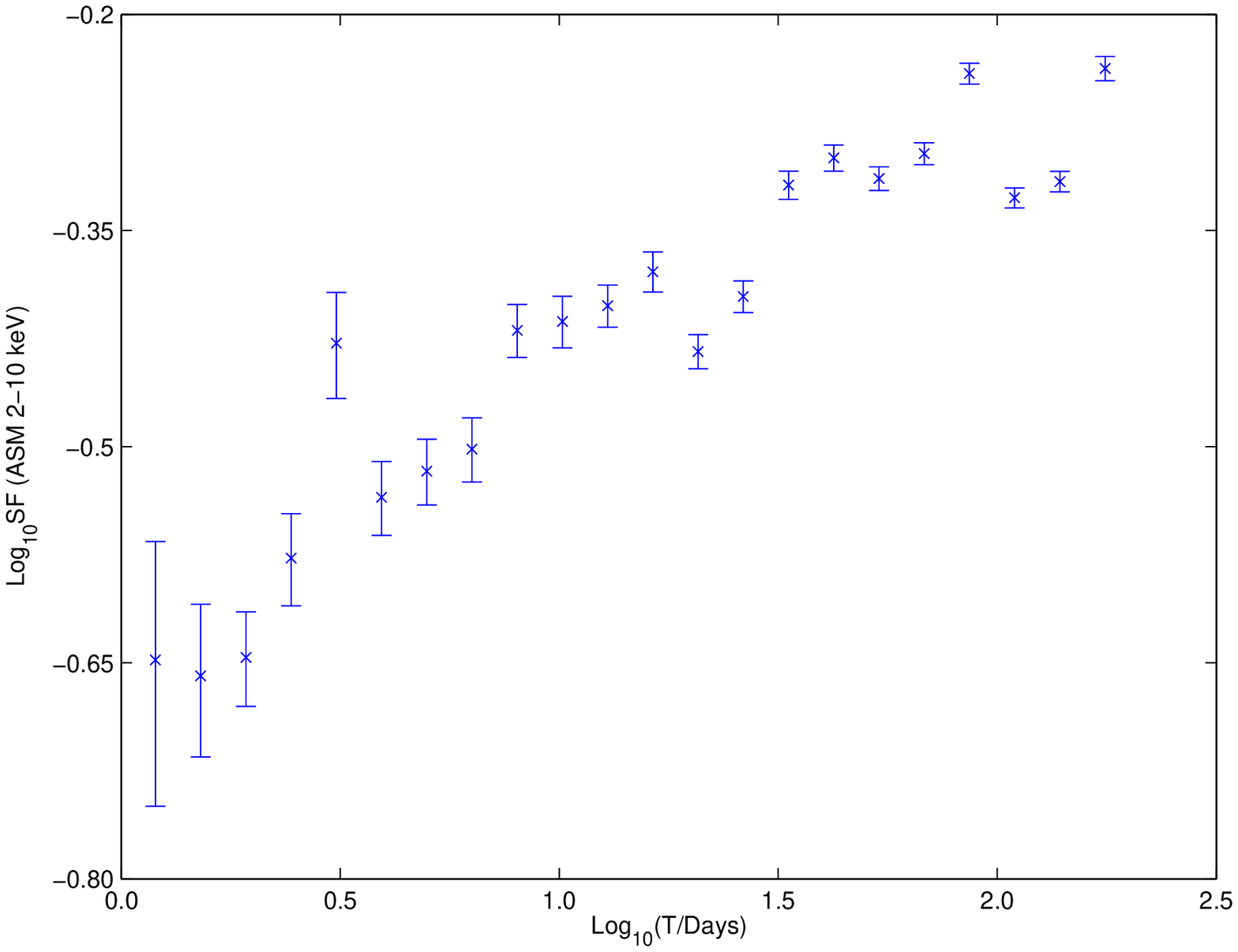}}%
\resizebox*{0.45\textwidth}{!}{\includegraphics[draft=false]{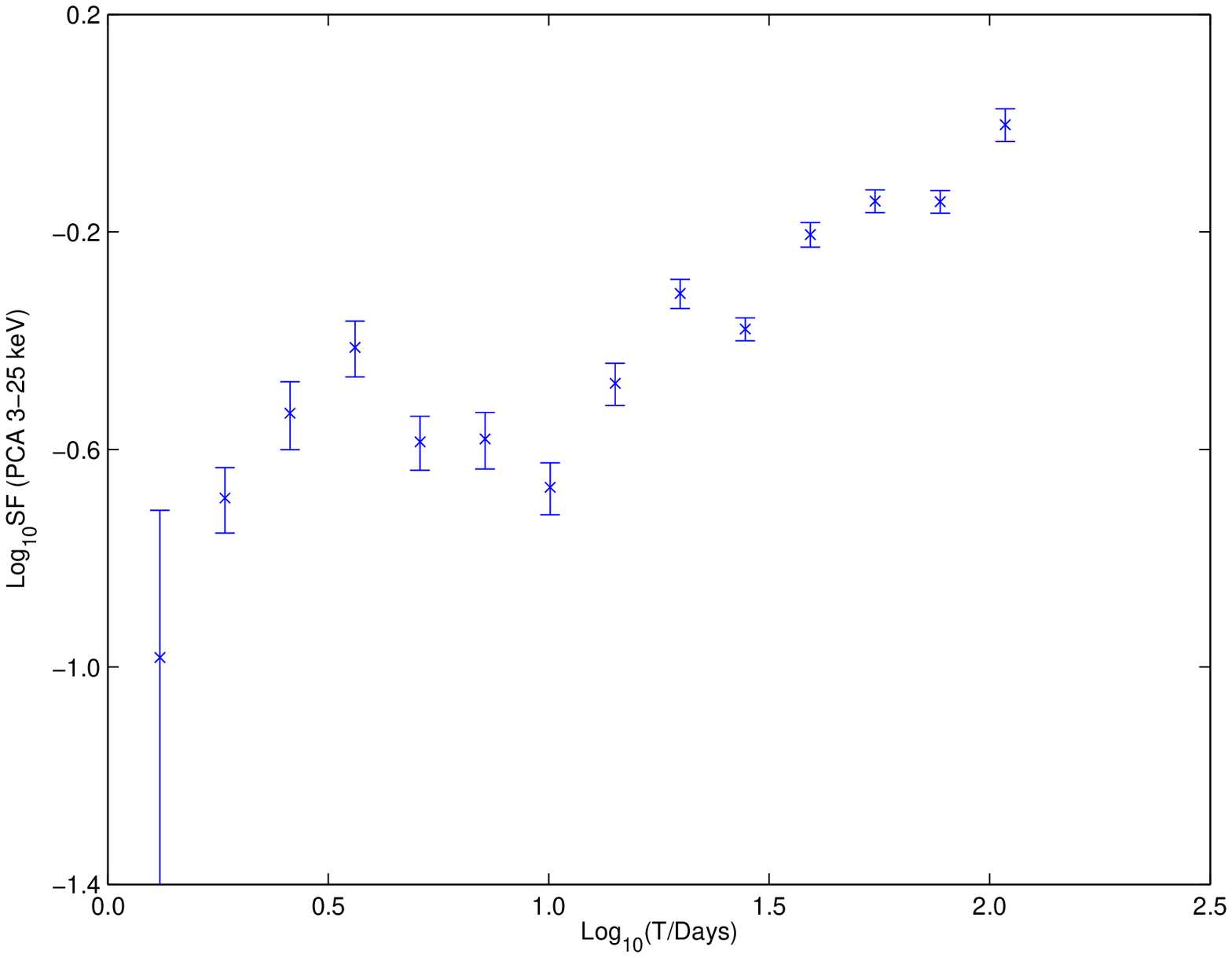}}\\
\resizebox*{0.45\textwidth}{!}{\includegraphics[draft=false]{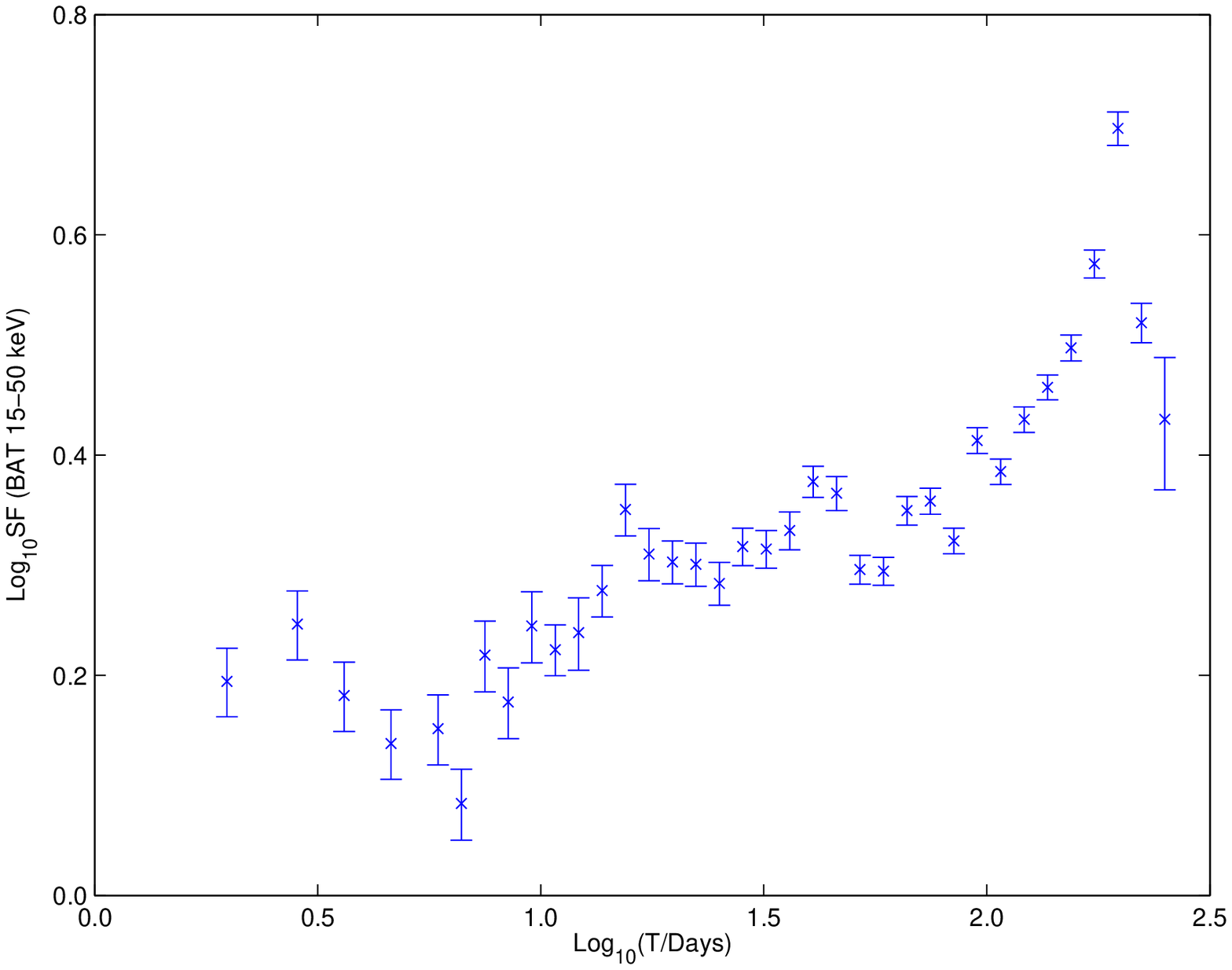}}%
\resizebox*{0.45\textwidth}{!}{\includegraphics[draft=false]{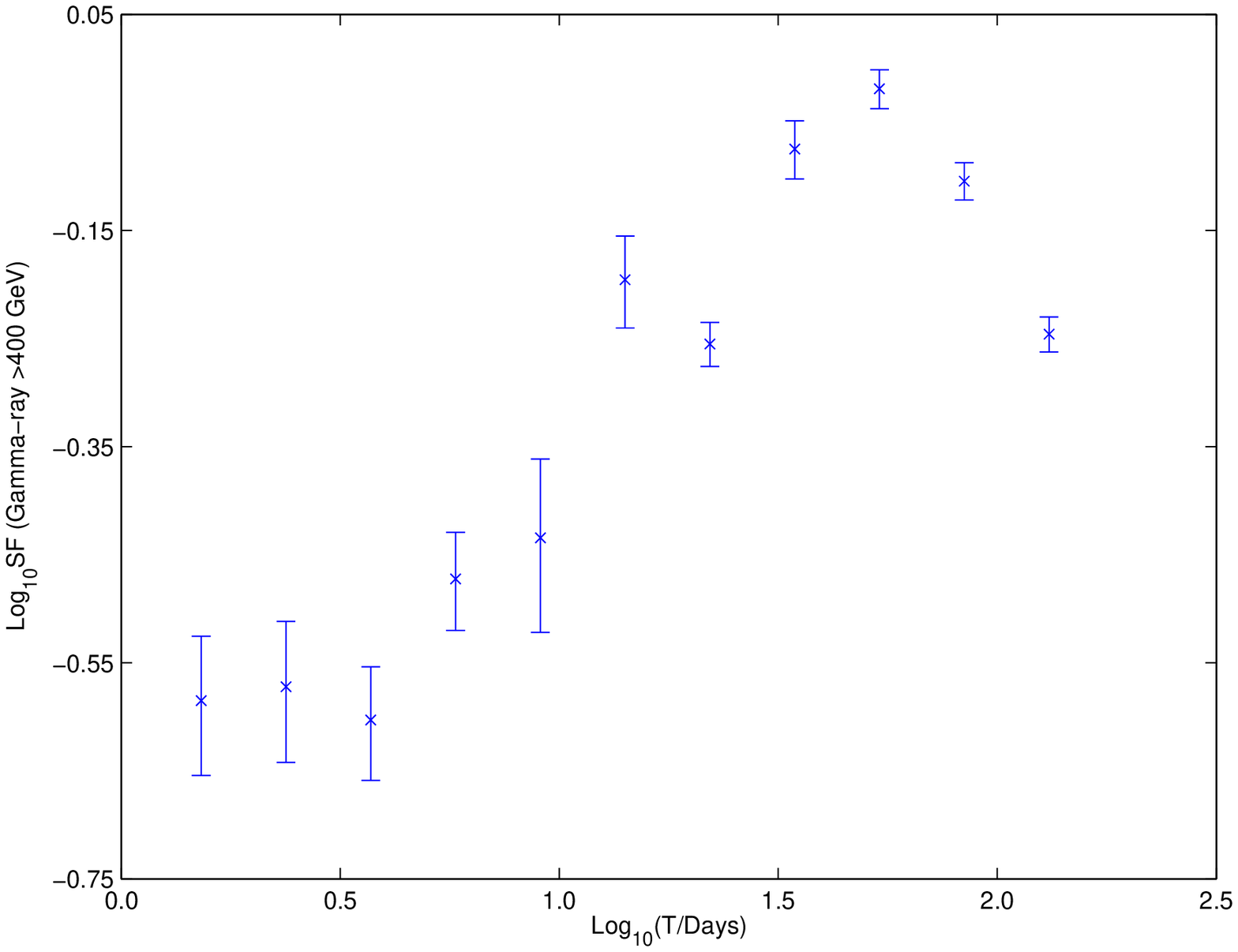}}\\[2ex]%

\caption{\label{FIG:SFS-X}The normalized, first-order structure
function for the ASM data({\it{Top Left}}); the PCA data ({\it{Top
Right}}); the BAT data ({\it{Bottom Left}}); the gamma-ray data
({\it{Bottom Right}}).}

\end{figure}

\noindent{For a discrete time series, we calculate the structure
function in bins of width $\Delta\tau$ such that the value of the
structure function for a given bin is found by summing over pairs of
observations separated by a time difference $\Delta$t satisfying
$\tau$ - $\Delta\tau$/2 $<$ $\Delta$t $<$ $\tau$ + $\Delta\tau$/2. The
structure functions for each of the X-ray bands in which substantial
data were collected and for the gamma-ray data are shown in
Figure~\ref{FIG:SFS-X}. In all cases, the structure function is found
to rise approximately linearly when plotted in log-log
representation. The data from the ASM show evidence for a break
between $\tau$ $\sim$ 20-25 days with the slope of the structure
function remaining approximately constant before and after this
break. The structure function of the PCA data shows evidence for a
break between $\tau$ $\sim$ 5-8 days. Again, no significant change in
the slope is seen after this break. There is clear evidence for a
break in the structure function of the BAT data between $\tau$ $\sim$
55-71 days after which it rises linearly once more but with a steeper
slope. There are no strong features in the structure function of the
gamma-ray data.}

\subsection{Optical Flux Variability}

The well-sampled optical lightcurves show evidence for variability
during this campaign.  The mean magnitudes for each band are shown in
Table~\ref{TAB:FVAR-FPP}.  When converted to linear fluxes, the
optical flux was found to peak at 1.43 times its mean value in both
the R-band and the V-band and to peak at 1.83 its mean flux in the
B-band.

Several R-band flare features with amplitudes of $~0.2$ mag are
resolved in the particularly well-sampled range of dates spanning
53800 $<$ MJD $<$ 53900. The fractional rms variability amplitude and
point-to-point fractional rms variability amplitude are 0.13 and 0.04
respectively for the R-band lightcurve, 0.14 and 0.05 respectively for
the V-band lightcurve and 0.19 and 0.06 respectively for the B-band
lightcurve. In all cases, the ratio of F$_{var}$ to F$_{pp}$ is $\sim$
3, indicating that, like the gamma-ray and X-ray bands, red-noise
processes are responsible for the optical variability. The measured
optical flux includes contributions from the host galaxy, estimated to
be $\sim 15\%$ \citep{Nilsson:99}. Since it affects only the mean
flux, not the variability, the optical fractional rms variability
amplitude should be about $15\%$ higher. The level of optical
variability seen in this analysis is slightly lower but is consistent
with the level of variability seen in other recent multiwavelength
studies of this source \citep{Rebillot:04}.

Variations in flux among the three optical bands are highly
correlated. There is weak evidence in this data set supporting a trend
of flattening of the optical spectrum with increasing R-band flux: a
trend seen more definitively by other authors in the long-term
lightcurves of Mrk\,421 and other BL Lac objects
(\citealp{Vagnetti:03}; \citealp{Hu:06}). This behavior, predicted by
several authors (\citealp{Li:00}; \citealp{Spada:01};
\citealp{Vagnetti:03:SSC}), is consistent with the expected spectral
variability behavior of a jet with emission dominated by the SSC
processes.

\begin{figure}
\resizebox*{0.95\textwidth}{!}{\includegraphics[draft=false]{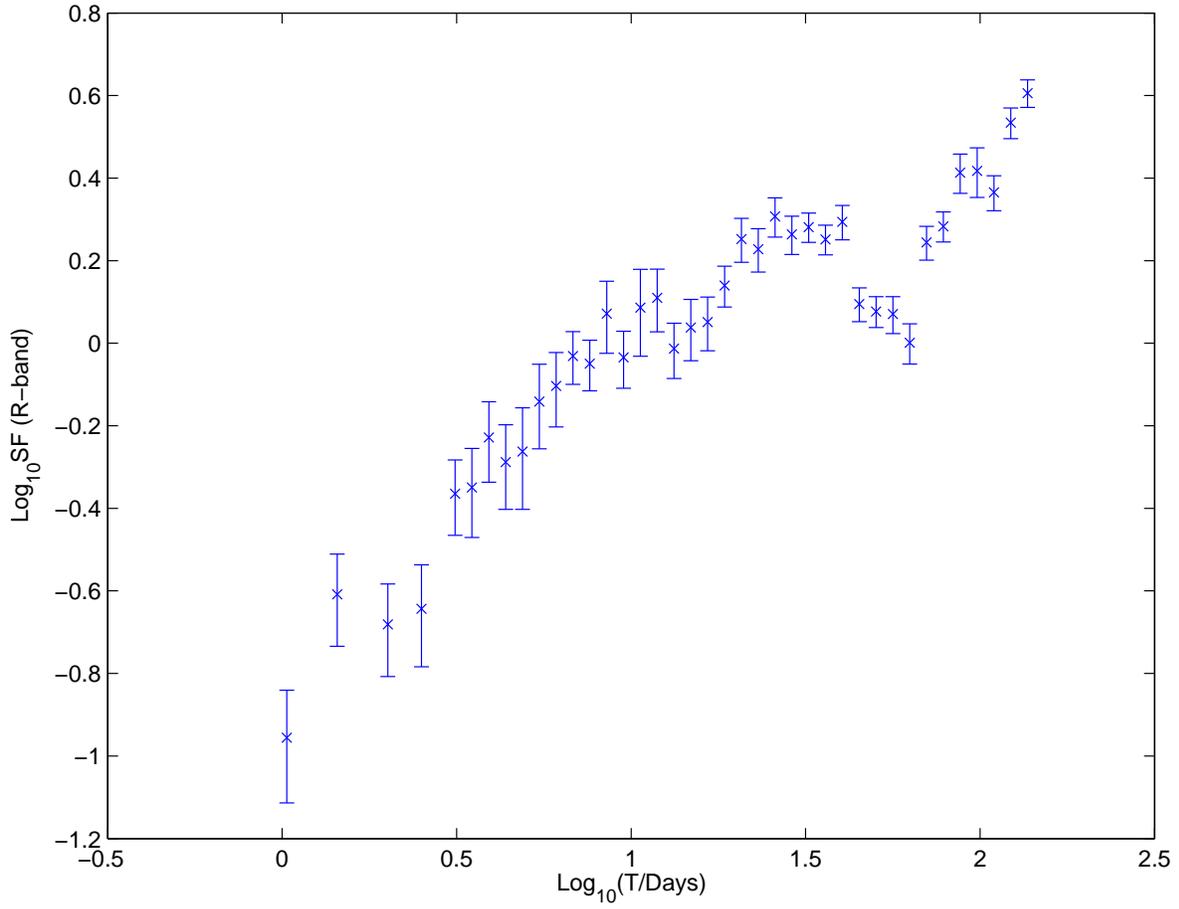}}%

\caption{\label{FIG:SFS-OPT}Structure function for the R-band optical
data.}

\end{figure}

The normalized, first-order structure function of the composite R-band
lightcurve has been computed and is shown in Figure
\ref{FIG:SFS-OPT}. The structure function rises approximately linearly
at short timescales up to a time-lag of between about 40 and 60 days
where a clear break is seen, above which the structure function rises
linearly again.  The break most likely indicates a characteristic
timescale of shot-like emission events. We note that structure
functions with similarly-shaped breaks have been observed in other
wavebands with different characteristic break frequencies, for
example, the EUV emission from Mrk\,421 recorded by the Extreme
Ultraviolet Explorer during a 1998 multiwavelength campaign
\citep{Takahashi:00}.

\subsection{Radio Flux Variability}

This is the waveband in which the lightcurves are least-sampled and
also, the waveband in which the least amount of variability is
seen. This is typical for blazars; the variations observed in the
radio regime tend to occur over longer timescales than those at
shorter wavelengths (\citealp{Kovalev:02}; \citealp{Blazejowski:05};
\citealp{Rebillot:06}; \citealp{Hovatta:07}). This is supported by the
fact that the lowest values of the fractional rms variability
amplitude, 0.04 to 0.27, were found for the radio data. The fact that
these data were less well-sampled than the data at other frequencies
would most likely lead to an underestimation of F$_{var}$ in this
waveband. The ratio of F$_{var}$ to F$_{pp}$ for each of the four
radio bands plotted in Figure~\ref{FIG:FVAR} is found to be $\sim$ 1,
indicative of white-noise variability.

\subsection{X-ray Correlations with Gamma-ray Data}\label{SEC:X-GAMMA}

\begin{figure}
\resizebox*{0.45\textwidth}{!}{\includegraphics[draft=false]{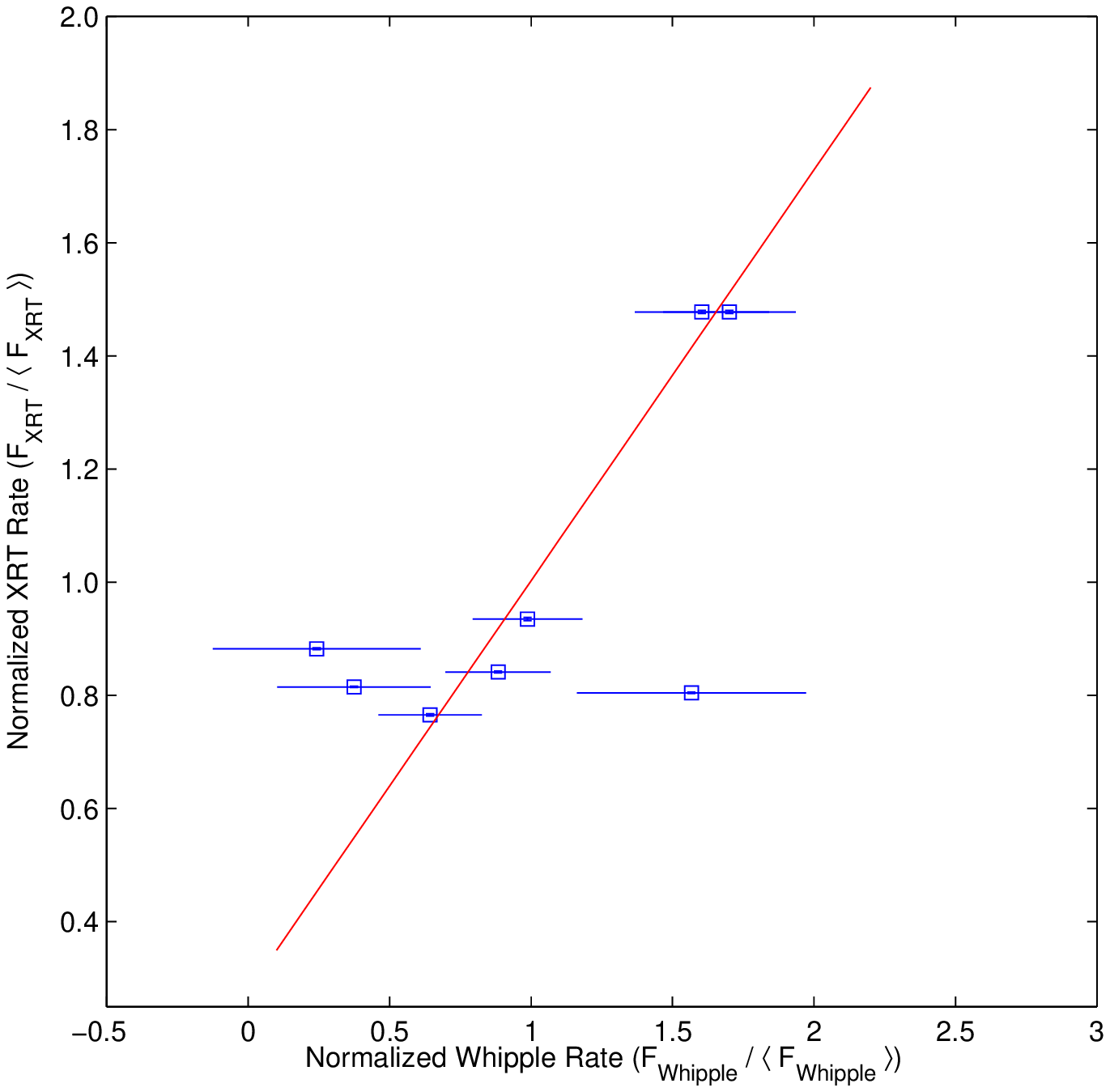}}%
\resizebox*{0.45\textwidth}{!}{\includegraphics[draft=false]{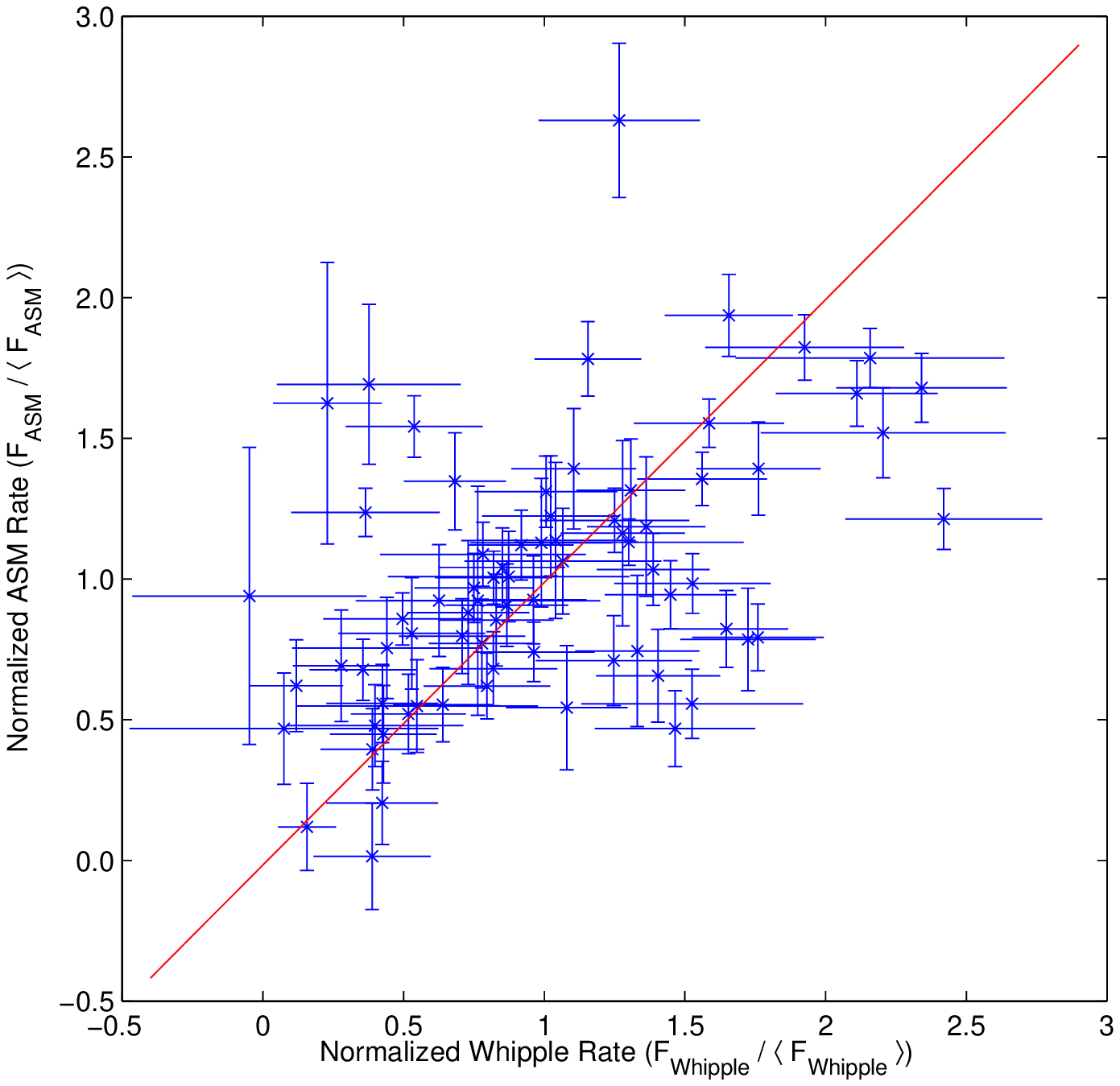}}\\
\resizebox*{0.45\textwidth}{!}{\includegraphics[draft=false]{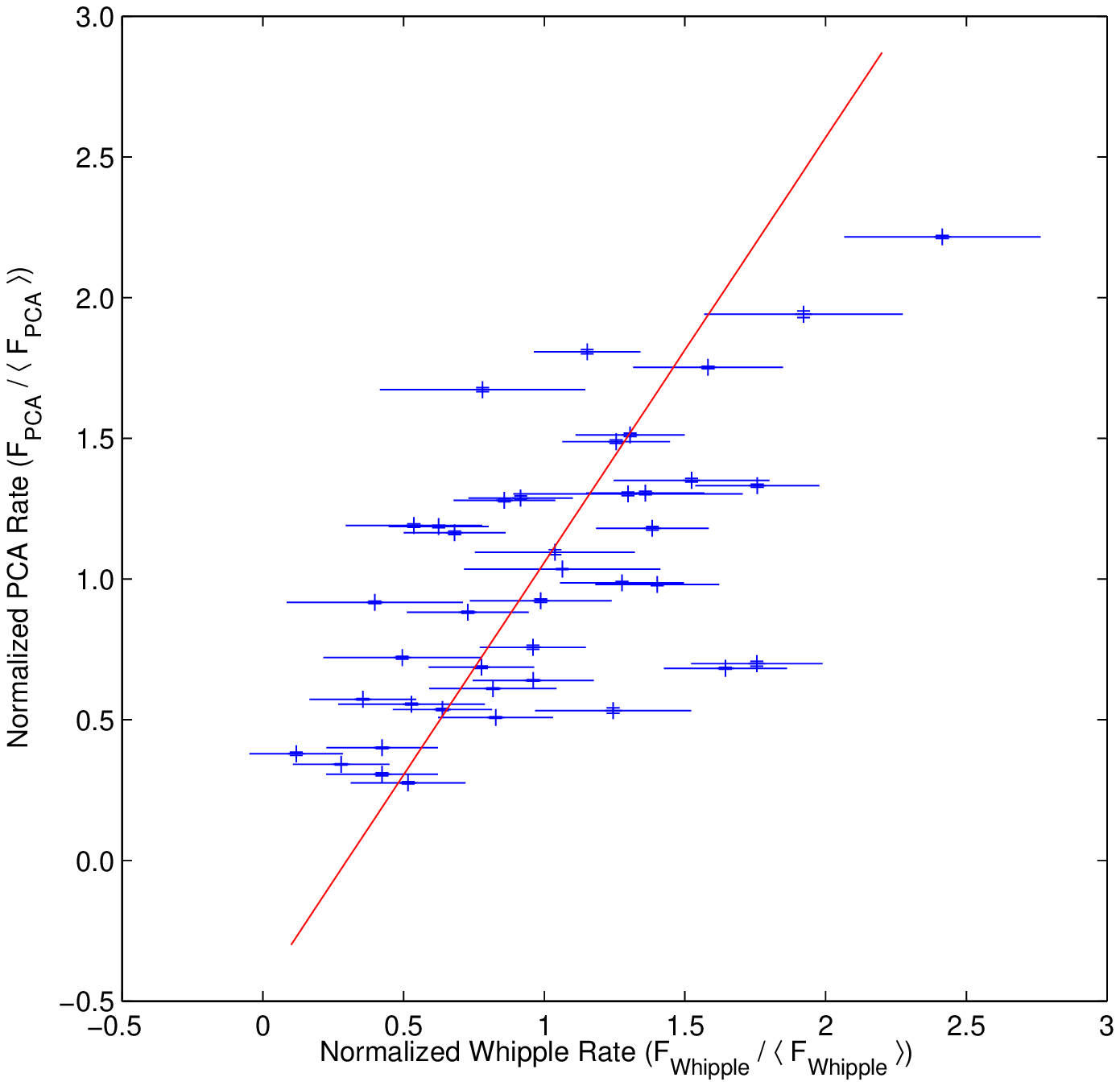}}%
\resizebox*{0.45\textwidth}{!}{\includegraphics[draft=false]{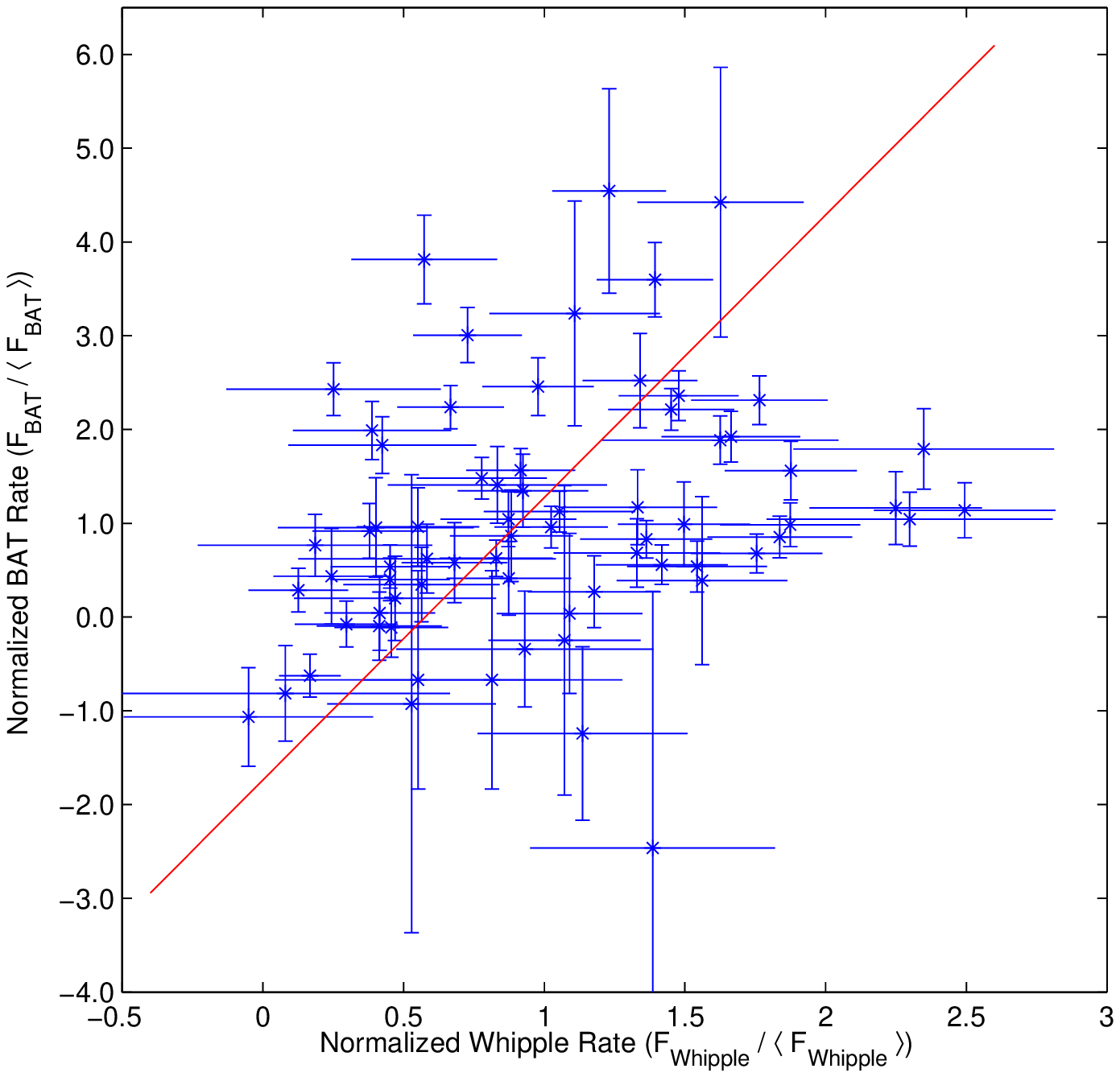}}\\[2ex]%

\caption{\label{FIG:X-GAMMA-CORRELATION}Flux-flux correlation for
  {\it{Top Left:}} the gamma-ray and XRT data (slope 0.73); {\it{Top
      Right:}} the gamma-ray and ASM data (slope 1.00); {\it{Bottom
      Left:}} the gamma-ray and PCA data (slope 1.51); {\it{Bottom
      Right:}} the gamma-ray and BAT data (slope 3.01). Note that for
  the ASM and PCA data, the scales on the x- and y-axes are equal
  while for the XRT data, the scale on the x-axis is twice that of the
  y-axis while for the BAT data, the scale on the y-axis is three
  times that of the x-axis.}

\end{figure}

\begin{figure}
\resizebox*{0.45\textwidth}{!}{\includegraphics[draft=false]{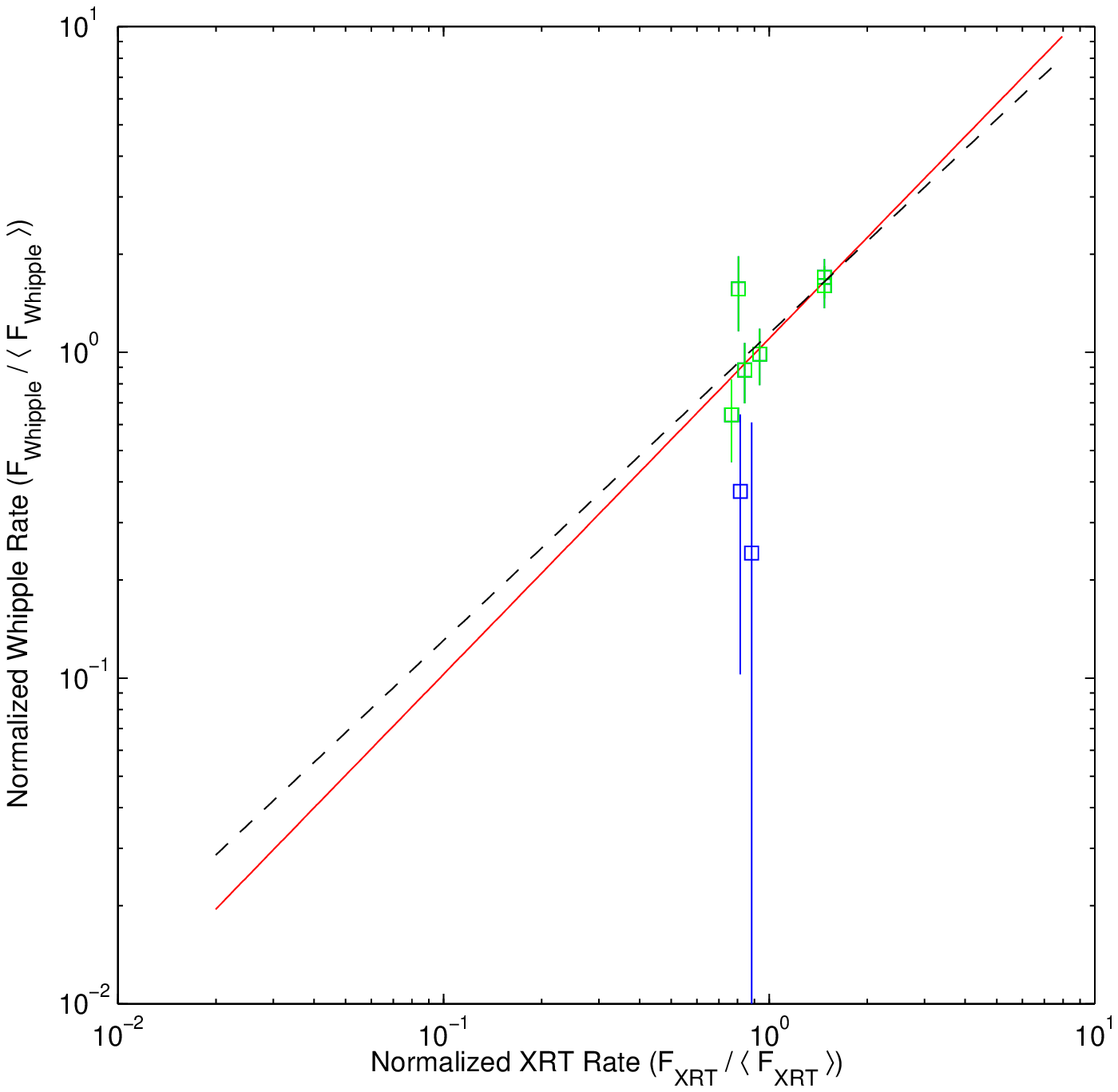}}%
\resizebox*{0.45\textwidth}{!}{\includegraphics[draft=false]{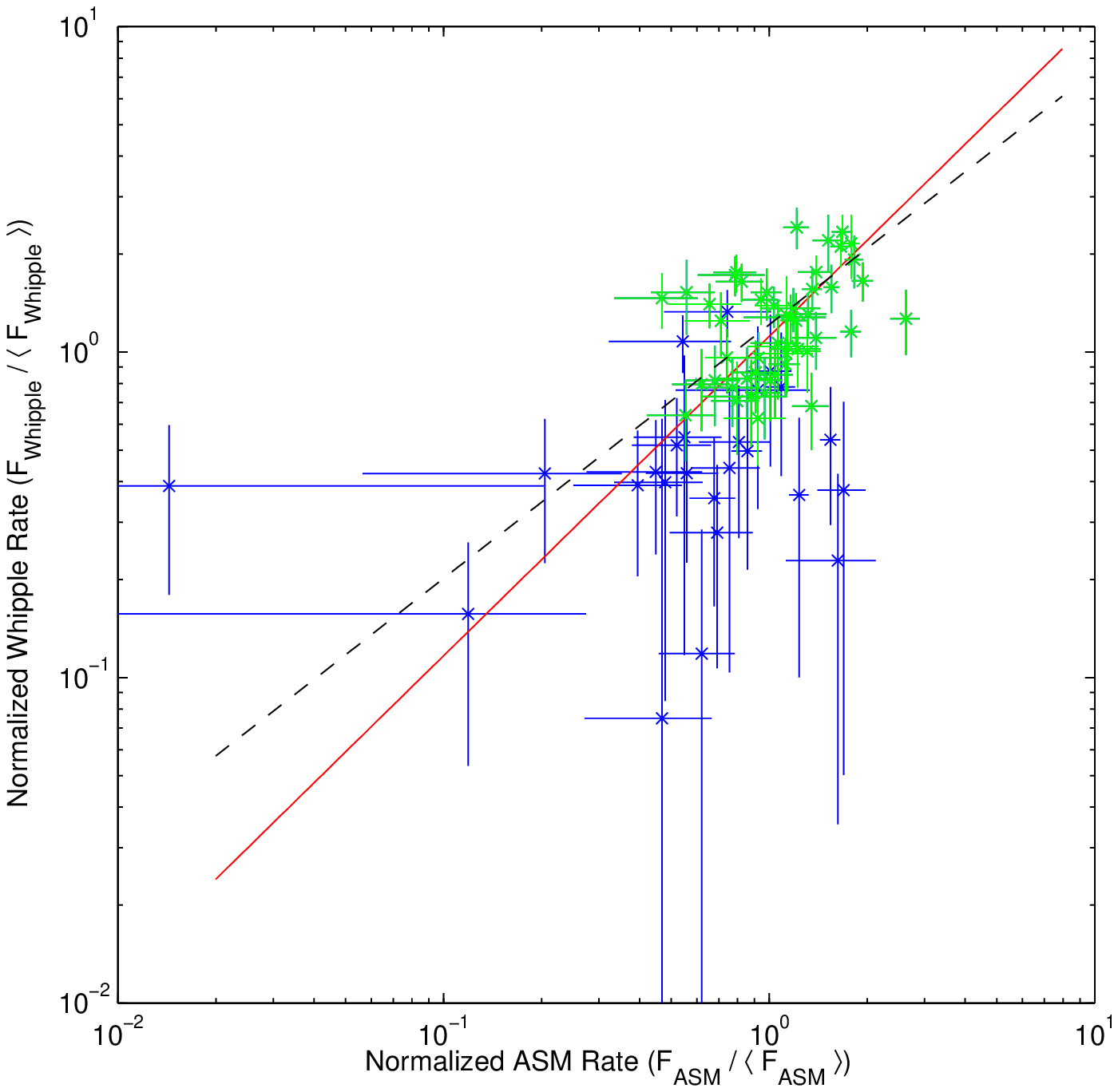}}\\
\resizebox*{0.45\textwidth}{!}{\includegraphics[draft=false]{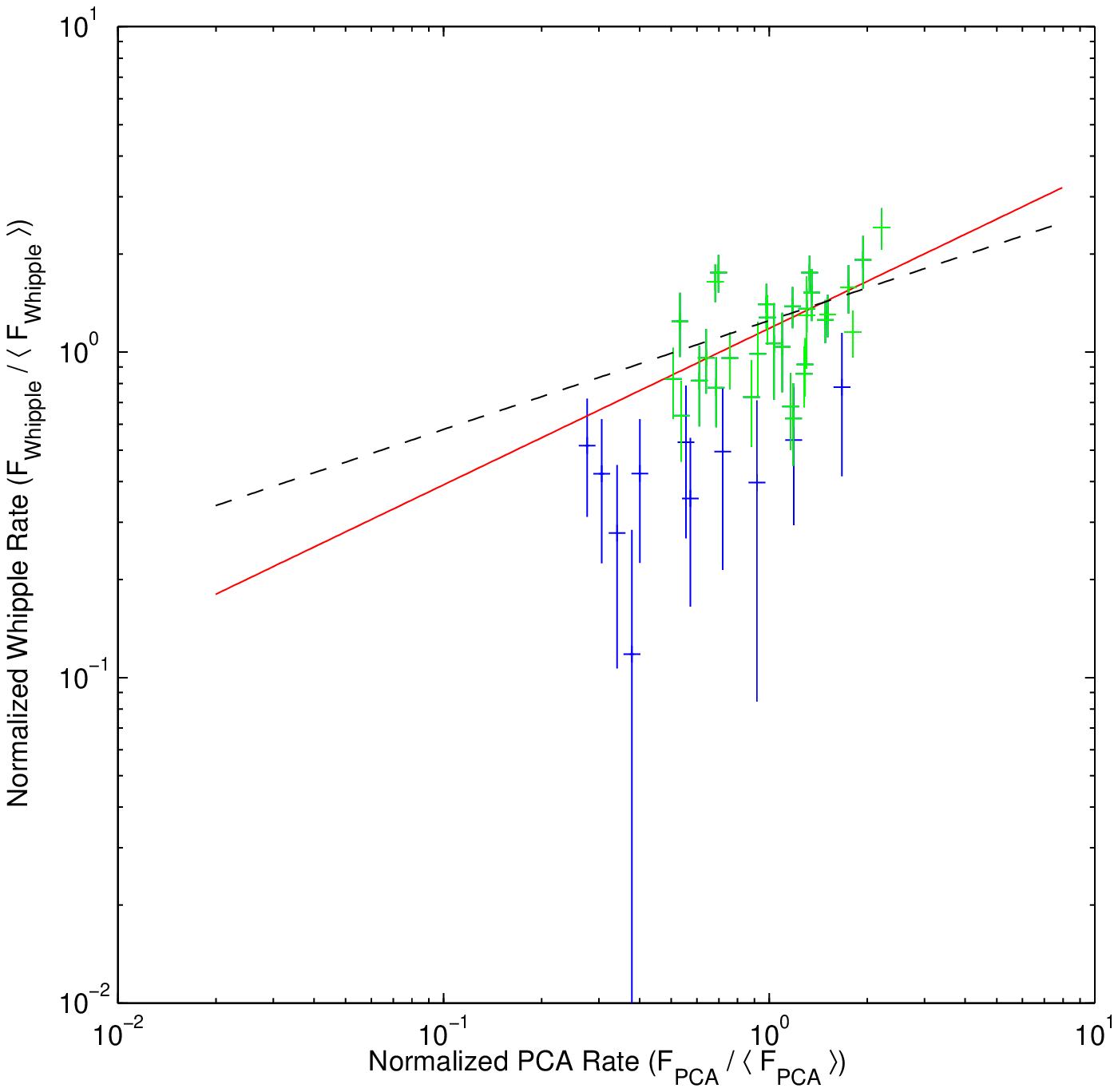}}%
\resizebox*{0.45\textwidth}{!}{\includegraphics[draft=false]{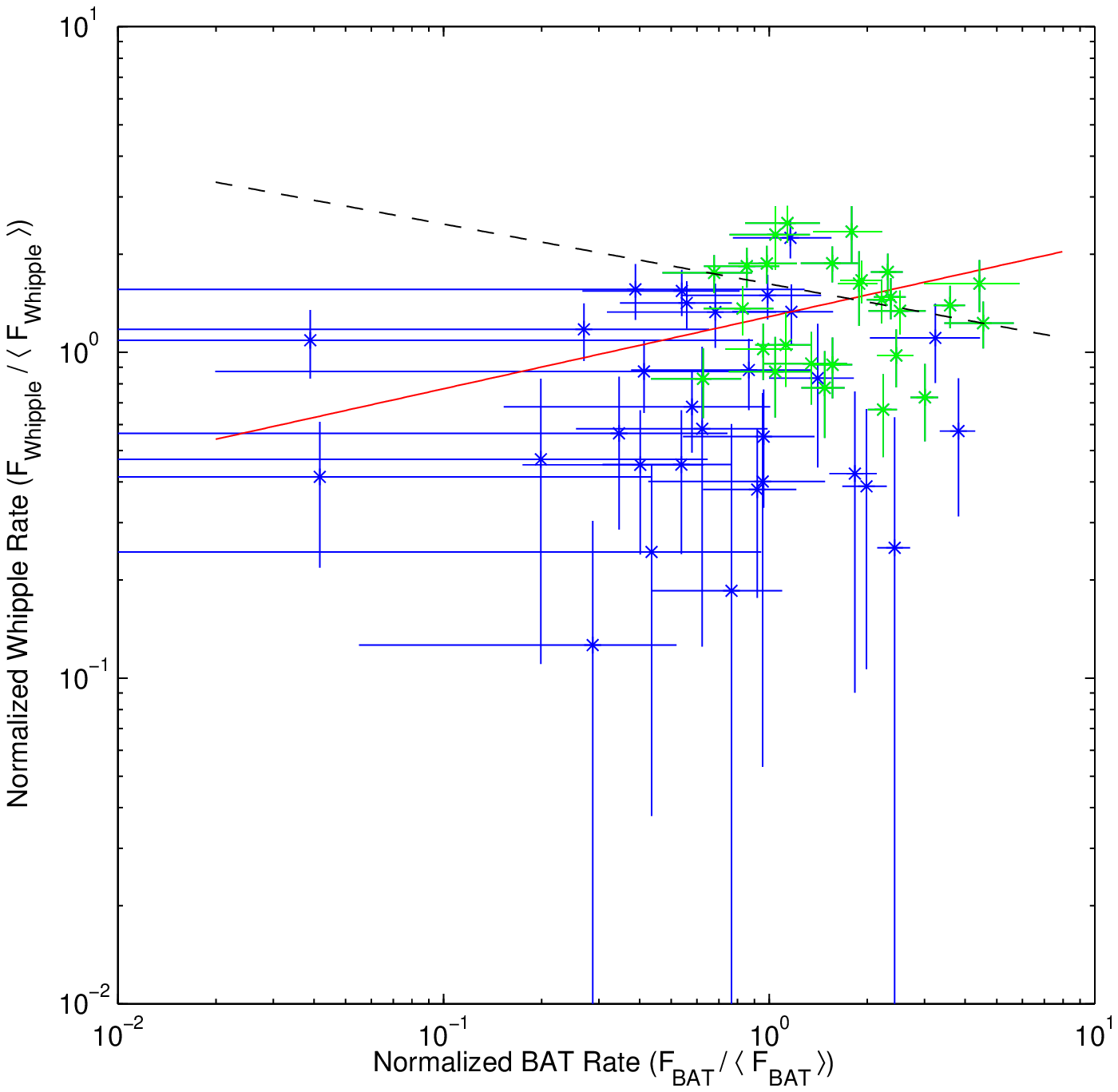}}\\[2ex]%

\caption{\label{FIG:X-GAMMA-LOG-CORRELATION}Flux-flux correlation on log
  scale for the X-ray and gamma-ray data. The solid, red lines show
  the slope when all the data are included while the black, dashed
  lines show the slope when only the data with significance $>$3 sigma
  (highlighted here in green) are included. {\it{Top Left:}} The
  gamma-ray and XRT data - slopes 1.03 (all), 0.94 ($>$3 sigma);
  {\it{Top Right:}} the gamma-ray and ASM data - slopes 0.98 (all),
  0.78 ($>$3 sigma); {\it{Bottom Left:}} the gamma-ray and PCA data -
  slopes 0.48 (all), 0.33 ($>$ 3 sigma); {\it{Bottom Right:}} the
  gamma-ray and BAT data - slopes 0.22 (all), -0.18 ($>$ 3 sigma).}

\end{figure}

\begin{figure}
\resizebox*{0.33\textwidth}{!}{\includegraphics[draft=false]{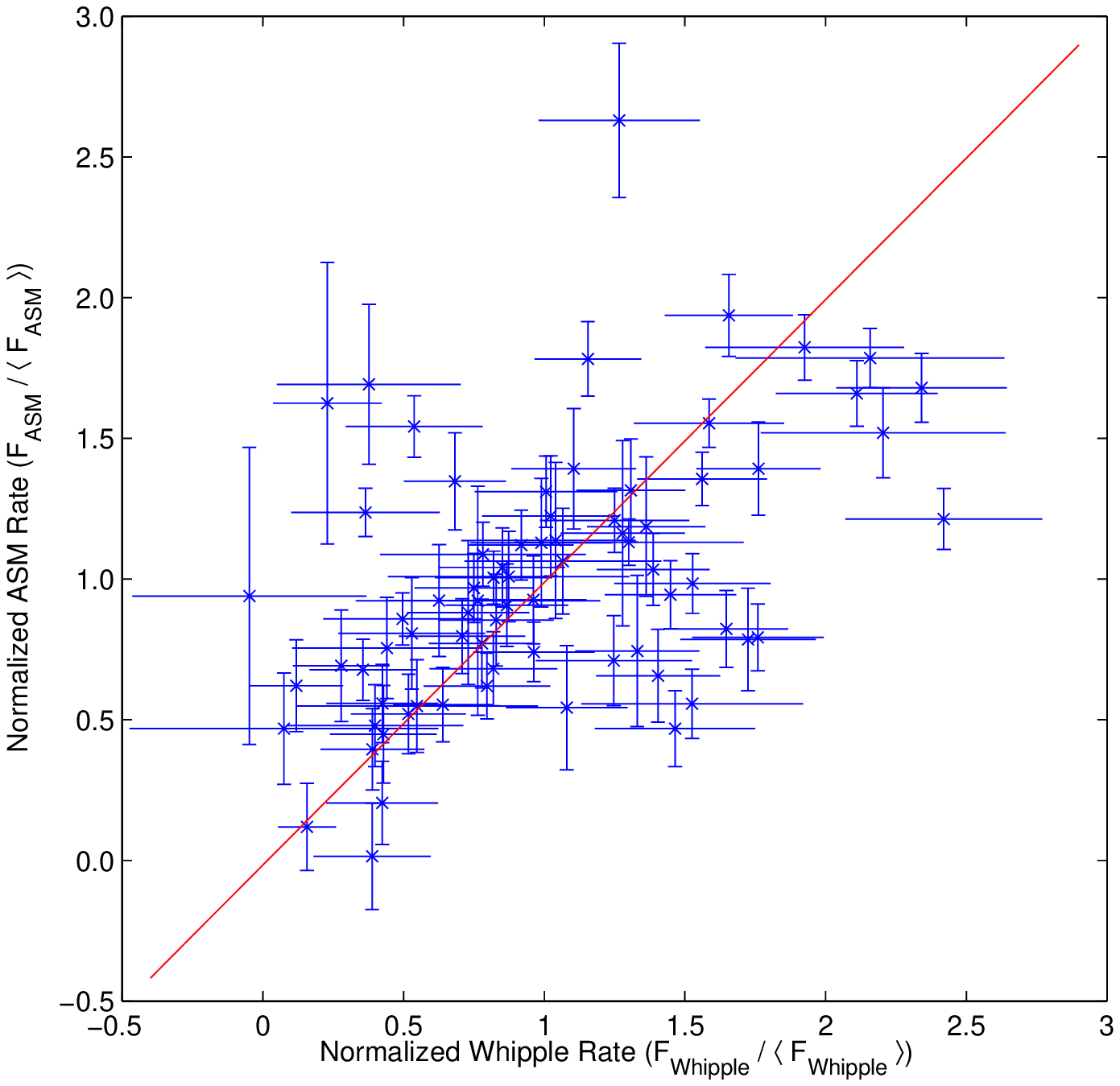}}%
\resizebox*{0.33\textwidth}{!}{\includegraphics[draft=false]{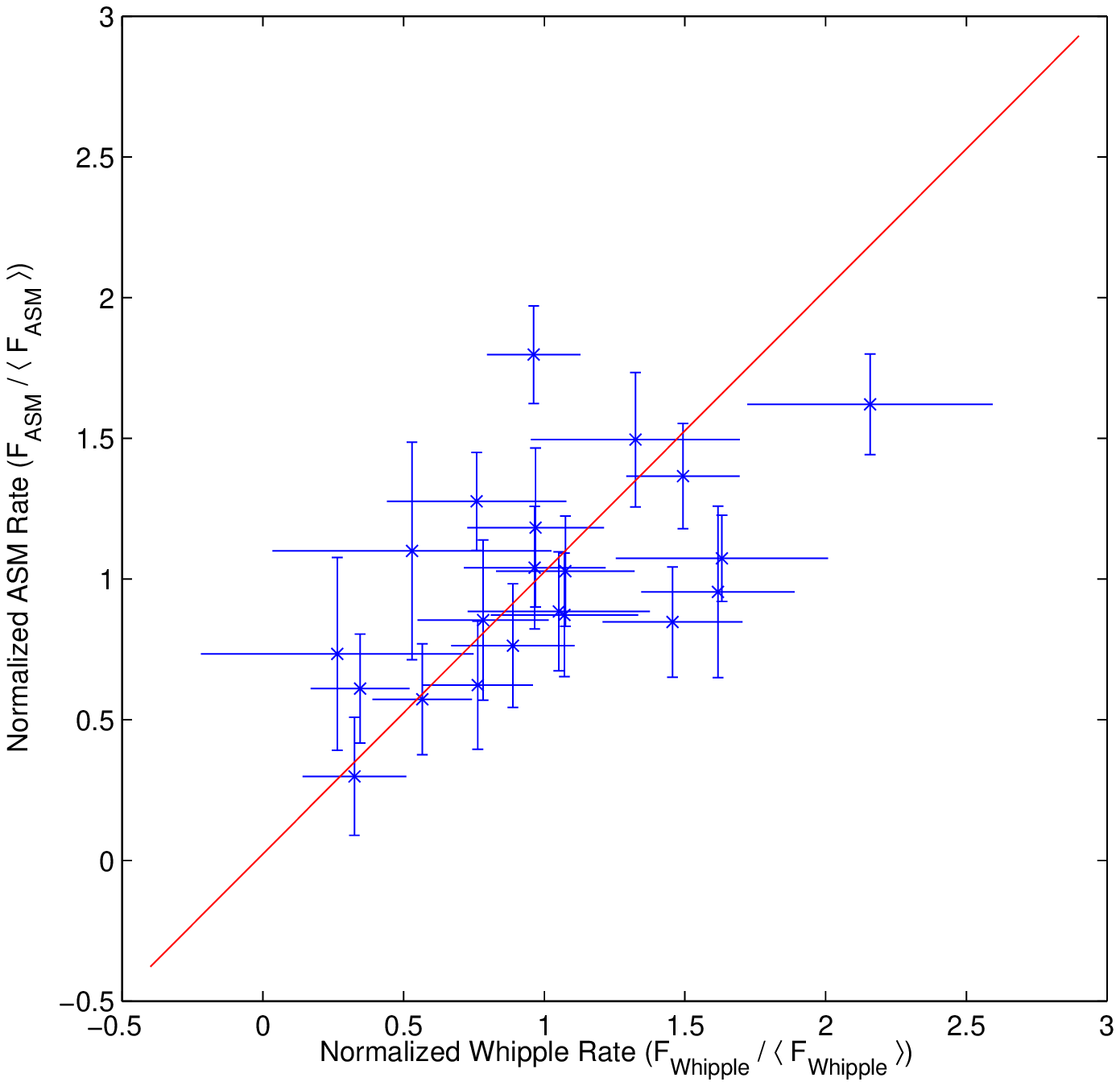}}%
\resizebox*{0.33\textwidth}{!}{\includegraphics[draft=false]{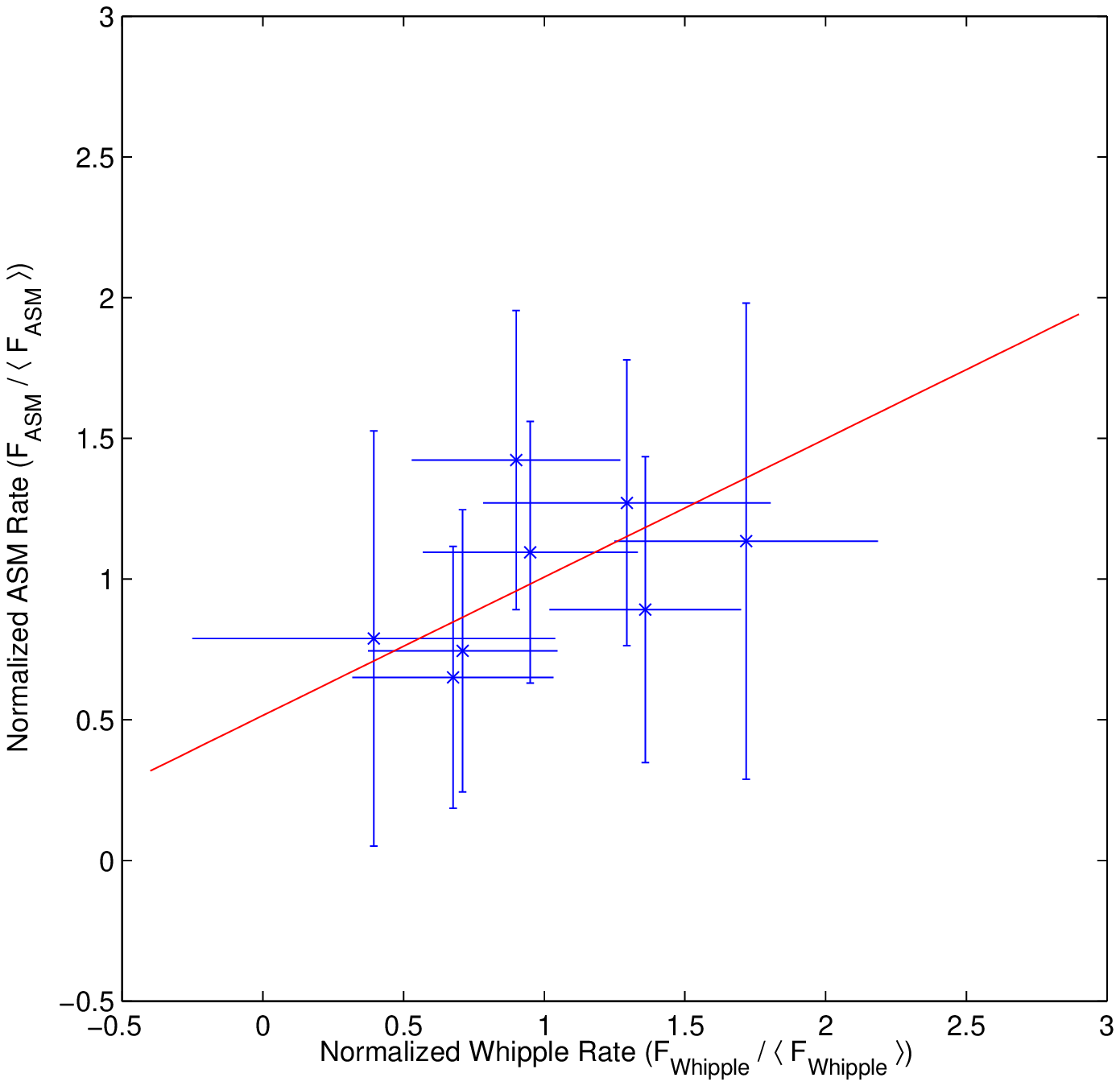}}\\

\caption{\label{FIG:X-GAMMA-ASM}Flux-flux correlation for the
  gamma-ray and ASM data on different timescales. {\it{Left:}} The
  daily-binned gamma-ray and ASM data (slope 1.00); {\it{Middle:}} the
  weekly binned gamma-ray and ASM data (slope 1.00); {\it{Right:}} the
  monthly-binned gamma-ray and ASM data (slope 0.49).}

\end{figure}

All of the X-ray data taken during this campaign are plotted with the
gamma-ray data in Figure~\ref{FIG:X-GAMMA}. For this plot, the XRT and
PCA data are plotted between 2--10\,keV, their counts having been
converted to fluxes using a log-parabolic model to fit the data. The
length of the exposure acquired varies from night to night, even for a
single instrument. There is evidence for correlated variability
throughout the campaign, in particular around the period of MJD 53875
to 53885.  The mean rate over the entire observing campaign for each
waveband is plotted as a horizontal, solid red line on these
plots. The data can be seen to follow similar trend in terms of their
mean flux level. Figure~\ref{FIG:X-GAMMA-CORRELATION} shows flux-flux
diagrams (for 1-day timescales) for each X-ray band with the gamma-ray
data. Each dataset was first normalized by dividing each flux point by
the mean flux in that waveband so that the flux-flux correlations
between the gamma-ray data and each X-ray band could be more easily
compared. Using a weighted total least-squares algorithm
\citep{Anton:07}, which took into account uncertainties in the X-ray
and the gamma-ray data, a straight line was fitted to each of the
flux-flux plots. The slopes these lines are found to increase with
X-ray energy and are 0.73, 1.00, 1.51 and 3.01 for the XRT, ASM, PCA
and BAT data respectively. As would be expected due to their overlap
in energy range, the slopes of the flux-flux relations between the
gamma-ray and the XRT (0.2\,-\,10\,keV), ASM (2\,-\,10\,keV) and PCA
(3\,-\,25\,keV) data are quite close to each other, their differences
presumably arising due to the slightly different energy ranges and
measurement uncertainties. Although the lower energy X-ray data
(i.e. not those from the BAT) and gamma-ray data do look somewhat
correlated, there are outliers on all plots. In general, the gamma-ray
and the lower-energy X-ray data exhibit low and high fluxes at similar
times. To compare the relationship between these X-ray and gamma-ray
data with that derived by \citet{Fossati:08}, the data were re-plotted
on a log-scale, with the gamma-ray data on the Y-axis
(Figure~\ref{FIG:X-GAMMA-LOG-CORRELATION}). The slopes of the lines
were found to be 1.03 for the XRT (0.2 - 10 keV), 0.98 for the ASM (2
- 10 keV), 0.48 for the PCA (3 - 25 keV) and 0.22 for the BAT (15 - 50
keV) data. In the energy range between 2 and 10 keV,
\citet{Fossati:08} found a slope of 0.88, similar to that found here
in that energy range, for the relationship between the PCA and
gamma-ray data from Mrk\,421. When only the significant detections
(defined here as $>$3$\,\sigma$ and highlighted in green on
Figure~\ref{FIG:X-GAMMA-LOG-CORRELATION}) were included in the slope
calculation, the higher-flux data remained and the slopes of the XRT,
ASM and PCA data decreased slightly to 0.94, 0.78 and 0.33
respectively while the relationship between the BAT and the gamma-ray
data became inverted (slope of -0.18), with little evidence for
correlation between the two bands. Flux-flux diagrams were also made
for weekly and monthly timescales. The ASM data in particular were
found to show evidence for correlated emission with the gamma-ray data
on weekly and on monthly timescales as shown in
Figure~\ref{FIG:X-GAMMA-ASM}. The slopes of the lines are 1.0, 1.0 and
0.5 for the daily, weekly and monthly lightcurves respectively.

\begin{figure}
\resizebox*{0.95\textwidth}{!}{\includegraphics[draft=false]{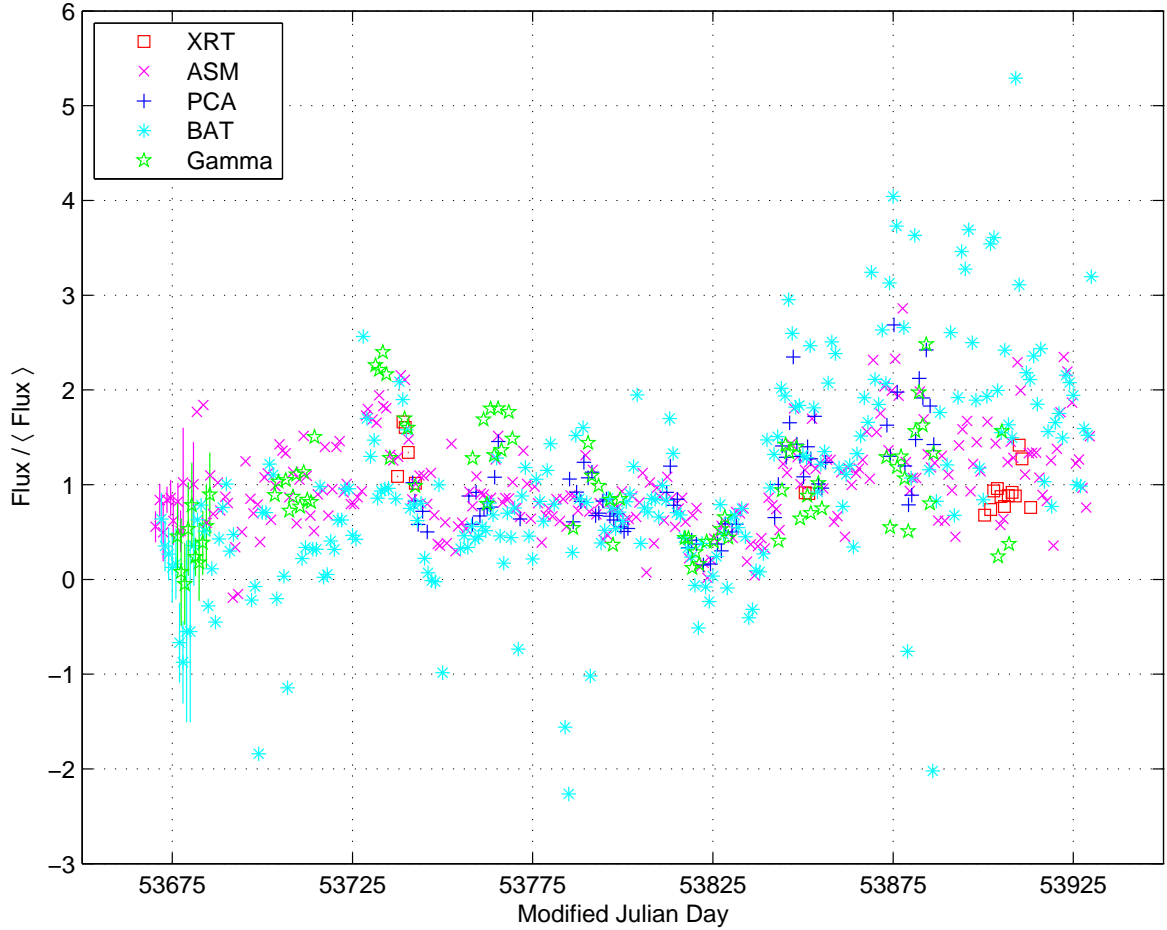}}
\caption{\label{FIG:X-GAMMA-FLUXOVERMEAN}For each of the four X-ray
datasets and for the gamma-ray data, the flux is plotted over the mean
flux in that band. For each dataset, the error-bars are plotted for
the first ten data points so that the typical magnitude of the
uncertainties for each waveband are conveyed.}

\end{figure}

Figure~\ref{FIG:X-GAMMA-FLUXOVERMEAN} shows the flux divided by the
mean flux in that band for each of the X-ray datasets and for the
gamma-ray data. A high degree of correlation can be seen, in
particular between the gamma-ray data and those from the ASM and the
PCA.

We searched for correlations between fluxes in the X-ray bands and the
gamma-ray band using the Discrete Correlation Function (DCF). The DCF,
introduced by \citet{Edelson:1988:dcf}, is an approximation to the
classical correlation function which is applicable to discrete
time-series. It gives the linear correlation coefficient ($R$) for two
light curves as a function of the time lag between them. The DCF is
well-suited to sparsely-sampled lightcurves and is less likely to give
rise to spurious results than a traditional correlation analysis with
interpolated light curves. To construct the DCF, we first collect the
set of unbinned discrete correlations, UDCF$_{ij}$:

\begin{center}
\begin{equation}\label{EQ:UDCF}
UDCF_{ij} = \frac{(a_i -\overline{a})(b_j - \overline{b})}{\sqrt{(\sigma^2_a - e^2_a)(\sigma^2_b - e^2_b)}}
\end{equation}
\end{center}

\noindent{for all pairs of observations a$_i$, b$_j$, with measurement
errors $\sigma_a$, $\sigma_b$, and mean measurement errors of e$_a$,
e$_b$, respectively. Each pair is associated with a time lag
$\Delta$t$_{ij}$ = t$_j$ - t$_i$. The DCF for a given time lag,
$\tau$, is then constructed as}

\begin{center}
\begin{equation}\label{EQ:DCF}
DCF(\tau) = \frac{1}{M} \sum UDCF_{ij}
\end{equation}
\end{center}

\noindent{where the sum runs over the M pairs of observations
separated by $\tau$ - $\delta$t/2 $<$ $\Delta$t$_{ij}$ $<$
$\tau$+$\delta$t/2 where $\delta$t is the chosen bin-width. The
uncertainty on the value of the DCF in a given bin is calculated as
the RMS variance of all the contributing UCDF$_{ij}$ about the value
DCF($\tau$).}

\begin{figure}
\resizebox*{0.95\textwidth}{!}{\includegraphics[draft=false]{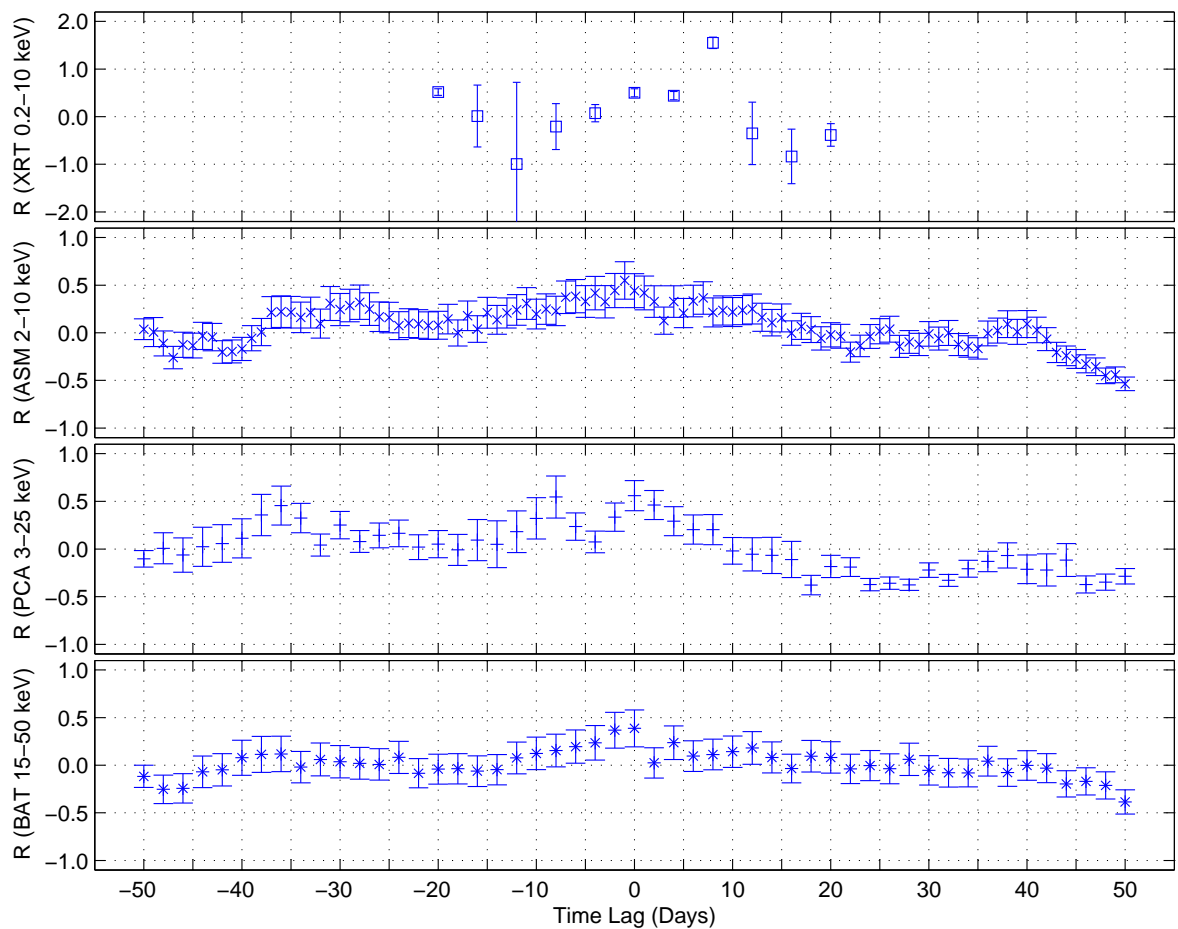}}

\caption{\label{FIG:DCFS-XrayGamma} The discrete correlation function
between {\it{Top:}} the TeV and XRT lightcurves with four-day bins due
to the sparsity of the XRT data points; {\it{Second from top:}} the
TeV and ASM lightcurves with one-day bins; {\it{Second from bottom:}}
the TeV and PCA lightcurves with two-day bins; {\it{Bottom:}} the TeV
and BAT lightcurves with two-day bins; The x-axes are defined such
that a positive value of R means that the X-rays precede the gamma
rays.}
\end{figure}

The DCF for each of the X-ray bands with the gamma-ray data is shown
in Figure~\ref{FIG:DCFS-XrayGamma}.  In these plots, the x-axis is
defined such that in the gamma-PCA DCF for example, a positive lag
means that the gamma-ray data lagged the X-ray data. For the
XRT-gamma-ray DCF, four-day binning is used due to the sparsity of the
XRT data.  For the ASM-gamma-ray DCF, one-day binning is used and, for
the PCA- and BAT-gamma-ray DCF, two-day binning is used. According to
both the SSC and EC models for blazar emission, the same population of
electrons is responsible for the X-ray and the gamma-ray
emission. Therefore, the X-ray and gamma-ray lightcurves should be
correlated. For the well-sampled ASM, PCA and BAT lightcurves, a
zero-lag correlation is seen clearly, with peak values of the DCF at
zero lag of 0.44, 0.56 and 0.39, respectively. The shape of the DCF
looks similar for the BAT and ASM data, as their lightcurves are
similar. The PCA data have the smallest uncertainties and also show
the highest DCF value at zero lag.

To measure the chance probability of accidental correlations in the
DCF, the Monte Carlo method described in \citet{Jordan:04} was
employed. Taking a correlation at zero lag as the a priori hypothesis,
the chance probability of obtaining the measured value of the DCF at
zero lag, with the assumption that the times of the flares are
randomly sequenced, was calculated. 100,000 lightcurves were simulated
with the same characteristic timescale as the PCA, determined by the
SF analysis (Section~\ref{SECTION:SF}). The DCF was calculated for
each of these simulated lightcurves and the gamma-ray lightcurve. The
chance probability of obtaining a DCF value at or above 0.56 from this
Monte Carlo method was found to be 0.1\%. We can state therefore, with
reasonable confidence that the gamma-ray and PCA X-ray data are
correlated within 1.5 days.

\subsection{Optical Correlations with other Wavelengths}

\begin{figure}
\resizebox*{0.33\textwidth}{!}{\includegraphics[draft=false]{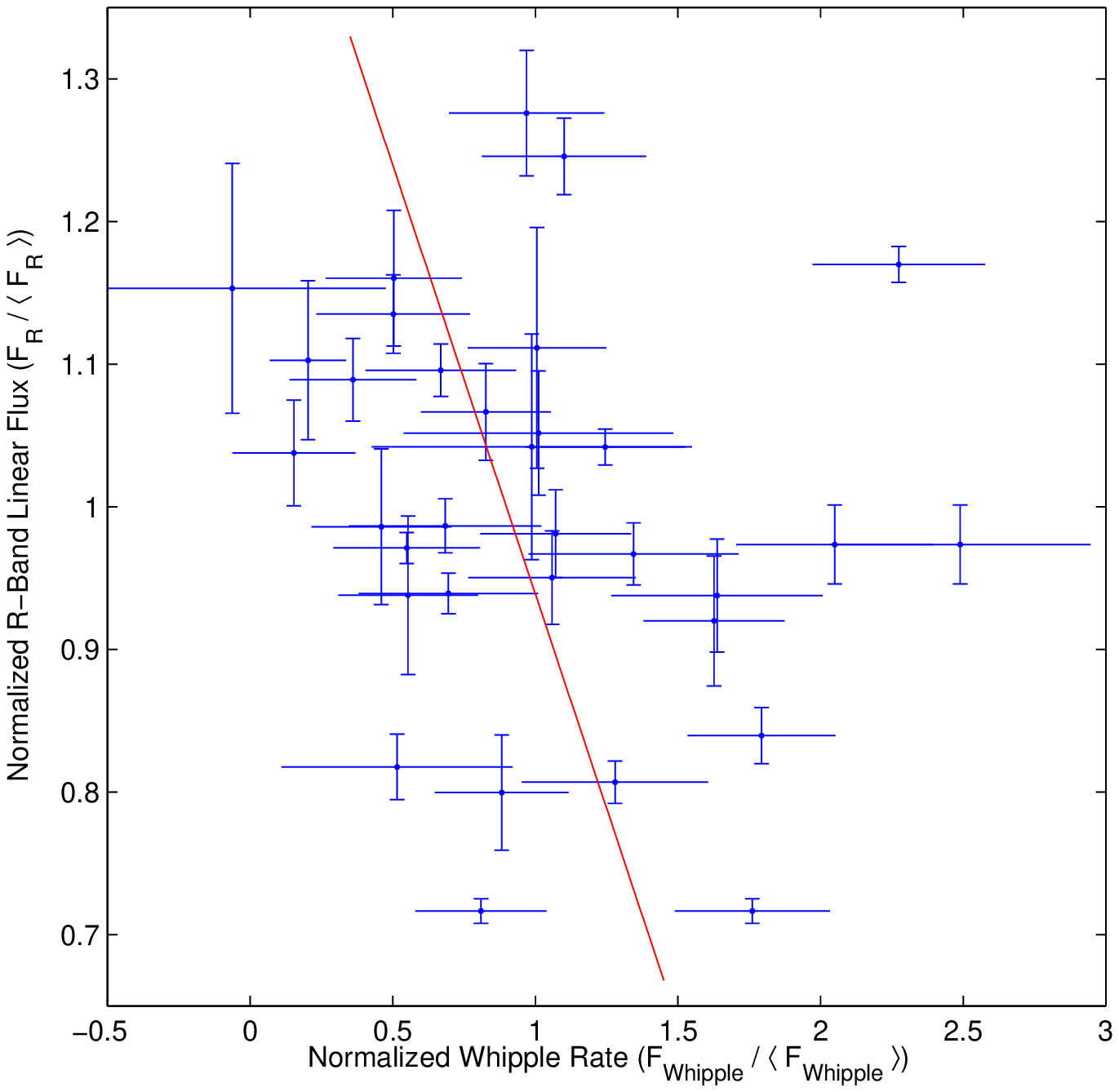}}%
\resizebox*{0.33\textwidth}{!}{\includegraphics[draft=false]{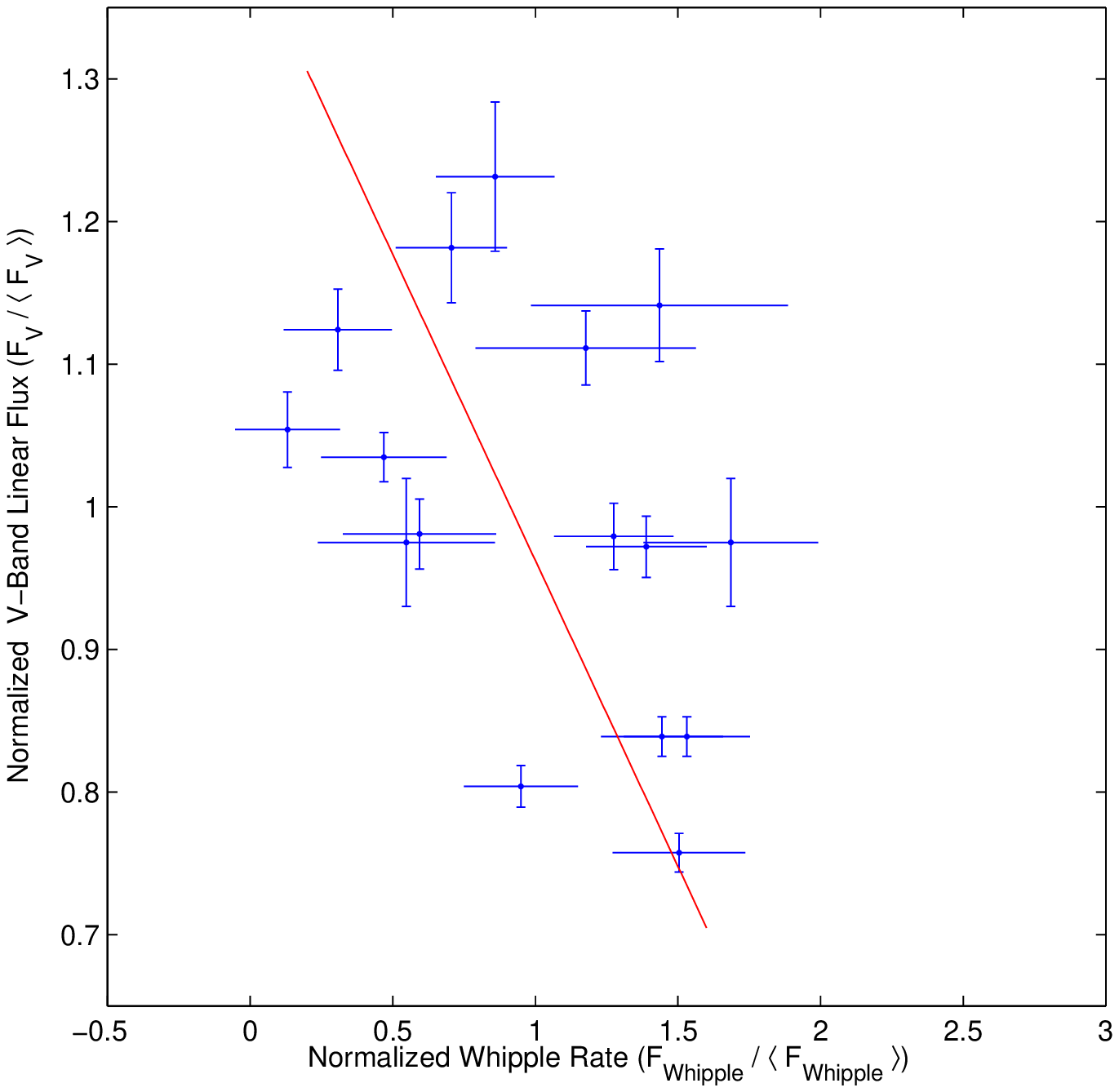}}%
\resizebox*{0.33\textwidth}{!}{\includegraphics[draft=false]{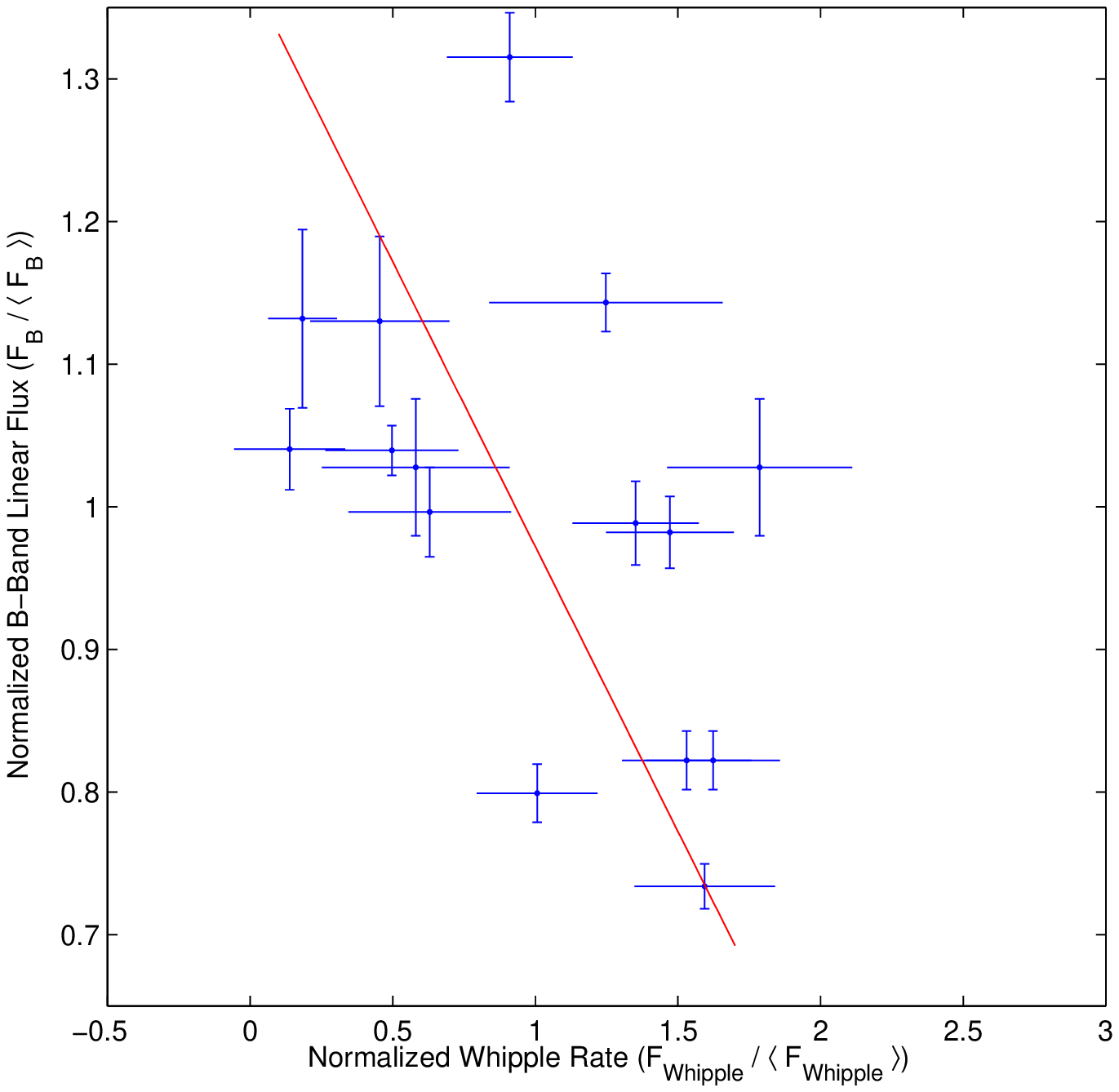}}\\[2ex]%

\caption{\label{FIG:OPT-GAMMA-CORRELATION}Flux-flux correlation for
  {\it{Left:}} gamma-ray and R-band optical data (slope -0.60);
  {\it{Middle:}} gamma-ray and V-band optical data (slope -0.43);
  {\it{Right:}} gamma-ray and B-band optical data (slope -0.40). Note
  that the range of the gamma-ray data is five times that of the
  optical data.}

\end{figure}

Figure~\ref{FIG:OPT-GAMMA-CORRELATION} shows the flux-flux correlation
for the gamma-ray and optical data (linear fluxes). Each dataset was
first normalized by dividing each flux point by the mean flux in that
waveband so that the flux-flux correlations between the gamma-ray and
optical data could be more easily compared. Using a weighted total
least-squares algorithm \citep{Anton:07}, which took into account
uncertainties in the optical and the gamma-ray data, a straight line
was fitted to each of the flux-flux plots. Although not
well-correlated, the data do show a somewhat negative correlative
trend, with the high-flux gamma-ray points tending to coincide with
the low-flux optical points and vice versa. The slopes of the lines
shown in Figure~\ref{FIG:OPT-GAMMA-CORRELATION} are found to decrease
marginally as energy increases, with a slope of -0.60 in the R-band,
-0.43 in the V-band and -0.40 in the B-band.

\begin{figure}
\resizebox*{0.95\textwidth}{!}{\includegraphics[draft=false]{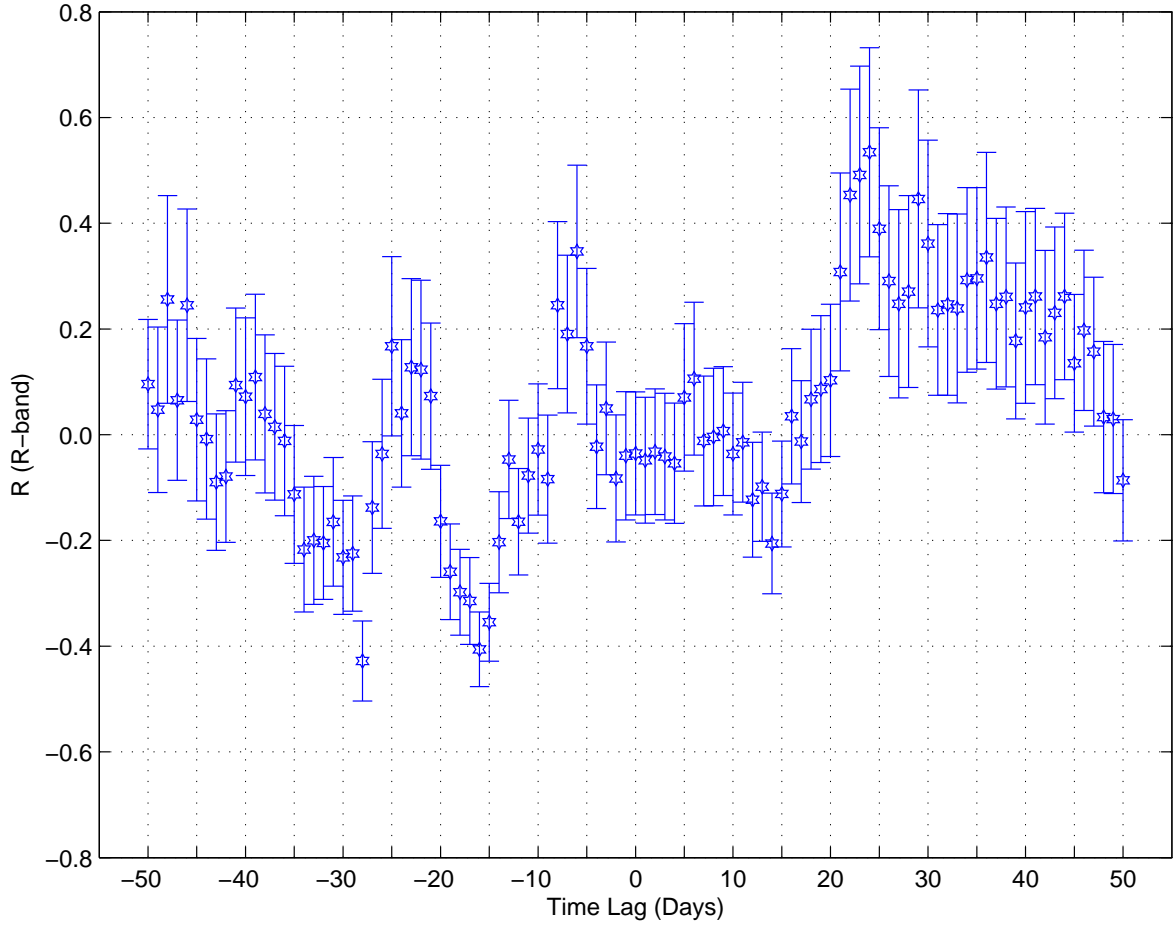}}\\%
\caption{\label{FIG:OPTICAL-GAMMA-DCF}The discrete correlation
function of the optical R-band and gamma-ray data. The x-axis is
defined such that a positive value of R means that the optical photons
precede the gamma rays.}
\end{figure}

We searched for correlations between fluxes in the optical R-band,
which was the best-sampled of the optical lightcurves, and all other
non-optical wavebands presented in this investigation using the
DCF. Of all the pairs of wavebands investigated, only the DCF between
the optical and TeV lightcurves, presented in
Figure~\ref{FIG:OPTICAL-GAMMA-DCF}, shows noteworthy features. The DCF
indicates a possible optical-TeV correlation with the optical flux
lagging the TeV flux by about 7 days.  An elevated level of
correlation is also seen with the optical leading the TeV by 25 to 45
days \citep{Steele:07}.

We investigated the significance of these possible correlations using
simulated optical lightcurves with the same variability properties as
the observed R-band lightcurve. The simulated optical lightcurves were
generated deterministically as a superposition of harmonic oscillators
with random phases whose frequencies are integer multiples of a
fundamental frequency corresponding to the total length of the
campaign. The amplitude of an oscillator at a given frequency is drawn
from a power spectral density (PSD) function derived from the
first-order structure function of the observed R-band
lightcurve. Assuming the process(es) responsible for the optical
variability are stationary over the period of observation, the slope
of the PSD, $\beta$, at a given frequency is related to the slope of
the first-order structure function, $\alpha$, by $\beta = 1 +
\alpha$. Using this relation, we derive the PSD of our simulated light
curves by a fit to the first-order structure function that rises
linearly with a slope $\alpha = 0.568$ until the beginning of the
break at $t = 39.8 d$, is flat ($\alpha = 0$) between $39.8 $ d $ < t
< 63.0$ d, and then rises again with slope $\alpha = 1.700$.

After multiple realizations of the campaign using simulated optical
lightcurves which are, by definition, not correlated with the TeV
lightcurve, we can assess the probability to have seen the observed
correlation features due to chance. For each realization, we compute
the DCF between the simulated optical and the real TeV
lightcurves. For each bin of the DCF, we record the values of the
correlation coefficient, $R$, its error, $\sigma_R$, and the test
statistic $S = R /\sigma_{R}$. After many realizations, we build the
distribution of $S$ seen in each bin of the DCF and compute the chance
probability to have found the observed value of $S = S_{obs}$ or
higher by integrating the distribution above $S_{obs}$. This gives us
the chance probability of having observed a given correlation in a
particular bin, but it does not speak to the likelihood of correlation
\emph{features} involving more than one bin. To assess the likelihood
of the observed DCF features as a whole, we construct two more
distributions, $\mathcal{L}_{\pm} = \prod_{i} p_{i}$ where for
$\mathcal{L}_{-}$ the product runs over all DCF bins, $i$, with $\tau
< 0$, and for $\mathcal{L}_{+}$ the product runs over bins with $\tau
>= 0$.  Using these two distributions, we find the likelihood to have
observed the optical precursor (postcursor) features by chance to be
$20\%$ ($60\%$). Further investigation reveals that the DCF feature
seen at $\tau = 7$ d probably arises due to contamination from the
R-band auto-correlation function, which also exhibits features at
multiples of 7 days. The possible reality of the optical precursor
emission is more difficult to assess since our method assesses all the
correlation features on a given side of the DCF at
once. Unfortunately, it would not have been possible to narrow the
likelihood computation to smaller range of $\tau$ in order to assess a
particular feature without subjecting the result to an unknown trials
penalty.

\subsection{Spectral Energy Distribution}
\label{SECTION:SED}

\begin{figure}
\rotatebox{270}{\resizebox*{0.65\textwidth}{!}{\includegraphics[draft=false]{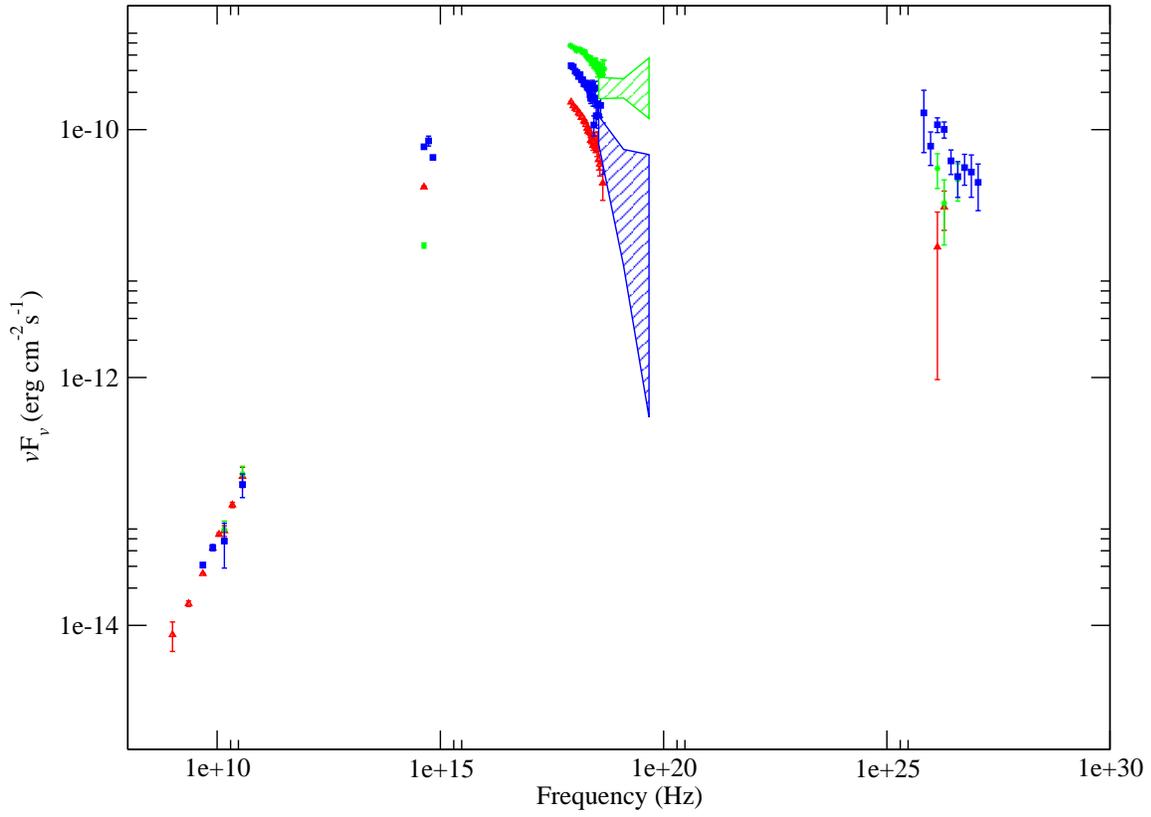}}}\\%
\caption{\label{FIG:SED-3NIGHTS} Spectral energy distribution for
  Mrk\,421 on three nights during this MWL campaign when the VHE
  emission, as defined by the gamma-ray data, was at high (MJD 53763;
  filled blue squares), medium (MJD 53852; green filled circles) and
  low (MJD 53820; closed red triangles) emission levels.}
\end{figure}

Spectral energy distributions (SEDs) are shown on three representative
dates in Figure~\ref{FIG:SED-3NIGHTS} when the VHE emission was at a
high (MJD 53763; filled blue squares), medium (MJD 53852; open green
circles) and low (MJD 53820; closed red triangles) emission level;
these dates were chosen because there were multiwavelength data
available over the full spectrum. They were not chosen where there was
the best correlation with the flux at X-ray wavelengths and hence the
SEDs are more typical. These nights are marked on the gamma-ray
lightcurve in Figure~\ref{FIG:GAMMA}. On MJD 53763, 135 minutes of
non-contiguous gamma-ray data were taken; on MJD 53852, 138 minutes of
non-contiguous gamma-ray data were taken while on MJD 53820, 90
minutes of non-contiguous gamma-ray data were taken. The optical data
shown on the SED has the galaxy contribution subtracted. There was no
significant variation in the radio fluxes over the three dates. The
X-ray spectrum for the PCA data was obtained using
XSPEC\footnote{http://heasarc.nasa.gov/xanadu/xspec/} while the TeV
spectrum was obtained using the forward-folding method described in
\citet{Rebillot:06}. Mrk\,421 was in the field of view of the BAT
during parts of each of the three observations. It was detected on MJD
53763 and MJD 53852 and upper limits were obtained for MJD 53820. A
spectral analysis of the BAT survey mode data on Mrk\,421 was
performed for two nights, MJD 53763 and MJD 53852. The BAT data from
these nights comprised 74 minutes and 44 minutes respectively and were
processed using standard Swift
FTOOLS\footnote{http://heasarc.gsfc.nasa.gov/docs/software/lheasoft}
to remove the diffuse background and the effects of other bright point
sources in the field of view.  Eight-channel spectral and response
files were produced for the position of Mrk\,421 for each observation
and were fit using XSPEC 11 to a simple power-law model to provide
fluxes in the energy range of 14-195 keV. The highest energy at which
a significant detection was made with the BAT was 75\,keV.

The data from the nights chosen to represent the high-state and
medium-state of Mrk\,421 are plotted in Figures~\ref{FIG:SED-MODEL1}
and ~\ref{FIG:SED-MODEL2} along with archival Mrk\,421 data
\citep{Buckley:00}. The dashed purple lines show the synchrotron and
self-Compton distributions for the parameters given in
Table~\ref{TAB:PARAMETERS}. The black dot-dashed curves show a
hypothetical black-body component peaked at 1 $\mu$m and the
corresponding external Compton component. The red solid line shows the
sum of the SSC and EC models fitting results, which, in both cases, is
in good agreement with the simultaneous optical, X-ray and gamma-ray
data.

\section{Discussion \& Conclusion}
\label{discussionAndConclusion}

The Mrk\,421 observations presented in this paper comprise one of the
most comprehensive datasets, both in terms of spectral and temporal
coverage, ever accumulated on a blazar. The well-sampled optical
lightcurves, in particular, differentiate this campaign from others
undertaken previously. Like some of these, this campaign was not
triggered by Mrk\,421 entering some pre-defined emission
state. Rather, data were gathered regardless of the flux level
observed from Mrk\,421 in any particular waveband. Great care was
taken to ensure that the data were calibrated properly in each
waveband so that the effects of changing atmospheric conditions,
nightly sky-brightness fluctuations or instrumental differences were
removed from the data.

The broadband temporal variability of Mrk\,421 was examined by
computing the fractional rms variability amplitude and the
point-to-point fractional rms variability amplitude for each waveband
in which it was detected consistently throughout the campaign. A
general trend of increasing F$_{var}$ towards higher frequencies was
observed, with the radio and optical data exhibiting very low
variabilities while the X-ray and gamma-ray data had high fractional
rms variability amplitudes. \citet{Blazejowski:05} probed the
fractional rms variability amplitude for Mrk\,421 data taken during
2003\,-\,2004. These data covered a subset of the wavebands presented
here (14.5 GHz radio, R-band optical, 3-25 keV X-ray and 0.2-20\,TeV
gamma ray) and, like the 2005\,-\,2006 data, were binned in one-day
bins. Except for the 14.5\,GHz radio data, F$_{var}$ was always larger
for the 2003-4 dataset. In general however, the differences between
the F$_{var}$ values for the 2005\,-\,2006 data and 2003\,-\,2004
data, are small, ranging between 15\% and 33\%.

\begin{figure}
\resizebox*{0.95\textwidth}{!}{\includegraphics[draft=false]{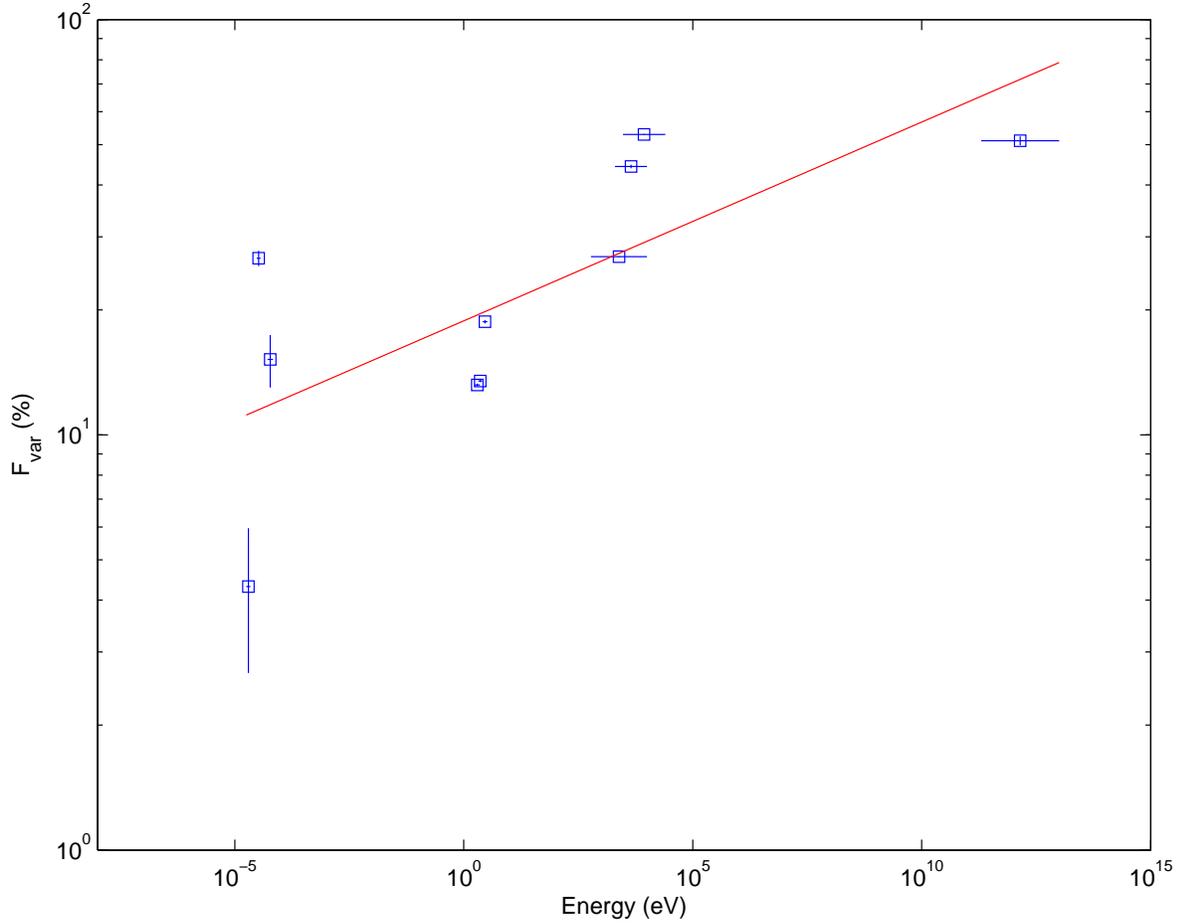}}
\caption{\label{FIG:FVAR-PWRLAW}The fractional rms variability
  amplitude as a function of Energy. The red line shows the best fit
  power law to these data. Unlike previous campaigns during which the
  variability was probed over shorter timescales, the data are not
  well-fit by a power-law (probability for this fit is 4.9\%).}
\end{figure}

\citet{Fossati:00} measured F$_{var}$ for Mrk\,421 in 1997 and 1998 in
the X-ray range from 0.26 to 4.76\,keV using data taken with
{\it{Beppo}}SAX. A trend of increasing F$_{var}$ with increasing
energy was observed. The fractional rms variability amplitudes
computed here for the X-ray data during 2005\,-\,2006 were all higher
than those obtained in 1997\,-\,1998. However, the 1997\,-\,1998 data
had much finer binning applied (200\,s\,-\,2500\,s) and covered a
narrower energy range (0.26 to 4.76 keV) than the lightcurves shown
here so, although the same trend of increasing F$_{var}$ with energy
was observed, the variability was being probed on different timescales
and over a different energy range than those shown
here. \citet{Giebels:07} found evidence for a power-law behavior of
F$_{var}$ over four decades of energy for Mrk\,421 on MJD
51991\,-\,51992, with F$_{var}$ $\propto$
E$^{0.24\pm0.01}$. Figure~\ref{FIG:FVAR-PWRLAW} shows F$_{var}$ as a
function of energy for Mrk\,421 during the campaign presented
here. The uncertainties on F$_{var}$ were calculated as described in
\citet{Vaughan:03}. Unlike the data from MJDs 51991 \& 51992, the
2005\,-\,2006 data are not well-described by a power-law fit. The
timescales being probed are much longer here however, and also, a
larger energy span (15 decades) is available so the engine is being
probed in quite a different regime.

The BAT data (15 - 50 keV) had the highest value of F$_{var}$ at 0.993
but, due to the fact that a statistically significant detection
(defined here as $>$3$\,\sigma$) was only obtained on 42.3\% of the
nights, the BAT data are not considered a consistently significant
enough detection to include their F$_{var}$ and F$_{pp}$ values in the
final analysis. The BAT energy band falls just at the upper end of the
synchrotron peak for Mrk\,421 so, presumably, given that the
synchrotron peaks of BL Lacs are known to shift in strength and in
energy on many timescales, there were many nights on which the X-ray
flux from Mrk\,421 was below the detection threshold of the BAT due to
the shifting synchrotron peak. X-ray photons were detected up to 75
keV with the BAT during this campaign with no significant detection
above these energies. The BAT energy regime was the waveband in which
the most dramatic increase (a factor of 5.3) over the mean flux level,
which was usually very low, was observed. This occurred on MJD 53909,
after the end of the gamma-ray observing campaign. The second and
third highest nights in the BAT energy regime were MJDs 53875 \&
53876, when fluxes of 4.0 and 3.7 the mean level (respectively) were
observed. No gamma-ray data were taken on MJD 53875 but on MJDs 53874
and 53876, the gamma-ray flux was at 0.55 and 1.30 its mean level. The
maximum nightly gamma-ray rate of 2.5 times the mean nightly rate was
recorded approximately 10 days later on MJD 53884. As can be seen in
Figure~\ref{FIG:X-GAMMA-FLUXOVERMEAN}, the X-ray and gamma-ray data
are, in general, correlated with similar levels of fluctuation being
present in the lower-energy X-ray data (XRT, ASM and PCA) and the
gamma-ray data but with larger variations often occurring at the higher
energies covered by the BAT.

As shown in Figures~\ref{FIG:X-GAMMA-CORRELATION} and
~\ref{FIG:X-GAMMA-LOG-CORRELATION}, where all data taken within one
day of each other are plotted, there is evidence for a correlation
between the gamma-ray data and the XRT, ASM and PCA data (i.e. 0.2 -
25 keV) in both linear and log-space but, no evidence for a positive
correlation with the BAT data (15 - 50 keV). The slopes of the
best-fit lines to the X-ray and gamma-ray data in
Figure~\ref{FIG:X-GAMMA-LOG-CORRELATION} are consistent with that
found by \citet{Fossati:08} in the 2 to 10 keV band when, unlike
during the campaign described here, Mrk\,421 was in an active state.

Figure~\ref{FIG:X-GAMMA-ASM} shows the flux-flux correlation for the
gamma-ray and ASM data on daily, weekly and monthly timescales. The
data show evidence for correlation on all of these timescales. The DCF
shows a similar shape for the ASM, PCA and BAT data with a peak
visible at zero-lag for each dataset
(Figure~\ref{FIG:DCFS-XrayGamma}). As described in
Section~\ref{SEC:X-GAMMA}, simulated lightcurves indicated that the
chance probability of this feature for the PCA data is low (0.1\%),
lending support to SSC and EC models, where the X-ray and gamma-ray
emission are produced by the same population of accelerated electrons.

The well-sampled optical lightcurve, in particular that in the R-band,
allowed a detailed correlation analysis to be carried out between the
optical and TeV data. As can be seen in
Figure~\ref{FIG:OPT-GAMMA-CORRELATION}, these data were not found to
be well-correlated. Although a peak at -7 days was seen in the
optical-gamma-ray DCF (Figure~\ref{FIG:OPTICAL-GAMMA-DCF}), indicating
that the optical lags the TeV, it was shown that the likelihood of
seeing such a peak by chance was 20\% and that a feature in the
autocorrelation function of the R-band data, which occurs at multiples
of 7 days, was likely responsible for this feature in the DCF.

\begin{figure}
\rotatebox{270}{\resizebox*{0.65\textwidth}{!}{\includegraphics[draft=false]{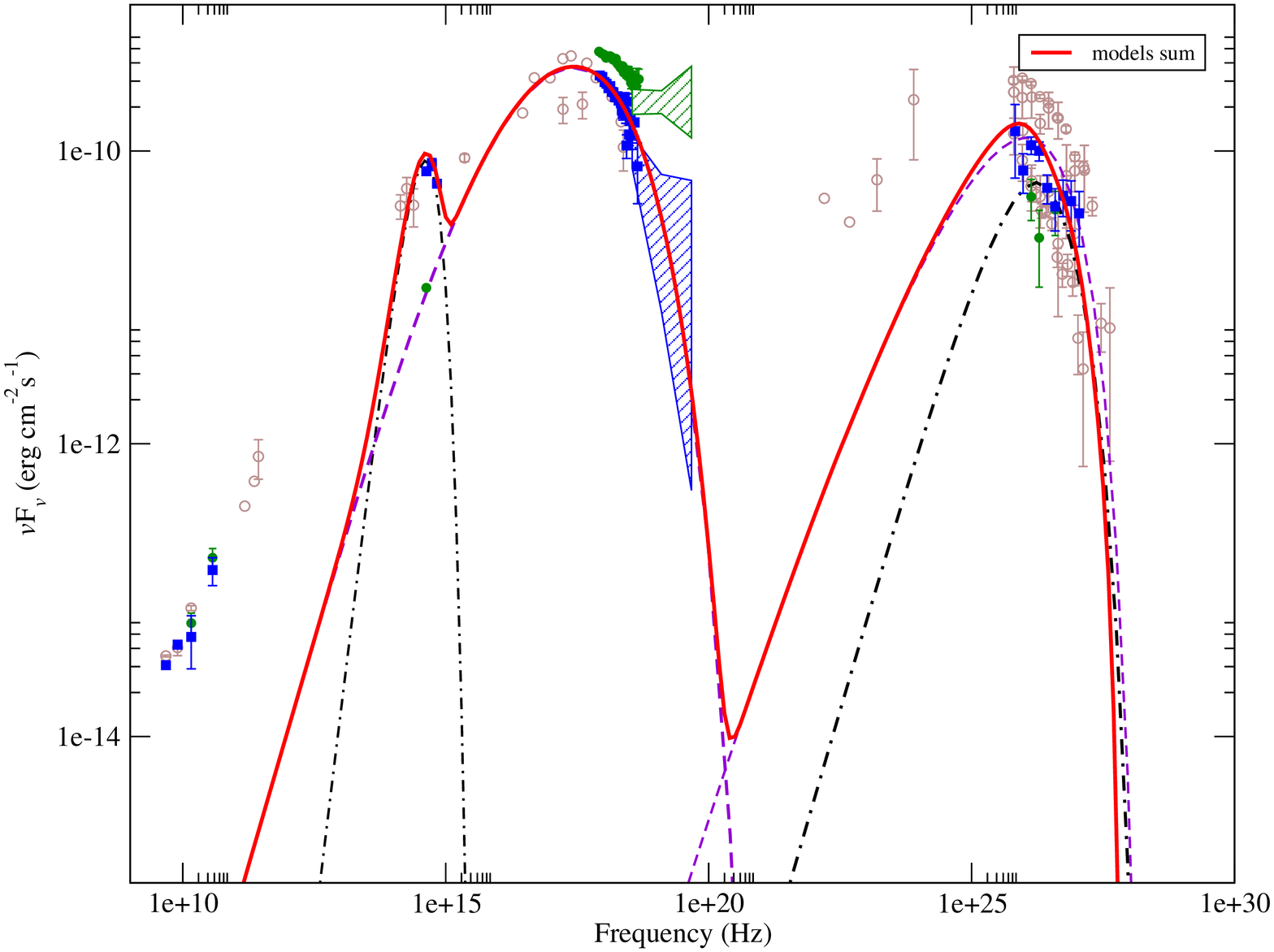}}}%
\caption{\label{FIG:SED-MODEL1}Fit of simple model to the spectral
  energy distribution for Mrk\,421 on MJD 53763, the high-state night
  (as defined by the gamma-ray data) are shown for all wavebands as
  blue filled squares. For reference, the data from the medium-state
  night (Figure~\ref{FIG:SED-MODEL2}) are also shown (green filled
  circles). Archival data are shown as brown open circles
  \citep{Buckley:00}. Details of the model fit to the data are given
  in the text. The data are fit with a combined SSC and EC model. The
  dashed purple line shows the synchrotron and self-Compton
  distributions for the parameters given in the text. The black
  dot-dashed curve shows a hypothetical black-body component peaked at
  1 $\mu$m and the corresponding external Compton component. The red
  solid line shows the sum of the SSC and EC models fitting results,
  which is in good agreement with the simultaneous optical, X-ray and
  gamma-ray data. The archival data on the SED show that the level of
  the radio emission did not change significantly over the past 10
  years. The 2005-6 radio data were taken within +6/-5 days of the
  low, medium and high state data.}
\end{figure}

\begin{figure}
\rotatebox{270}{\resizebox*{0.65\textwidth}{!}{\includegraphics[draft=false]{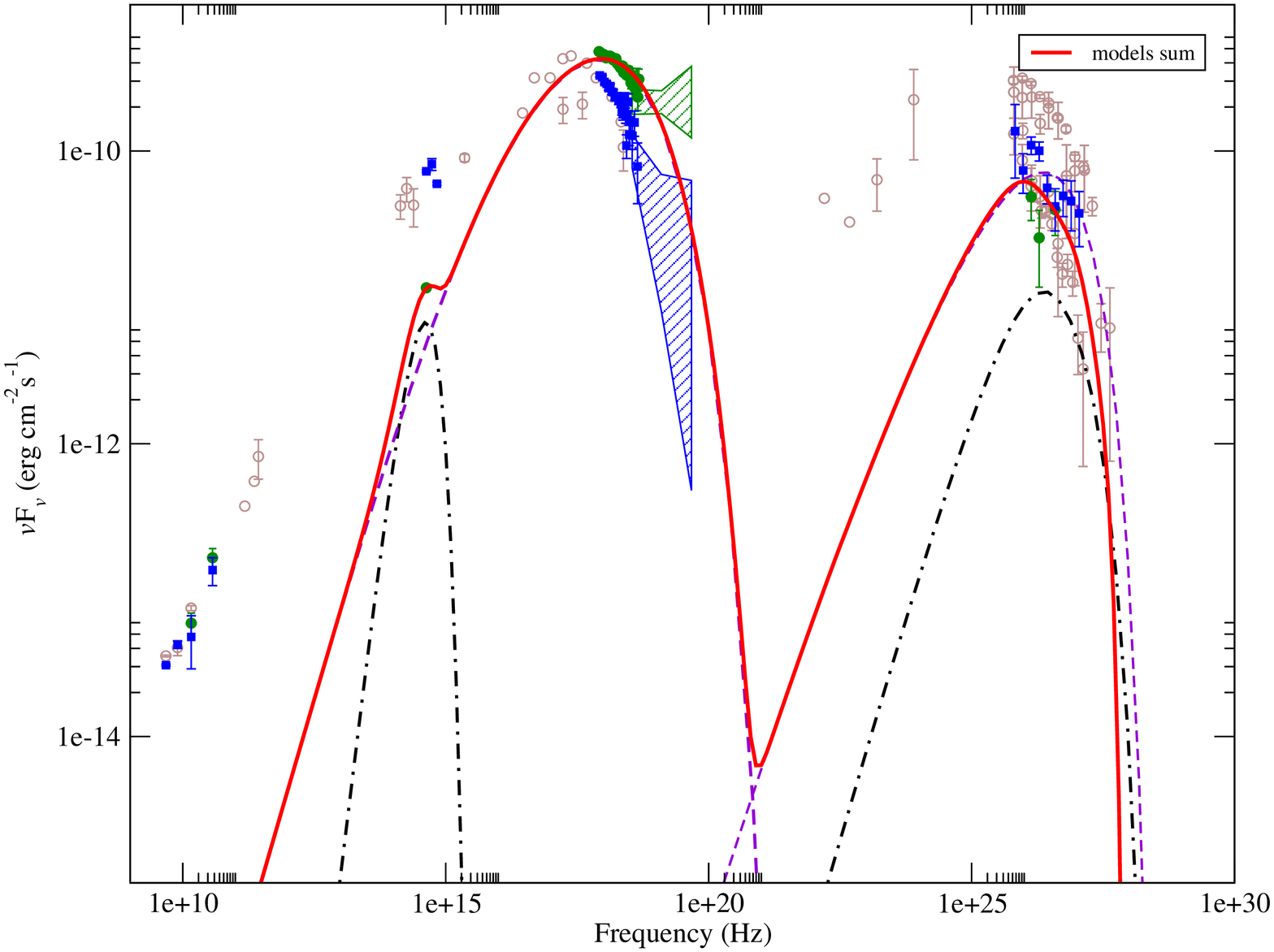}}}\\%
\caption{\label{FIG:SED-MODEL2}Fit of simple model to the spectral
  energy distribution for Mrk\,421 on MJD 53852, the medium-state
  night (as defined by the gamma-ray data) are shown for all wavebands
  as green filled circles. For reference, the data from the high-state
  night (Figure~\ref{FIG:SED-MODEL1}) are also shown (blue filled
  squares). Archival data are shown as brown open circles
  \citep{Buckley:00}. Details of the model fit to the data are given
  in the text. The data are fit with a combined SSC and EC model. The
  dashed purple line shows the synchrotron and self-Compton
  distributions for the parameters given in the text. The black
  dot-dashed curve shows a hypothetical black-body component peaked at
  1 $\mu$m and the corresponding external Compton component. The red
  solid line shows the sum of the SSC and EC models fitting results,
  which is in good agreement with the simultaneous optical, X-ray and
  gamma-ray data. The archival data on the SED show that the level of
  the radio emission did not change significantly over the past 10
  years. The 2005-6 radio data were taken within +6/-5 days of the
  low, medium and high state data.}
\end{figure}

For the high- and medium-state nights (MJD 53763 and 53852,
respectively) we compare our results to a simple one-zone SSC model
with an external Compton component. We make an ad-hoc assumption of
Comptonization of a one micron emission-component reflected back into
the jet. Using the prescription of \citet{Inoue:96} we determine the
break energy and maximum energy of the electron spectrum using a
simple model of diffusive shock acceleration. While the fraction of
light scattered back into the jet is a completely free parameter, the
maximum electron energy, pair absorption in the source, and shape of
the IC peak are all determined in a self-consistent way. The resulting
model fits to the SED are shown for the high-state night in
Figure~\ref{FIG:SED-MODEL1} and for the low-state night in
Figure~\ref{FIG:SED-MODEL2}. We note that the lightcurves of Mrk\,421
in this campaign as well as in previous studies show roughly symmetric
flares with comparable rise and fall times. By fitting the measured
optical and X-ray structure functions to those generated from
simulated lightcurves (composed of a random sequence of triangular
flares) we derive a characteristic rise-time of approximately 0.6 days
in the X-ray variability, and between 10 and 20 days in the
optical. This is in general agreement with the difference in cooling
times for the electron populations generating the optical and X-ray
synchrotron emission, where we expect: 

\begin{center}
\begin{displaymath}
\tau_{cool,opt} = \sqrt{\frac{\omega_X}{\omega_{opt}}}\cdot\tau_{cool,X} \approx \sqrt{1000}\, \cdot 0.6\,\rm{days} = 19\,\rm{days}
\end{displaymath}
\end{center}

As detailed in the following paragraph, with the simple model adopted
here, it is difficult to obtain a good fit to the rather hard
power-law TeV emission without invoking extreme values for some of the
model parameters (high Doppler factor, low magnetic field and small
emission region) perhaps indicating that a more detailed theoretical
analysis is warranted. We find a low value for the magnetic field (B
$\sim$ 0.12\,Gauss) similar to the value of 0.1\,Gauss found by
\citet{Lichti:08} during June 2006. Their best fits however, gave a
significantly lower value for the Doppler factor ($\delta=15$),
compared with the value ($\delta=90$) found here, which we note is
similar to that found in some previous studies
(\citealp{Krawczynski:01}; \citealp{Rebillot:06}). This high value for
the Doppler factor results in some inconsistencies in the shock
acceleration parameters namely, a very hard electron spectrum ($\sim
E^{-1.5}$) contrary to the predictions for non-relativistic diffusive
shock acceleration (giving E$^{-2.0}$) or ultra-relativistic shock
acceleration (giving E$^{-2.2}$). The acceleration efficiency would
have to be quite low, with the mean-free-path for scattering $\sim$20
times the Bohm limit and a shock velocity of 0.02$c$, resulting in
relatively few shock crossings. If the extreme parameters suggested by
the simple model adopted here are correct, these observations lend
some support to a new model for relativistic shock acceleration
\citep{Stecker:07}, which predicts a spectral index of 1.5 or, they
could indicate an even more dramatic departure from the
shock-acceleration model, such as the Poynting jet model described in
\citet{Krawczynski:07}. It is clear however, that more detailed
observations and modeling are necessary before stronger conclusions on
such models can be drawn. The parameters used for both fits shown in
Figures~\ref{FIG:SED-MODEL1} and ~\ref{FIG:SED-MODEL2} are listed in
Table~\ref{TAB:PARAMETERS}. The only differences between the
medium-state and high-state fits in these fits are the fraction of the
optical light and the luminosity of the optical emitting region. We
note that the flux in the X-ray band was higher and the spectrum
harder on the 'medium' state gamma-ray night and was lower and softer
than this on the 'high' state gamma-ray night.

For Markarian\,501 \citet{Ghisellini:05} used external seed photons
from a slow jet component to reproduce the TeV emission with lower
Doppler factors. However, when we attempt to fit the Mrk\,421 data in
this way, we find that we can no longer fit the detailed shape of the
declining X-ray and TeV spectrum. While there are too many free
parameters to find a unique solution, or to draw definitive
conclusions, it is clear that a large contribution from an external
Compton component is a strong possibility for Mrk\,421. While the data
do not demand an external Compton component, the additional cooling
may be present and can reduce the maximum energy for shock
acceleration, somewhat mitigating the aforementioned
problems. Increasing the cooling further quickly results in a
Compton-catastrophe, overproducing the observed gamma-ray emission.

The radio portion of the SED is not well-fit by this single-zone
model. Although the TeV blazars are weak radio sources, the data have
sufficiently good statistics to indicate that strong variability,
characteristic of blazars at higher energies, is not present in the
radio band so, it is likely that a different population of particles
is responsible for the synchrotron emission at these energies.

The AGN monitoring campaign is ongoing at the Whipple 10m
telescope. In addition to providing long-term lightcurves on TeV
blazars, which are even more relevant now that the {\it{Fermi}} Space
Telescope (formally GLAST) is providing longterm monitoring of the sky
in the MeV - GeV energy range, these observations can also be used to
trigger VERITAS and other VHE detectors when any of the AGN bring
monitored enter a high emission state \citep{Swordy:08}.

\section{Acknowledgments}
The authors would like to thank Emmet Roache, Joe Melnick, Kevin
Harris, Edward Little, and all of the staff at the Whipple Observatory
for their support. We also thank the anonymous referee for providing
us with constructive suggestions, which were implemented to improve
this paper. This research was supported in part by the
U. S. Department of Energy, the National Science Foundation, the
Smithsonian Institution, Science Foundation Ireland, by PPARC in the
U.K. and by NSERC in Canada. The ASM quicklook results were provided
by the ASM/RXTE teams at MIT and at the RXTE SOF and GOF at NASA's
GSFC. The UMRAO team acknowledges the support of the NSF (AST 0607523)
and the support of the University of Michigan.  YYK is a Research
Fellow of the Alexander von Humboldt Foundation. \mbox{RATAN--600}
observations are partly supported by the Russian Foundation for Basic
Research (projects 01-02-16812, 05-02-17377,
08-02-00545). \facility[NRAO(VLBA)]{The National Radio Astronomy
  Observatory is a facility of the National Science Foundation
  operated under cooperative agreement by Associated Universities,
  Inc.}  This research has made use of the MOJAVE \citep{Lister:05}
and 2 cm Survey \citep{Kellermann:04} programs database. The
Mets\"ahovi team acknowledges support from the Academy of
Finland. Acquisition and analysis of Swift data was supported at PSU
through NASA grant NNX08AC38G, as part of the Swift Cycle 3 GI
program.

\begin{deluxetable}{llccc}
\tablewidth{0pt}
\tablecaption{The Mrk\,421 dataset presented in this paper. \label{TAB:ALL}}

\tablehead{       &                    &\colhead{Energy}     &\colhead{No. Data} & \colhead{MJD}\\
\colhead{Waveband}&\colhead{Instrument}&\colhead{Range (eV)} &\colhead{points}   & \colhead{Range}}
\startdata
Radio             & RATAN 1 GHz      & 3.7 - 4.6  x 10$^{-6}$     &   2 & 53821 \& 53890 \\
                  & RATAN 2.3 GHz    & 8.6 - 10.5 x 10$^{-6}$     &   2 & 53821 \& 53890 \\
                  & RATAN 4.8 GHz    & 1.8 - 2.2  x 10$^{-5}$     &   2 & 53821 \& 53890 \\
                  & UMRAO 4.8 GHz    & 1.8 - 2.2  x 10$^{-5}$     &  12 & 53706 - 53918  \\
                  & RATAN 8 GHz      & 3.0 - 3.6  x 10$^{-5}$     &   2 & 53821 \& 53890 \\
                  & UMRAO 8 GHz      & 3.0 - 3.6  x 10$^{-5}$     &  16 & 53650 - 53899  \\
                  & RATAN 11 GHz     & 4.1 - 5.0  x 10$^{-5}$     &   2 & 53821 \& 53890 \\
                  & UMRAO 14.5 GHz   & 5.2 - 6.8  x 10$^{-5}$     &  22 & 53664 - 53907  \\
                  & VLBA 15 GHz      & 5.6 - 6.8  x 10$^{-5}$     &   1 & 53830          \\
                  & RATAN 22 GHz     & 8.2 - 10.0 x 10$^{-5}$     &   2 & 53821 \& 53890 \\
                  & Metsahovi 37 GHz & 1.5 - 1.6  x 10$^{-4}$     &  31 & 53650 \& 53905 \\
\hline
Optical           & Abastumani       & 1.7 - 2.3          &  31 & 53801 - 53896  \\
                  & Bell             & 1.8 - 2.2          &  19 & 53824 - 53909  \\
                  & Bradford         & 1.7 - 2.2          &  35 & 53771 - 53922  \\
                  & FLWO             & 1.8 - 2.3          &  64 & 53663 - 53898  \\
                  & Tenagra          & 1.7 - 2.2          &  13 & 53726 - 53876  \\
                  & Torini           & 1.7 - 2.3          &   4 & 53757 - 53838  \\
                  & Bordeaux         & 2.1 - 2.5          &   9 & 53819 - 53861  \\
                  & Bradford         & 2.1 - 2.5          &  36 & 53771 - 53922  \\
                  & Tenagra          & 2.1 - 2.5          &  13 & 53726 - 53876  \\
                  & WIYN             & 2.1 - 2.5          &  19 & 53842 - 53872  \\
                  & Bradford         & 2.7 - 3.2          &  31 & 53771 - 53896  \\
                  & Tenagra          & 2.6 - 3.2          &  13 & 53726 - 53876  \\
                  & WIYN             & 2.6 - 3.3          &  19 & 53842 - 53872  \\
\hline
X-ray             & XRT              & 0.2 - 10 x 10$^3$    &  24 & 53737 - 53914  \\ 
                  & ASM              & 2 - 10 x 10$^3$      & 256 & 53670 - 53930  \\
                  & PCA              & 3 - 25 x 10$^3$      &  75 & 53741 - 53887  \\
                  & BAT              & 15 - 50 x 10$^3$     & 234 & 53670 - 53931  \\
\hline
Gamma-ray         & Whipple          & 0.2 - 10 x 10$^{12}$ &  80 & 53676 - 53908  \\
\enddata

\end{deluxetable}

\begin{deluxetable}{lcccccc}
\tablewidth{0pt} \tablecaption{\label{TAB:OPTICAL} Listed are the
participating optical observatories and instruments, camera type,
filter system, number of nightly data points contributed per filter,
and the range of dates where observing took place. Some information is
unavailable for the Bradford and Bordeaux observatories.  For the FLWO
1.2m, the filter system is "SDSS" for Sloan Digital Sky Survey.}

\tablehead{                      &                   & \colhead{Filter}         & \multicolumn{3}{c}{No. Nights}          & \colhead{MJD}   \\
           \colhead{Observatory} & \colhead{Camera}  & \colhead{System}         &  R & V & B                              & \colhead{Range}}
\startdata
Abastumani 0.7 m                 &  ST-6             &  Cousins                 &  31 &     &          & 53801 - 53896 \\
Bell 0.6 m                       &  Appogee AP2P     &  Bessel                  &  19 &     &          & 53824 - 53909 \\
Bordeaux                         &  Th7896M          &  GC495+BG38              &     &   9 &          & 53819 - 53861 \\
Bradford 0.36 m                  &  FLI MaxCam ME2   &  Johnson                 &  35 &  36 &  31      & 53771 - 53922\tablenotemark{a} \\
FLWO 1.2 m                       &  Keplercam        &  SDSS                    & 64\tablenotemark{b}&&& 53663\tablenotemark{c} - 53898 \\
Tenagra 0.8 m                    &  SITe based       &  Johnson/Cousins         &  13 &  13 &  13      & 53726 - 53876 \\
Torino 1.05 m                    &  Loral CCD        &  Cousins                 &  4  &     &          & 53757 - 53838 \\
WIYN 0.9 m                       &  S2KB             &  Harris                  &     &  19 &  19      & 53842 - 53872 \\
\enddata

\tablenotetext{a}{The B-band data from Bradford were taken between MJDs 53771-53896.}
\tablenotetext{b}{These data were taken in the r-band}
\tablenotetext{c}{Data-taking on Mrk\,421 at FLWO began before the gamma-ray data-taking.}

\end{deluxetable}

\begin{deluxetable}{lcccccc}
\tablewidth{0pt}

\tablecaption{\label{TAB:FVAR-FPP} The fractional rms variability
amplitude, F$_{var}$, and the point-to-point fractional variability
amplitude, F$_{pp}$ for each of the wavebands explored in this paper.}

\tablehead{        & \colhead{Energy Range}& \colhead{No.}    & \colhead{\% Nights}            & \colhead{Mean Flux\tablenotemark{b}} &                     & \\
\colhead{Waveband} & \colhead{(eV)}        & \colhead{Nights} & \colhead{$>$3$\sigma$\tablenotemark{a}}& \colhead{(Min/Max)}          & \colhead{F$_{var}$} & \colhead{F$_{pp}$}}
\startdata
{\it{Radio}}&                         &     &      &       &       &       \\
4.8 GHz     & 1.8 - 2.2 x 10$^{-5}$   & 12  & 100  & 6.46 (5.8/7.5) x 10$^{-1}$     & 0.043 & 0.066 \\
8.0 GHz     & 3.0 - 3.6 x 10$^{-5}$   & 16  & 93.8 & 5.29 (1.5/9.9) x 10$^{-1}$     & 0.266 & 0.330 \\
14.5 GHz    & 5.2 - 6.8 x 10$^{-5}$   & 22  & 86.4 & 4.94 (3.3-9.0) x 10$^{-1}$     & 0.152 & 0.134 \\
37.0 GHz    & 1.5 - 1.6 x 10$^{-4}$   & 31  & 80.7 & 3.46 (1.4/5.3) x 10$^{-1}$     & N/A   & N/A   \\
{\it{Optical}} &                      &     &      &       &       &       \\
R-band      & 1.7 - 2.3               & 97  & 100  & 13.0 (12.6/13.3)               & 0.132 & 0.043 \\
V-Band      & 2.1 - 2.5               & 46  & 100  & 13.4 (13.0/13.7)               & 0.135 & 0.045 \\
B-Band      & 2.6 - 3.3               & 45  & 100  & 13.8 (13.1/14.2)               & 0.187 & 0.065 \\
{\it{X-ray}}&                         &     &      &       &       &       \\
XRT         & 0.2 - 10 x 10$^3$       & 24  & 100  &  47.2 (32.1/78.5)              & 0.269 & 0.175 \\
ASM         & 2 - 10  x 10$^3$        & 256 & 81.6 &  1.99 (-0.38/5.68)             & 0.443 & 0.227 \\
PCA         & 3 - 25  x 10$^3$        & 75  & 100  &  5.18 (78.3/0.14) x 10$^{-10}$ & 0.529 & 0.273 \\
BAT         & 15 - 50  x 10$^3$       & 234 & 42.3 &  2.72 (-6.15/14.4) x 10$^{-3}$ & 0.993 & 0.360 \\
{\it{Gamma ray}} &                    &     &      &       &       &       \\
Whipple     & 0.2 - 10 x 10$^{12}$    & 80  & 66.3 & 1.04 (-0.05/2.59)              & 0.511 & 0.246 \\
\enddata
\tablenotetext{a}{This is the percentage of nights on which a $>$3$\sigma$ detection was observed.}
\tablenotetext{b}{The mean, minimum and maximum nightly-averaged
signal observed in each waveband during this campaign. The flux units
are the same as those quoted in Figure~\ref{FIG:ENTIRE-DATASET}.}
\end{deluxetable}

\begin{deluxetable}{ll}
\tablewidth{0pt} \tablecaption{\label{TAB:PARAMETERS} The model
parameters used for fitting the broadband SED of Mrk\,421, shown in
Figures~\ref{FIG:SED-MODEL1} and ~\ref{FIG:SED-MODEL2}.}

\tablehead{\colhead{Parameter} & \colhead{Value}}
\startdata
Doppler Factor & 90 \\
Magnetic field in the bulk frame & 0.12 Gauss \\
Electron energy density compared with equipartition value & 0.3 \\
Electron spectral index & 1.5 \\
Size of emission region & 70R$_{Sch}$ \\
Electron mean-free-path compared with gyroradius (Bohm limit) & 20 \\
Cooling time at maximum electron energy & 2.25 min \\
Acceleration time at maximum electron energy & 2.25 min \\
Shock velocity in bulk frame & 0.02c \\
Soft photon peak wavelength & 1.0$\mu$m \\
Mass of black hole & 10$^8$M$_\sun$ \\
\enddata

\end{deluxetable}

\clearpage

\clearpage

\clearpage

\clearpage

\clearpage

\clearpage

\clearpage

\clearpage

\clearpage

\clearpage

\clearpage

\clearpage

\clearpage

\clearpage

\clearpage

\clearpage

\clearpage

\clearpage

\clearpage

\clearpage

\end{document}